\documentclass[paper]{JHEP3}
\usepackage{amsmath,amssymb,amsfonts,epic,epsfig,psfrag}
\DeclareMathAlphabet{\mathW}{OT1}{pnc}{m}{n}
\DeclareMathAlphabet{\mathsfsl}{OMS}{pplcm}{b}{n}
\author{V.~A.~Fateev$^{1,2}$ and A.~V.~Litvinov$^{1,3}$\\
$^1$~Landau Institute for Theoretical Physics, 142432 Chernogolovka, Russia.\\
$^2$~Laboratoire de Physique Th\'eorique et Astroparticules, UMR5207 CNRS-UM2, Universit\'e
Montpellier~II, Pl.~E.~Bataillon, 34095 Montpellier, France\\
$^3$~NHETC, Department of Physics and Astronomy, Rutgers University,\\ Piscataway, NJ 08855-0849, USA}
\abstract{Two-dimensional $\mathfrak{sl}(n)$ quantum Toda field theory on a sphere is considered. This theory provides an important example of conformal field theory with higher spin symmetry. We derive the three-point correlation functions of the exponential fields if one of the three fields has a special form. In this case it is possible to write down and solve explicitly the differential equation for the four-point correlation function if the fourth field is completely degenerate. We give also expressions for the  three-point correlation functions in the cases, when they can be expressed in terms of known functions. The semiclassical and minisuperspace approaches in the conformal Toda field theory are studied and the results coming from these approaches are compared with the proposed analytical expression for the three-point correlation function. We show, that in the framework of semiclassical and minisuperspace approaches general three-point correlation function can be reduced to the finite-dimensional integral.}
\title{Correlation functions in conformal Toda field theory I}
\preprint{RUNHETC-2007-16\\PTA/07-42}
\dedicated{September 24, 2007}
\keywords{Integrable Field Theories, Conformal and W Symmetry}
\begin{document}
\section*{Introduction}\addcontentsline{toc}{section}{Introduction}
It is well known, that the problem of integrating over all metrics modulo diffeomorphism on a two-dimensional surface can be reduced to studying the quantum Liouville field theory \cite{Polyakov:1981rd}. First attempts to solve this theory were transformed into a beautiful and complete theory known as the two-dimensional conformal field theory \cite{Belavin:1984vu}. This theory is exactly solvable because the algebra of generators of the conformal symmetry in two dimensions, which governs the theory, is infinite dimensional. It coincides with the Virasoro algebra, which is the central extension of the algebra of vector fields on a circle. It is well known, that Virasoro algebra can be obtained as a quantum Drinfeld-Sokolov reduction of the affine $\mathfrak{\hat{sl}}(2)$ algebra. The same construction can be generalized to the case of general affine simple Lie algebra $\mathfrak{\hat{g}}$. As a result, after reduction one obtains associative  algebra ($\mathW{W}$ algebra), as an additional infinite dimensional symmetry consistent with conformal symmetry, i.~e. as a direct extension of the Virasoro algebra \cite{Zamolodchikov:1985wn}. Two-dimensional Toda field theory (TFT) associated with simple Lie algebra $\mathfrak{g}$ generalizes Liouville field  theory in a similar sense. The algebra of the generators of the symmetry, which governs TFT dynamics, coincides with $\mathW{W}$ algebra (associated with the corresponding Lie algebra $\mathfrak{g}$).

Due to its geometric interpretation \cite{Gervais:1993yh,Razumov:1993pv}, TFT is relevant in the investigation of the $W$ strings  and $W$ gravity (see for example Refs \cite{Pope:1992mi,West:1993np}). It also provides an important example of non-rational conformal field theory with higher spin symmetry and hence has its own interest. This higher spin symmetry manifests  itself also in rational conformal field theories, which describe the critical behavior of many interesting statistical systems, like for example $Z_n$ Ising models (parafermionic CFT \cite{Fateev:1985mm}), tricritical Ising and $Z_3$ Potts models, Ashkin-Teller models and also in the large variety of integrable statistical systems studied and solved in Refs \cite{Date:1987vf,Jimbo:1987ra}. The results derived in conformal Toda field theory can be applied to study of the short-distance asymptotics of the correlation functions in the massive integrable quantum field theory, which is known as affine Toda field theory, as well as to calculation of the vacuum expectation values of the exponential fields in this theory (see for example Refs \cite{Ahn:1999dz,Ahn:2000ki,Fateev:2000pi}). As conformal TFTs appear by the quantum Hamiltonian reduction of the WZNW models (see for example \cite{Feher:1992yx}), they can be also applied to study WZNW models with non-compact Lie algebras. 

There has been much progress in understanding Liouville field theory ($\mathfrak{sl}(2)$ TFT) and hence in the conformal field theory itself in the middle of 90's. In particular, the three-point correlation function was found explicitly for arbitrary exponential fields \cite{Dorn:1992at,Dorn:1994xn,Zamolodchikov:1995aa,Teschner:2003en}. Known three-point correlation functions, together with the fact, that conformal blocks are completely determined by the conformal symmetry, solve the conformal bootstrap problem in Liouville field theory.

Conformal Toda field theory is much more complicated than the Liouville field theory. One of the main reasons is that in TFT we need in general case more data to solve the conformal bootstrap problem \cite{Bowcock:1993wq}. In particular, this difficulty manifests itself in the fact that contrary to the Liouville field theory it is impossible to write down the differential equation for the four-point correlation function, which contains one completely degenerate and three arbitrary fields \cite{Bowcock:1993wq,Fateev:2005gs}. In Liouville field theory it allows to write down functional relation for the general three-point correlation function, which in some domain of parameters has a unique solution (see for example \cite{Teschner:1995yf}). In TFT this procedure fails (see section \ref{DIFF_EQN} for details). It means that other methods should be applied. It is interesting, that the difficulty of such a type appears already at the classical level, where the problem of finding the solution to the $\mathfrak{sl}(n)$ classical Toda equation for $n>2$ with three singular points (which determines so called "heavy" semiclassical limit of the three-point correlation function) reduces to the problem of studying Fuchsian ordinary differential equation with accessory parameters (see section \ref{CL-heavy}). Accessory parameters are absent in the Liouville case ($\mathfrak{sl}(2)$ TFT) and this is  the reason why this theory is rather simpler.

This paper is the first of two papers, devoted to study the correlation functions in the $\mathfrak{sl}(n)$ TFT, which can be found analytically (may be in terms of finite dimensional integrals). It is organized as follows: in section \ref{TFT} we briefly remind some basic facts about conformal TFT and propose an analytical expression for the three-point correlation function of the exponential fields in the case, when parameters of one of the field take the special values (see Eq \eqref{C}). We give also several another examples of correlation functions, which can be expressed in terms of known functions. In section \ref{DIFF_EQN} we present the derivation of the proposed three-point correlation function \eqref{C} by using the special properties of the operator algebra of degenerate fields. In sections \ref{CL-heavy} and \ref{CL-light}  the semiclassical analysis of the theory is developed.  In the section \ref{CL-heavy} we study the case, when all exponential fields in correlation function are "heavy" (i.~e. have parameters proportional to the opposite coupling constant) and in the section \ref{CL-light} we study the case, when all exponential fields are "light" (i.~e. have parameters proportional to the coupling constant). In section \ref{MSSL} we study the minisuperspace approach to the $\mathfrak{sl}(n)$ TFT. We show, that in the case of light exponential fields, as well as in the minisuperspace limit, three-point correlation function can be expressed in terms of finite dimensional integrals. In both cases,  semiclassical and minisuperspace asymptotic is in complete agreement with the proposed quantum results. The calculation details and useful formulae are given in the appendices.

In the second part of this paper \cite{Part-Deux} we will give more detailed description of the correlation functions in conformal TFT, which can be expressed in terms of finite dimensional Coulomb integrals. 
\section{Toda Field Theory}\label{TFT}
We start by recalling some basic facts and notions.
The Lagrangian of the $\mathfrak{sl}(n)$ conformal TFT has the form
\begin{equation}\label{Lagrangian}
    \mathcal{L}=\frac{1}{8\pi}(\partial_a\varphi)^2+
    \mu\sum_{k=1}^{n-1} e^{b(e_k,\varphi)},
\end{equation}
here $\varphi$ is the two-dimensional $(n-1)$ component scalar field $\varphi=(\varphi_1\dots\varphi_{n-1})$, $b$ is the dimensionless coupling constant, $\mu$ is the scale parameter called the cosmological constant and $(e_k,\varphi)$ denotes the scalar product, where vectors $e_k$ are the simple roots of the Lie algebra $\mathfrak{sl}(n)$ with the matrix of the scalar products $K_{ij}=(e_i,e_j)$ (Cartan matrix)
\begin{equation}\label{K}
     K_{ij}=
   \begin{pmatrix}
    \;2  & -1 & 0 & \hdotsfor{2} & 0\\
    -1 &  \;2 & -1 & \hdotsfor{2} & 0\\
    0 & -1 & \hdotsfor{4}\\
    \hdotsfor{4} & -1 & 0\\
    0 & \hdotsfor{2} & -1 & \;2 & -1\\
    0 & \hdotsfor{2} & 0 & -1 & \;2
   \end{pmatrix}.
\end{equation}
In the following we will use standart for the two-dimensional physics complex notations:
\begin{equation}
  z=x_1+ix_2,\;\;\bar{z}=x_1-ix_2,\;\;\;\;\partial=\frac{\partial}{\partial z}
  ,\;\;\bar{\partial}=\frac{\partial}{\partial\bar{z}}
\end{equation}
and introduce the notation for the measure
\begin{equation}
  d^2z=dx_1dx_2.
\end{equation}
Total normalization of the Lagrangian \eqref{Lagrangian} is chosen in such a way, that
\begin{equation}
   \varphi_i(z,\bar{z})\varphi_j(0,0)=
    -\delta_{ij}\log\vert z\vert^2+\dots\quad
   \text{at}\;z\rightarrow 0.
\end{equation}
In is useful to write TFT action explicitly in reference metric $\hat{g}_{ab}$ on a surface
\begin{equation}\label{TFT_Action}
  \mathcal{A}_{TFT}=\int\left(
  \frac{1}{8\pi}\hat{g}^{ab}(\partial_a\varphi,\partial_b\varphi)+
   \frac{(Q,\varphi)}{4\pi}\hat{R}+
   \mu\sum_{k=1}^{n-1} e^{b(e_k,\varphi)}
  \right)\sqrt{\hat{g}}\;d^2x,
\end{equation}
here $\hat{R}$ is the scalar curvature of the background metric.\footnote{Bellow we consider mainly the case of sphere, in order to avoid the problem with moduli. It is useful to choose the metric $\hat{g}_{ab}=\delta_{ab}$ everywhere except the north pole ($z=\infty$), where the curvature is located. Such a choice prescribes the asymptotic $\varphi=-Q\log\vert z\vert+\dots$ at $z\rightarrow\infty$.}
If the background charge $Q$ is related with the parameter $b$ as
\begin{equation}\label{Qandb}
    Q=\left(b+\frac{1}{b}\right)\rho
\end{equation}
with $\rho$ being a Weyl vector (half of the sum of all positive roots), then the theory \eqref{TFT_Action} is conformaly invariant.\footnote{More strictly, it becomes to be invariant under the combined Weyl transformation: $\hat{g}_{ab}\rightarrow\Omega(x)\hat{g}_{ab}$ and $\varphi\rightarrow\varphi-Q\log\Omega(x)$.} Moreover it ensures  higher-spin symmetry: there are $n-1$ holomorphic currents $\mathbf{W^k}(z)$ with the spins $k=2,3,\dots,n$, which are expressed through the field $\varphi$ via the Miura transformation \cite{Fateev:1987zh}
  \begin{equation}\label{WCurrents} 
    \prod\limits_{i=0}^{n-1}(q\partial+(h_{n-i},\partial\varphi)
    )=\sum_{k=0}^{n}\mathbf{W^{n-k}}(z)(q\partial)^{k}, 
  \end{equation}
where 
\begin{equation}
  q=b+1/b
\end{equation}
and vectors $h_{k}$ are the weights of the first fundamental representation $\pi _{1}$ of the Lie algebra $\mathfrak{sl}(n)$ with the highest weight $\omega_{1}$ (first fundamental weight)
\begin{equation}
    h_{k}=\omega_{1}-e_{1}-\dots-e_{k-1}.
\end{equation} 
In particular, it follows from Eq \eqref{WCurrents}, that the currents $\mathbf{W^{0}}(z)=1,\;\mathbf{W^1}(z)=0$ and the current
\begin{equation*}
     \mathbf{W^2}(z)=T(z)=-
     \frac{1}{2}(\partial\varphi)^{2}+(Q,\partial^2\varphi) 
\end{equation*}
is the stress-energy tensor of the theory, which ensures local conformal invariance of TFT. The currents $\mathbf{W^k}(z)$ form  closed $\mathW{W_n}$ algebra, which contains as subalgebra the Virasoro algebra with the central charge
\begin{equation}
    c=n-1+12Q^2=(n-1)(1+n(n+1)(b+b^{-1})^2).
\end{equation}
This $\mathW{W_n}$ algebra represents only the chiral part of the algebra of generators of the symmetry, which governs the theory. Total algebra is a tensor product of the both holomorphic and antiholomorphic algebras 
$\mathW{W_n}\otimes\overline{\mathW{W}}_{\mathW{n}}$. 

Basic objects of conformal Toda field theory are the exponential fields parameterized by a $(n-1)$ component vector parameter $\alpha$
\begin{equation}\label{field}
    V_{\alpha}=e^{(\alpha,\varphi)},
\end{equation}
which are the spinless primary fields. They have the simple operator product expansion (OPE) with the currents $\mathbf{W^k}(\xi)$. Namely,
\begin{equation}
    \mathbf{W^k}(\xi)V_{\alpha}(z,\bar{z})=
    \frac{w^{(k)}(\alpha)V_{\alpha}(z,\bar{z})}{(\xi-z)^k}+\dots,
\end{equation}
here $\dots$ means the contribution of less singular terms. Similar OPE's with antiholomorphic currents 
$\overline{\mathbf{W}}^{\mathbf{k}}(\bar{\xi})$ are also valid. The quantum numbers $w^{(k)}(\alpha)$ possess the symmetry under the action of the Weyl group $\mathsfsl{W}$ of the Lie algebra $\mathfrak{sl}(n)$ (which is generated by reflections in the hyperplanes perpendicular to the simple roots $e_k$) \cite{Fateev:1987zh}
\begin{equation}\label{WeylSymmetry}
     w^{(k)}(\alpha)=w^{(k)}_{\hat{s}}(\alpha)\equiv
     w^{(k)}(Q+\hat{s}(\alpha-Q)),\quad\hat{s}\in\mathsfsl{W}.
\end{equation}
In particular,
\begin{equation}
    w^{(2)}(\alpha)=\Delta(\alpha)=\frac{(\alpha,2Q-\alpha)}{2}
\end{equation}
is the conformal dimension of the field $V_{\alpha}$. Equation \eqref{WeylSymmetry} suggests the idea, that the fields related via the action of the Weyl group should coincide up to a multiplicative factor. One of the important properties of TFT is that it is really true
\begin{equation}\label{R_s}
    V_{Q+\hat{s}(\alpha-Q)}=R_{\hat{s}}(\alpha)V_{\alpha}\,,  
\end{equation}
where $R_{\hat{s}}(\alpha)$ is the reflection amplitude, which was found in \cite{Fateev:2001mj}
\begin{equation}\label{ReflAmp}
   \begin{gathered}
    R_{\hat{s}}(\alpha)=A(Q+\hat{s}(\alpha-Q))/A(\alpha),\\
    A(\alpha)=(\pi\mu\gamma(b^2))^{\frac{(\alpha-Q,\rho)}{b}}
    \prod_{e>0}\Gamma(1-b(\alpha-Q,e))
    \Gamma(-b^{-1}(\alpha-Q,e)).
   \end{gathered}
\end{equation}
In Eq \eqref{ReflAmp} the product goes over all positive roots.

Multipoint correlation functions of the exponential fields 
\begin{equation}\label{CorrFunc}
    \langle V_{\alpha_1}(z_1,\bar{z}_1)
    \dots V_{\alpha_l}(z_l,\bar{z}_l)\rangle=
    \int[\mathcal{D}\varphi]e^{-\mathcal{A}_{TFT}}
    V_{\alpha_{1}}(z_1,\bar{z}_1)\dots V_{\alpha_l}(z_l,\bar{z}_l)
\end{equation}
are the main objects of the theory. One of the most important problems in TFT is to find  these quantities. This problem is nontrivial due to the exponential interaction term in the Lagrangian \eqref{Lagrangian}. One can try naively to explore perturbation theory in cosmological constant $\mu$. However, pertubatively, correlation functions \eqref{CorrFunc} are equal to zero unless the "on-shell" condition
\begin{equation}\label{on-shell}
  \sum_{j=1}^l\alpha_j+b\sum_{k=1}^{n-1}s_ke_k=2Q
\end{equation}
with some non-negative integers $s_k$ is satisfied. Alternatively, one can perform zero mode integration \cite{Goulian:1990qr}. Namely, let us define a zero mode $\varphi_0$ of the field $\varphi$:
\begin{math}
       \varphi=\varphi_0+\tilde{\varphi}
\end{math} 
with the condition that
\begin{math}
       \int d^2\,x\;\tilde{\varphi}=0.
\end{math}
The integral in Eq \eqref{CorrFunc} over the zero mode $\varphi_0$ can be transformed to the Euler integral. As a result, after integration we obtain 
\begin{multline}\label{GoulLieFormula}
    \langle V_{\alpha_1}(z_1,\bar{z}_1)\dots V_{\alpha_l}(z_l,\bar{z}_l)\rangle=\\=
    \frac{1}{b^{n-1}}
    \int[\mathcal{D}\tilde{\varphi}]e^{-S_0}
    \left[\,\prod_{k=1}^{n-1}\Gamma(-s_k)
    \left(\mu\int e^{b(e_k,\tilde{\varphi})}\right)^{s_k}\right]
    V_{\alpha_1}(z_1,\bar{z}_1)\dots V_{\alpha_l}(z_l,\bar{z}_l),
\end{multline}
with 
\begin{equation*}
   s_k=\frac{(2Q-\sum\alpha_j,\omega_k)}{b}.
\end{equation*}
Here vectors $\omega_k$ are the fundamental weights of the Lie algebra $\mathfrak{sl}(n)$\footnote{They are defined as a dual basis to the simple roots $(e_i,\omega_j)=\delta_{ij}$.}. Integration in Eq \eqref{GoulLieFormula} is performed in the theory of a free massless $(n-1)$ component scalar field with the action 
\begin{equation*}
 S_0=\frac{1}{8\pi}\int(\partial_a\varphi)^2d^2x.
\end{equation*}
Equation \eqref{GoulLieFormula} has no meaning if all numbers $s_k$ are general. However in the resonance situation, when all numbers $s_k$ are non-negative integers, the gamma functions in the right hand side of Eq \eqref{GoulLieFormula} have  simple poles in each of the variables $(2Q-\sum\alpha_j,\omega_k)$ for $k=1,\dots,n-1$ and we can treat the main residue in these poles (the residue in each of these poles) as the corresponding free field integrals. Namely
\begin{multline}\label{GoulLieFormula1}
    \underset{(2Q-\sum\alpha_j,\omega_1)=bs_1}{\text{res}}\dots
    \underset{(2Q-\sum\alpha_j,\omega_{n-1})=bs_{n-1}}{\text{res}}
    \langle V_{\alpha_1}(z_1,\bar{z}_1)\dots 
    V_{\alpha_l}(z_l,\bar{z}_l)\rangle=\\=
    \frac{(-\mu)^{s_1+\dots s_{n-1}}}{s_1!\dots s_{n-1}!}
    \langle V_{\alpha_1}(z_1,\bar{z}_1)\dots 
    V_{\alpha_l}(z_l,\bar{z}_l)
    (\mathcal{Q}_1)^{s_1}\dots (\mathcal{Q}_{n-1})^{s_{n-1}}\rangle_0,
\end{multline}
here $\langle\dots\rangle_0$ means average over the free massless fields.
In Eq \eqref{GoulLieFormula1} we have introduced the notations for the so called screening charges
\begin{equation}\label{screening}
  \mathcal{Q}_{k}=\int e^{b(e_k,\varphi_k)}d^2\xi,\;\;k=1,\dots,n-1.
\end{equation}
Correlation function in the r.~h.~s. of Eq \eqref{GoulLieFormula1} can be calculated using the Wick rules in the free field theory together with integration over the position of all screening fields $e^{b(e_k,\varphi)}$.

Equation \eqref{GoulLieFormula1}, which was obtained from the classical arguments, modifies in quantum case. Namely, if the screening conditions
\begin{equation}\label{General-Scr}
   (2Q-\sum_{j=1}^l\alpha_j,\omega_k)=bs_k+b^{-1}\tilde{s}_k
\end{equation}
are satisfied for any two sets $(s_1,\dots,s_{n-1})$ and $(\tilde{s}_1,\dots,\tilde{s}_{n-1})$ of non-negative integers, then the correlation function \eqref{CorrFunc} admits a pole in each of the variable $(2Q-\sum\alpha,\omega_k)$ with the main residue being expressed in terms of free field correlation function 
\begin{multline}\label{General-GL}
    \underset{(2Q-\sum\alpha_i,\omega_1)=bs_1+b^{-1}\tilde{s}_1}{\text{res}}\dots
    \underset{(2Q-\sum\alpha_i,\omega_{n-1})=bs_{n-1}+b^{-1}\tilde{s}_{n-1}}{\text{res}}
    \langle V_{\alpha_1}(z_1,\bar{z}_1)\dots 
    V_{\alpha_l}(z_l,\bar{z}_l)\rangle=\\=
    \frac{(-\mu)^{s_1+\dots s_{n-1}}}{s_1!\dots s_{n-1}!}
    \frac{(-\tilde{\mu})^{\tilde{s}_1+\dots\tilde{s}_{n-1}}}{\tilde{s}_1!\dots\tilde{s}_{n-1}!}
    \times\\\times
    \langle V_{\alpha_1}(z_1,\bar{z}_1)\dots 
    V_{\alpha_l}(z_l,\bar{z}_l)
    (\mathcal{Q}_1)^{s_1}\dots (\mathcal{Q}_{n-1})^{s_{n-1}}
(\tilde{\mathcal{Q}}_1)^{\tilde{s}_1}\dots (\tilde{\mathcal{Q}}_{n-1})^{\tilde{s}_{n-1}}\rangle_0.
\end{multline}
In Eq \eqref{General-GL} we have introduced the notation for the dual screening charges
\begin{equation}
  \tilde{\mathcal{Q}}_{k}=\int e^{b^{-1}(e_k,\varphi_k)}d^2\xi
\end{equation}
and for the dual cosmological constant
\begin{equation}\label{mu-dual}
   \tilde{\mu}=\frac{1}{\pi\gamma(1/b^2)}\left(
   \pi\mu\gamma(b^2)\right)^{1/b^2}.
\end{equation}  
Operators $\mathcal{Q}_k$ and $\tilde{\mathcal{Q}}_k$ have an important property, that they commute with
all generators of the both holomorphic and antiholomorphic $\mathW{W}$ algebras. In this paper we will consider for simplicity the case, when all numbers $\tilde{s}_k=0$. It is reasonable to suppose, that the screening condition \eqref{General-Scr}  defines up to the Weyl transformation \eqref{WeylSymmetry} all possible poles of the correlation function \eqref{CorrFunc}, as a function of the parameters $\alpha_k$. One should emphasize, that a simple pole in the correlation function \eqref{CorrFunc} appears if at least one screening condition \eqref{General-Scr} is satisfied.

Knowledge of two-point and three-point correlation functions of the primary fields $V_{\alpha}$  is the first step  for the calculation of the higher multipoint correlation functions of the theory. In the $\mathfrak{sl}(2)$ case (Liouville field theory), this knowledge together with the statement, that conformal blocks are completely determined by the conformal symmetry, allows us, in principle, to compute any multipoint correlation functions in this theory \cite{Belavin:1984vu}. In the $\mathfrak{sl}(n)$ TFT case for $n>2$ the situation is more complicated and we need more data (see for example \cite{Bowcock:1993wq}).

Two-point correlation function in TFT normalized by the condition
\begin{equation}
  \langle V_{\alpha}(z)V_{2Q-\alpha}(0)\rangle=|z|^{-4\Delta(\alpha)}.
\end{equation}
All other non-zero two-point correlation functions can be obtained from this correlation function by the Weyl reflection \eqref{R_s}. For example
\begin{equation}\label{MaxRefl-def}
  \langle V_{\alpha}(z)V_{\alpha^{*}}(0)\rangle=\frac{R^{-1}(\alpha)}{|z|^{4\Delta(\alpha)}},
\end{equation}
here $R(\alpha)$ is the maximal refrection amplitude defined as
\begin{equation}\label{MaxRefl}
  R(\alpha)=\frac{A(2Q-\alpha)}{A(\alpha)}
\end{equation}
with $A(\alpha)$ given by Eq \eqref{ReflAmp} and conjugated vector parameter $\alpha^{*}$ defined as
\begin{equation}\label{Conj-charge}
  (\alpha,e_k)=(\alpha^{*},e_{n-k}).
\end{equation}

Much more complicated object -- three-point correlation function has  standart coordinate dependence due to the conformal invariance of the theory
\begin{equation}\label{Ctotal}
    \langle V_{\alpha_1}(z_1,\bar{z}_1)
    V_{\alpha_2}(z_2,\bar{z}_2)V_{\alpha_3}(z_3,\bar{z}_3)\rangle=\frac
    {C(\alpha_1,\alpha_2,\alpha_3)}
    {\vert z_{12}\vert^{2(\Delta_1+\Delta_2-\Delta_3)}
    \vert z_{13}\vert^{2(\Delta_1+\Delta_3-\Delta_2)}
    \vert z_{23}\vert^{2(\Delta_2+\Delta_3-\Delta_1)}}.
\end{equation}
All non-trivial information about the operator algebra of the primary fields $V_{\alpha}$ of the model is encoded in the constants $C(\alpha_1,\alpha_2,\alpha_3)$. According to Eq \eqref{GoulLieFormula1} if the parameters $\alpha_1$, $\alpha_2$ and $\alpha_3$ satisfy the screening condition 
\begin{equation*}
  \alpha_1+\alpha_2+\alpha_3+bs_1e_1+\dots+bs_{n-1}e_{n-1}=2Q,
\end{equation*}
function $C(\alpha_1,\alpha_2,\alpha_3)$ will have a pole in each of the variables $(2Q-\alpha_1-\alpha_2-\alpha_3,\omega_k)$ and
we can define the main residue in these poles in terms of Coulomb integral\footnote{In the integral \eqref{GoulLieFormula1} we can set using the projective invariance $z_1=0$, $z_2=1$ and $z_3=\infty$.}
\begin{multline}\label{ResCond}
    \underset{(2Q-\sum\alpha_i,\omega_1)=bs_1}{\text{res}}\dots
    \underset{(2Q-\sum\alpha_i,\omega_{n-1})=bs_{n-1}}{\text{res}}
     C(\alpha_1,\alpha_2,\alpha_3)=\\=
     (-\pi\mu)^{s_1+\dots+s_{n-1}}I_{s_1\dots s_{n-1}}(\alpha_1,\alpha_2,\alpha_3) 
\end{multline}
with
\begin{multline}\label{I}
    I_{s_1\dots s_{n-1}}(\alpha_1,\alpha_2,\alpha_3)=
    \int d\mu_{s_1}(t_1)\dots\,d\mu_{s_{n-1}}(t_{n-1})
    \times\\\times
    \prod_{k=1}^{n-1}\mathcal{D}_{s_k}^{-2b^2}(t_k)\prod_{j=1}^{s_k}
    \vert t_k^{(j)}\vert^{-2b(\alpha_1,e_k)}
    \vert t_k^{(j)}-1\vert^{-2b(\alpha_2,e_k)}
    \prod_{l=1}^{n-2}\mathcal{A}_{s_ls_{l+1}}^{b^2}(t_l,t_{l+1}),
\end{multline}
here $t_k^{(j)}$ is the coordinate of the $j$-th screening field $e^{b(e_k,\varphi)}$
and quantities $\mathcal{D}_{s_k}(t_k)$  and $\mathcal{A}_{s_ls_m}(t_l,t_m)$ for $l\neq m$ are defined as
\begin{equation}\label{D-A}
    \mathcal{D}_{s_k}(t_k)=\prod_{i<i'}^{s_k}|t_k^{(i)}-t_k^{(i')}|^{2}\qquad\text{and}\qquad
    \mathcal{A}_{s_ls_m}(t_l,t_m)=\prod_{i=1}^{s_l}\prod_{i'=1}^{s_m}|t_l^{(i)}-t_m^{(i')}|^2.
\end{equation}
Throughout this paper we use the notation for the measure of integration
\begin{equation}
 d\mu_{s_k}(t_k)=\frac{1}{\pi^{s_k}s_k!}\prod_{i=1}^{s_k}d^2t_k^{(i)}.
\end{equation}

In the case of algebra $\mathfrak{sl}(2)$ Coulomb integral \eqref{I} is known also as two-dimensional generalization of Selberg integral. It  can be calculated explicitly in terms of $\Gamma$-functions \cite{Selberg,Dotsenko:1984nm,Dotsenko:1984ad} (see also Refs \cite{Fateev:2006:JETP,Fateev:2007:TMPH}). Unfortunately, it is not clear  how to calculate integral $I_{s_1\dots s_{n-1}}(\alpha_1,\alpha_2,\alpha_3)$ for arbitrary parameters $\alpha_k$ in the case of general $n>2$, but if one of the parameters $\alpha_k$ satisfy the special condition, for example
\begin{equation}\label{CrucialCondition}
  \alpha_3=\varkappa\omega_{n-1},
\end{equation}
then the integral \eqref{I} can be carried out explicitly in terms of $\Gamma$-functions (see appendix \ref{AppIntegrals}). Namely, the integral $I_{s_1\dots s_{n-1}}(\alpha_1,\alpha_2,\varkappa\omega_{n-1})$ is non-zero only if $s_1\leq s_2\leq\dots\leq s_{n-1}$.
In order to write down an answer we define an auxiliary function
\begin{equation*}
    R_k^l=\prod_{i=1}^{l}\gamma(-ib^2)\prod_{j>k}^n
    \gamma(b(Q-\alpha_1,h_j-h_k)-ib^2)
    \gamma(b(Q-\alpha_2,h_j-h_k)-ib^2),
\end{equation*}
with 
\begin{equation}
   \gamma(x)=\frac{\Gamma(x)}{\Gamma(1-x)}.
\end{equation}
Integral \eqref{I} equals in this case (one has to remember, that parameters $\alpha_1$, $\alpha_2$ and $\varkappa$ are subject to the condition $\alpha_1+\alpha_2+\varkappa\omega_{n-1}=2Q-bs_1e_1-\dots-bs_{n-1}e_{n-1}$)
\begin{multline}\label{I-equals}
    I_{s_1\dots s_{n-1}}(\alpha_1,\alpha_2,\varkappa\omega_{n-1})=\\=
    \left[\frac{-1}{\gamma(-b^2)}\right]^{s_1+\dots+s_{n-1}}
     \prod_{j=0}^{s_{n-1}}\left[\frac{1}{\gamma(b\varkappa+jb^2)}\right]
     R_1^{s_1}R_2^{s_2-s_1}\dots R_{n-1}^{s_{n-1}-s_{n-2}}. 
\end{multline}
It is easy to check, that function
\begin{multline}\label{C}
    C(\alpha_1,\alpha_2,\varkappa\omega_{n-1})=
    \left[\pi\mu\gamma(b^2)b^{2-2b^2}\right]^
    {\frac{(2Q-\sum\alpha_i,\rho)}{b}}\times\\\times
    \frac{\left(\Upsilon(b)\right)^{n-1}\Upsilon(\varkappa)
    \prod\limits_{e>0}\Upsilon\Bigl((Q-\alpha_1,e)\Bigr)
    \Upsilon\Bigl((Q-\alpha_2,e)\Bigr)}
    {\prod\limits_{ij}\Upsilon\Bigl(\frac{\varkappa}{n}+
    (\alpha_1-Q,h_i)+(\alpha_2-Q,h_j)\Bigr)},
\end{multline}
which was proposed in \cite{Fateev:2005gs}, satisfies the condition \eqref{ResCond} at this special case. Here $\Upsilon(x)$ is the entire selfdual function (with respect to transformation $b\rightarrow1/b$), which was defined in \cite{Zamolodchikov:1995aa} by the integral representation
\begin{equation}
    \log\Upsilon(x)=\int_{0}^{\infty}\frac{dt}{t}
    \left[\left(\frac{b+b^{-1}}{2}-x\right)^2e^{-t}-\frac
    {\sinh^2\left(\frac{b+b^{-1}}{2}-x\right)\frac{t}{2}}
    {\sinh\frac{bt}{2}\sinh\frac{t}{2b}} 
    \right].
\end{equation} 
This function satisfies functional relations
\begin{equation}
  \begin{aligned}
     &\Upsilon(x+b)=\gamma(bx)b^{1-2bx}\Upsilon(x),\\
     &\Upsilon(x+1/b)=\gamma(x/b)b^{2x/b-1}\Upsilon(x).
  \end{aligned}
\end{equation} 
and in fact is completely determined by them for the general real values of the parameter $b$ up to a multiplicative constant, which is fixed by the condition
\begin{equation*}
  \Upsilon\left(\frac{b+b^{-1}}{2}\right)=1.
\end{equation*}
This function was firstly introduced by Barnes \cite{Barnes}, as a generalization of ordinary Gamma function and  in the  semiclassical limit ($b\rightarrow0$) it has an asymptotic
\begin{equation}
   \frac{\Upsilon(by)}{\Upsilon(b)}\rightarrow\frac{b^{1-y}}{\Gamma(y)}\;\;
   \text{as}\;\;b\rightarrow0.
\end{equation}
One can easily check, that the  correlation function \eqref{C} is consistent with the reflection identification of the exponential fields \eqref{R_s}. Due to the symmetry reason, formula similar to \eqref{C}, but with $\alpha_3=\varkappa\omega_1$ is also valid\footnote{One has to change $h_k\rightarrow h_k^{*}=-h_{n+1-k}$ in \eqref{C}.}. Note that the condition \eqref{CrucialCondition} is crucial at this point and the general formula for the three-point correlation function is much more complicated.

One of the important sets of fields in TFT form so called completely degenerate fields \cite{Fateev:1987zh}. Completely degenerate fields  $V_{\alpha}$ in TFT are parameterized by two highest weights $\Omega_1$ and $\Omega_2$ of the finite dimensional representations of the Lie algebra $\mathfrak{sl}(n)$ and correspond to the value of the parameter $\alpha$ (up to Weyl transformation \eqref{WeylSymmetry})
\begin{equation}\label{Completely_degenerate}
  \alpha=-b\Omega_1-\frac{1}{b}\Omega_2.
\end{equation}
These fields posses an important property that in their operator product expansion with general primary field $V_{\alpha}$ appear only a finite number of primary fields $V_{\alpha'}$ with their descendant fields
\begin{equation}\label{Fusion_completely_degenerate}
   V_{-b\Omega_1-b^{-1}\Omega_2}V_{\alpha}=
   \sum_{s,p}C_{-b\Omega_1-b^{-1}\Omega_2,\alpha}^
   {\alpha'_{sp}}
   \left[V_{\alpha'_{sp}}\right],
\end{equation}
here by square brackets we denote the contribution of the descendant fields and introduce the parameter $\alpha'_{sp}$ as
\begin{equation}\label{Fusion_completely_degenerate-def}
  \alpha'_{sp}=\alpha-bh_s^{\Omega_1}-b^{-1}h_p^{\Omega_2}.
\end{equation}
In Eq \eqref{Fusion_completely_degenerate-def} $h_s^{\Omega}$ are the weights of the representation $\Omega$ and $C_{-b\Omega_1-b^{-1}\Omega_2,\alpha}^{\alpha'_{sp}}$ denotes the structure constant of the operator algebra. During this paper we will consider for simplicity the case $\Omega_2=0$. 

General structure constant of OPE $C_{\alpha_1,\alpha_2}^{\alpha_3}$
defined as
\begin{equation}\label{Structure_Constant}
  C_{\alpha_1,\alpha_2}^{\alpha_3}\overset{\text{def}}{=}C(\alpha_1,\alpha_2,2Q-\alpha_3)=
  R(\alpha_3)C(\alpha_1,\alpha_2,\alpha_3^{*}),
\end{equation}
where $R(\alpha_3)$ is the maximal reflection amplitude given by Eq \eqref{MaxRefl} and conjugated parameter $\alpha_3^{*}$ is defined by Eq \eqref{Conj-charge}\footnote{We remind, that parameters $\alpha^{*}$ and $2Q-\alpha$ are connected via Weyl transformation \eqref{MaxRefl}.}.
Strictly speaking, structure constant with completely degenerate field defined by Eq \eqref{Structure_Constant} as
\begin{equation}\label{Structure_Constant-1}
 C_{-b\Omega_1,\alpha}^{\alpha-bh_s^{\Omega_1}}=C(-b\Omega_1,\alpha,2Q-\alpha+bh_s^{\Omega_1})
\end{equation}
will be infinite because general weight $h_s^{\Omega_1}$ of the representation $\Omega_1$ has a form
\begin{equation}
  h_s^{\Omega_1}=\Omega_1-\sum_{j=1}^{n-1}s_je_j
\end{equation}
with some non-negative integers $s_j$ and hence the sum of all parameters in the three-point correlation function in the r.~h.~s. of Eq \eqref{Structure_Constant-1} satisfies the screening condition \eqref{on-shell}. In this case one should treat the structure constant $C_{-b\Omega_1,\alpha}^{\alpha-bh_s^{\Omega_1}}$ as the main residue of the corresponding three-point correlation function. This residue is given by the Coulomb integral \eqref{I}. Namely
\begin{equation}\label{Structure_Constant_degenerate}
  C_{-b\Omega_1,\alpha}^{\alpha-bh_s^{\Omega_1}}=(-\pi\mu)^{s_1+\dots+s_{n-1}}I_{s_1\dots s_{n-1}}
     (-b\Omega_1,\alpha,2Q-\alpha+bh_s^{\Omega_1}). 
\end{equation}
The complexity of these structure constants\footnote{We study these structure constants in more details in forthcoming paper \cite{Part-Deux}.} depend drastically on the multiplicities of the corresponding weights $h_s^{\Omega_1}$.

To illustrate this fact we give here some basic structure constants, which can be expressed in terms of known functions. Let us consider, for example, the case $\Omega_1=\omega_k$ corresponding to the $k$-th fundamental representation of the Lie algebra $\mathfrak{sl}(n)$. We denote as $H_k$ the set of weights $h_s^{(k)}$  of the fundamental representation $\pi_k$ with highest weight $\omega_k$ ($h_s^{(k)}\in H_k$). Then the operator product expansion of the field $V_{-b\omega_k}$ with arbitrary field $V_{\alpha}$ due to Eq \eqref{Fusion_completely_degenerate} has a form
\begin{equation}\label{Fusion_completely_degenerate-k}
  V_{-b\omega_k}V_{\alpha}=\sum_{s}C_{-b\omega_k,\alpha}^{\alpha-bh_s^{(k)}}
  \left[ V_{\alpha-bh_{s}^{(k)}}\right].
\end{equation}
To describe the structure constants $C_{-b\omega_k,\alpha}^{\alpha-bh_s^{(k)}}$ in this expansion
we denote as $\mathcal{R}_{s}^{k}$ the set of positive roots $e$ such that $e+h_{s}^{(k)}$ $\in H_{k}.$ Then
\begin{equation}\label{StrConst_k}
    C_{-b\omega_k,\alpha}^{\alpha-bh_s^{(k)}}=
    \left(-\frac{\pi \mu }{\gamma (-b^{2})}\right)
     ^{(\omega_k-h_s^{(k)},\rho)}
     \prod\limits_{e\in 
     \mathcal{R}_{s}^{k}}\frac{\gamma\bigl(b(\alpha-Q,e)\bigr)}
     {\gamma\bigl(1+b^2+b(\alpha-Q,e)\bigr)}. 
\end{equation}
This result has been derived from the free field integral \eqref{Structure_Constant_degenerate} using the same technique, which was used  in appendix \ref{AppIntegrals} to derive Eq \eqref{I}.

Another interesting situation, when the integral \eqref{I} can be calculated exactly is the structure constants with degenerate field $V_{-be_{0}}$ (\begin{math}e_{0}=\sum_{k=1}^{n-1}e_k\end{math} is the maximal root corresponding to highest weight of adjoint representation). This operators plays a role of integrable perturbation of the theory, which moves conformal TFT to the massive affine TFT. The operator product expansion of the field $V_{-be_{0}}$ with general primary field $V_{\alpha}$ has a form
\begin{equation}\label{OPE_adj}
   V_{-be_{0}}V_{\alpha}=C_{-be_0,\alpha}^{\alpha}\left[ V_{\alpha}\right]
   +\sum_{e}C_{-be_0,\alpha}^{\alpha-be}\left[ V_{\alpha-be}\right],
\end{equation}
where the sum goes over all roots of $\mathfrak{sl}(n)$.
The diagonal structure constant $C_{-be_0,\alpha}^{\alpha}$ can be represented as
\begin{equation}\label{DiagStrConst}
   C_{-be_0,\alpha}^{\alpha}=\sum_{i=1}^{n}
   \prod\limits_{j\neq i}^{n}\frac{\pi\mu\gamma 
   \bigl(b(\alpha-Q,h_j-h_i)\bigr)}{\gamma (-b^{2})\gamma 
   \bigl(1+b^2+b(\alpha-Q,h_j-h_i)\bigr)}\mathcal{F}_{i}^{2}(\alpha), 
\end{equation}
where functions $\mathcal{F}_{i}(\alpha)$ can be expressed through the higher hypergeometric functions at unity $_{n}F_{n-1}(1)$ as
\begin{equation}
  \mathcal{F}_{i}(\alpha)=1+\sum_{k=1}^{\infty}
  \prod\limits_{j=1}^{n}
  \frac{(b(Q-\alpha,h_j-h_i)-b^2)_{k}}
  {(1+b(Q-\alpha,h_j-h_i))_{k}}, 
\end{equation}
where 
\begin{equation}
(x)_{k}=x(x+1)...(x+k-1).
\end{equation}
For the positive roots $e=h_j-h_i$ with $i>j$ the structure constant $C_{-be_0,\alpha}^{\alpha-be}$ is
given by the product of $\gamma$-functions
\begin{multline}\label{Str-positive}
    C_{-be_0,\alpha}^{\alpha-be}=\left(\frac{-\pi\mu}
    {\gamma(-b^{2})}\right)^{(n-i+j-1)}\times\\\times
    \prod_{k=1}^{j-1}\frac{\gamma\bigl(b(Q-\alpha,h_k-h_j)-b^2\bigr)}
    {\gamma\bigl(1+b(Q-\alpha,h_k-h_j)\bigr)}
    \prod_{k=i+1}^{n-1}\frac{\gamma\bigl(b(Q-\alpha,h_i-h_k)-b^2\bigr)}
    {\gamma\bigl(1+b(Q-\alpha,h_i-h_k)\bigr)}.
\end{multline}
While the structure constants for the negative roots can be expressed through the structure constants for the positive roots \eqref{Str-positive} as
\begin{equation}
   C_{-be_0,\alpha}^{\alpha+be}=R^{-1}(\alpha)R(\alpha+be)C_{-be_0,\alpha'}^{\alpha'-be},
\end{equation}
where $R(\alpha)$ is the maximal reflection amplitude \eqref{MaxRefl} and $\alpha'=\alpha+be$.

An important point should be emphasized here. It follows from Eq \eqref{DiagStrConst}, that  the structure constant $C_{-be_0,\alpha}^{\alpha}$ is expressed in terms of higher hypergeometric functions $_{n}F_{n-1}$ at unity \eqref{DiagStrConst}, while the structure constants $C_{-be_0,\alpha}^{\alpha-be}$ have more simple form and are expressed in terms of product of $\gamma$-functions \eqref{Str-positive}. The difference between these two cases is related with the fact that field with zero weight in the adjoint representation with highest weight $e_0$ appears with multiplicity $(n-1)$, while the weights corresponding to the roots $e$ appear with multiplicity equal to $1$. The same is true for the fundamental representation with the highest weight $\omega_k$, where all weights $h_s^k$ of this representation also appear with multiplicity $1$ and as a result the structure constants \eqref{StrConst_k} are expressed in terms of $\gamma$-functions. We see, that the fact that some weights have multiplicity more than one makes the situation more difficult. In the $\mathfrak{sl}(2)$ case (Liouville field theory) it does not happen because all weights appear with multiplicity one. 
\section{Differential equation}\label{DIFF_EQN}
Three-point correlation function \eqref{C}, which was derived in section \ref{TFT} by the calculation of the Coulomb integrals, can be obtained also from rather different arguments. The idea is to explore the associativity condition of the operator algebra and to use the special properties of degenerate fields. This approach was proposed in Ref \cite{Teschner:1995yf} in order to find the structure constants in the Liouville field theory ($\mathfrak{sl}(2)$ TFT). Here we will consider in details the case of $\mathfrak{sl}(3)$ TFT, as the next step of complexity.  

The chiral part of the algebra of symmetries in this case consists of two currents of the spin two and three\footnote{This basis of currents is slightly differs from the basis defined by Miura transformation \eqref{WCurrents}. Basis \eqref{WCurrents} is more convenient, because commutation relations of the $W$ algebra are bilinear. In Eq \eqref{currents} the current  $W$ is primary field with respect to Virasoro algebra and differs from the corresponding current in Eq \eqref{WCurrents} by adding term proportional to $T'$.}
\begin{equation}\label{currents}
    \mathbf{W^2}(z)=T(z)=
    \sum_{n=-\infty}^{\infty}\frac{L_n}{z^{n+2}}\qquad
    \text{and}\qquad
    \mathbf{W^3}(z)=W(z)=
    \sum_{n=-\infty}^{\infty}\frac{W_n}{z^{n+3}}.
\end{equation}
The Laurent componets $L_k$ and $W_k$ form closed $\mathW{W_3}$ algebra with the commutation relations  \cite{Zamolodchikov:1985wn,Fateev:1987vh}
\begin{subequations}\label{W3algebra}
\begin{equation}\label{LunderL}
    \left[L_n,L_m\right]=(n-m)L_{n+m}+\frac{c}{12}(n^3-n)
    \delta_{n,-m},
\end{equation}
\begin{equation}\label{WunderL}
    \left[L_n,W_m\right]=(2n-m)W_{n+m},
\end{equation}
\begin{multline}\label{WunderW}
    \left[W_n,W_m\right]=\frac{c}{3\cdot5!}(n^2-1)(n^2-4)n
    \delta_{n,-m}+\frac{16}{22+5c}(n-m)\Lambda_{n+m}+\\+
    (n-m)\left(\frac{1}{15}(n+m+2)(n+m+3)-\frac{1}{6}(n+2)(m+2)
    \right)
    L_{n+m},
\end{multline}
\end{subequations}
here
\begin{equation*}
    \Lambda_n=\sum_{k=-\infty}^{\infty}:L_kL_{n-k}:+\frac{1}{5}x_n
    L_n,
\end{equation*}
\begin{equation*}
    x_{2l}=(1+l)(1-l)\qquad x_{2l+1}=(2+l)(1-l).
\end{equation*}
This algebra is not Lie algebra due to the quadratic terms in the r.~h.~s. of Eq \eqref{WunderW}, however, as was noticed by A. Zamolodchikov \cite{Zamolodchikov:1985wn}, the Jacoby identities are satisfied. 

The operator product expansions of the holomorphic currents \eqref{currents} with the primary fields $V_{\alpha}$ has the form
\begin{equation}\label{OPE}
  \begin{aligned}
    &T(\xi)V_{\alpha}(z)=\frac{\Delta(\alpha)V_{\alpha}(z)}
    {(\xi-z)^2}+
    \frac{\partial V_{\alpha}(z)}{(\xi-z)}+\dots\\
    &W(\xi)V_{\alpha}(z)=\frac{w(\alpha)V_{\alpha}(z)}{(\xi-z)^3}+
    \frac{W_{-1}V_{\alpha}(z)}{(\xi-z)^2}+
    \frac{W_{-2}V_{\alpha}(z)}{(\xi-z)}+\dots
   \end{aligned}
\end{equation}
here\addtocounter{equation}{-1}
\begin{subequations}
\begin{equation}\label{delta}
  \Delta(\alpha)=\frac{(2Q-\alpha,\alpha)}{2}
\end{equation}
is the conformal dimension and
\begin{equation}\label{omega}
  w(\alpha)=i\sqrt{\frac{48}{22+5c}}\;
  (\alpha-Q,h_1)(\alpha-Q,h_2)(\alpha-Q,h_3)
\end{equation}
\end{subequations}
is the quantum number associated to the $W(z)$ current. Along this section we omit sometimes (where it is not important) the $\bar{z}$ dependence of the fields $V_{\alpha}$. In Eq \eqref{OPE} we introduce the notations $W_{-1}V_{\alpha}(z)$ and $W_{-2}V_{\alpha}(z)$ for the $W$ descendant fields. Using Eq \eqref{OPE} one can obtain  Ward identities
\begin{subequations}
\begin{align}\label{LWard}
     &\langle T(z)V_1(z_1)
     \dots V_N(z_N)\rangle=\sum_{k=1}^N
     \left(\frac{\Delta_k}{(z-z_k)^2}+\frac{\partial_k}{(z-z_k)}
     \right)
     \langle V_1(z_1)\dots V_N(z_N)\rangle,\\\label{WWard}
     &\langle W(z)V_1(z_1)\dots V_N(z_N)\rangle=\sum_{k=1}^N
     \left(\frac{w_k}{(z-z_k)^3}+\frac{W_{-1}^{(k)}}{(z-z_k)^2}+
     \frac{W_{-2}^{(k)}}{(z-z_k)}\right)
     \langle V_1(z_1)\dots V_N(z_N)\rangle.
\end{align}
\end{subequations}
Let us explain our notations. For example
\begin{equation*}
  W_{-1}^{(k)}\langle V_1(z_1)\dots V_N(z_N)\rangle
  \stackrel{\text{def}}{=}
  \langle V_1(z_1)\dots W_{-1}V_k(z_k)\dots V_N(z_N)\rangle.
\end{equation*}
One should emphasize, that contrary to the Virasoro generators operators $W_{-k}$ generally speaking  do not act on correlation functions as some differential operators. This important difference explains the essential complication, which appear in the analysis of the $\mathfrak{sl}(n)$ conformal TFT for $n>2$.

The transformation laws for the currents $T(z)$ and $W(z)$ under the holomorphic substitution $z\rightarrow f(z)$ have a form \footnote{$T(z)$ does  not transform like a tensor, but is shifted by the Schwartz derivative, which is defined as $\{f,z\}=f'''/f'-3/2(f''/f')^2$, while $W(z)$ is really a tensor as follows from Eq \eqref{WunderL}. In the case of $\mathfrak{sl}(n)$ algebra for $n>3$ it is also possible to choose currents $\mathbf{W^{k}}(z)$ is such a way, that they will be primary with respect to stress-energy tensor, i.~e. will transform like a tensors under the change of variables.}
\begin{equation}\label{TrLaw}
    T(z)\rightarrow\left(\frac{df}{dz}\right)^2T(f)+
    \frac{c}{12}\{f,z\},
   \qquad
    W(z)\rightarrow\left(\frac{df}{dz}\right)^3W(f).
\end{equation}
The condition, that infinity is a regular point leads to the following asymptotic condition for the 
currents $T(z)$  and $W(z)$
\begin{subequations}
 \begin{align}
    &T(z)\sim\frac{1}{z^4}\quad\text{at}\quad z\rightarrow\infty
    \label{DecreaseLawL}\\
\intertext{and} 
    &W(z)\sim\frac{1}{z^6}\quad\text{at}\quad z\rightarrow\infty.
    \label{DecreaseLawW}
  \end{align}
\end{subequations}
It follows from the asymptotic \eqref{DecreaseLawL} of the current $T(z)$, that correlation functions of the primary fields satisfy certain set of differential equations, which restrict their possible coordinate dependence  \cite{Belavin:1984vu}. In the particular cases of two and three points, correlation functions are completely determined by them up to a numerical factor. For the case of two-point correlation function 
these differential equations put the limitation on it. Namely, two-point correlation function
is non-zero only if the dimensions of two fields are equal. One can consider the asymptotic \eqref{DecreaseLawW} of the current $W(z)$ in a similar way. Applying \eqref{DecreaseLawW} to the Ward identity \eqref{WWard}, we obtain five algebraic equations, which also restrict possible form of the correlation functions. Let us illustrate, how does it work in the case of two-point correlation function. These five algebraic equations connect different correlation functions, which enter in Ward identity \eqref{WWard}. Namely, we obtain a system of equations
\begin{equation}
    \left(%
\begin{array}{ccccc}
  0 & 0 & 0 & 1 & 1 \\
  0 & 1 & 1 & z_1 & z_2 \\
  w_1+w_2 & 2z_1 & 2z_2 & z_1^2 & z_2^2 \\
  3(w_1z_1+w_2z_2) & 3z_1^2 & 3z_2^2 & z_1^3 & z_2^3 \\
  6(w_1z_1^2+w_2z_2^2) & 4z_1^3 & 4z_2^3 & z_1^4 & z_2^4 \\
\end{array}%
\right)
\left(%
\begin{array}{c}
  \langle V_1(z_1)\;V_2(z_2)\rangle \\
  \langle W_{-1}V_1(z_1)\;V_2(z_2)\rangle \\
  \langle V_1(z_1)\;W_{-1}V_2(z_2)\rangle \\
  \langle W_{-2}V_1(z_1)\;V_2(z_2)\rangle \\
  \langle V_1(z_1)\;W_{-2}V_2(z_2)\rangle \\
\end{array}%
\right)=0\;.
\end{equation}
This algebraic system has a non-zero solution if the determinant of the matrix above equals to zero for any points $z_1$ and $z_2$. A simple calculation leads to
\begin{equation}
  \det=-(w_1+w_2)(z_{12})^6.
\end{equation}
It means that the correlation function $\langle V_1(z_1)\;V_2(z_2)\rangle$ is zero unless $w_1=-w_2$. As a result, we obtain the following form of the two-point correlation function
\begin{equation}\label{2point}
\langle V_1(z_1,\bar{z}_1)V_2(z_2,\bar{z}_2)\rangle\sim
\frac{\delta_{\Delta_1,\Delta_2}\delta_{w_1,-w_2}}{|z_{12}|^{4\Delta_1}}
\end{equation}
Omitted multiplicative constant in \eqref{2point} depends only on the particular normalization of the fields.

To extract the information about the fusion rules it is reasonable to study completely degenerate representations of the $\mathW{W_3}$ algebra \eqref{W3algebra}. Namely, if parameters $(\Delta(\alpha),w(\alpha))$ corresponding to the field $V_{\alpha}$ take one of the four values
\begin{subequations}
\begin{equation}\label{deg1}
    \Delta=-\frac{4b^2}{3}-1\qquad
    w^2=-\frac{2\Delta^2}{27}\frac{5b+\frac{3}{b}}{3b+\frac{5}{b}},
\end{equation}
\begin{equation}\label{deg2}
    \Delta=-\frac{4}{3b^2}-1\qquad
    w^2=-\frac{2\Delta^2}{27}\frac{3b+\frac{5}{b}}{5b+\frac{3}{b}},
\end{equation}
\end{subequations}
or in terms of parameter $\alpha$ (modulo Weyl transformation \eqref{WeylSymmetry})
\begin{equation}\label{Alpha_deg}
   \alpha=-b\omega_k\qquad\text{or}\qquad
   \alpha=-\frac{1}{b}\omega_k\qquad\qquad
   k=1,2.
\end{equation}
Then this field exhibits three null-vectors \cite{Fateev:1987vh,Bajnok:1992nj,Bowcock:1992gt}
\begin{subequations}\label{null}
\begin{equation}\label{1null}
   \chi_1= \left(W_{-1}-\frac{3w}{2\Delta}L_{-1}\right)V_{\alpha}=0,
\end{equation}
\begin{equation}\label{2null}
    \chi_2=\left(W_{-2}-\frac{12w}{\Delta(5\Delta+1)}L_{-1}^2+
    \frac{6w(\Delta+1)}{\Delta(5\Delta+1)}L_{-2}\right)V_{\alpha}=0,
\end{equation}
\begin{equation}\label{3null}
    \chi_3=\left(W_{-3}-\frac{16w}{\Delta(\Delta+1)(5\Delta+1)}
    L_{-1}^3+\frac{12w}{\Delta(5\Delta+1)}L_{-1}L_{-2}+
    \frac{3w}{2\Delta}\frac{(\Delta-3)}{(5\Delta+1)}L_{-3}\right)
    V_{\alpha}=0.
\end{equation}
\end{subequations}
The next natural step is to investigate, how  equations \eqref{1null}, \eqref{2null} and \eqref{3null} put the limitations on the three-point correlation functions, i.~e. we want to define the  fusion  rules. Let us consider three-point correlation function $\langle V(z)V_1(z_1)V_2(z_2)\rangle$, where field $V(z)$ is degenerate field with parameter \eqref{Alpha_deg} and fields $V_1(z_1)$ and $V_2(z_2)$ are some arbitrary primary fields. In the  Ward identity \eqref{WWard} for this case participate seven functions:
\begin{equation}
  \langle V(z,\bar{z}) V_1 (z_1,\bar{z}_1)V_2(z_2,\bar{z}_2)\rangle
\end{equation}
and also six functions, which can be obtained by the application of the operators $W_{-1}$ and $W_{-2}$ to the fields $V$, $V_1$ and $V_2$. Due to conformal invariance, the coordinate dependence of the three-point correlation function is known explicitly
\begin{equation}\label{3}
    \langle V(z) V_1 (z_1)V_2 (z_2)\rangle\sim
    (z-z_1)^{(\Delta_2-\Delta_1-\Delta)}(z-z_2)^{(\Delta_1-
    \Delta_2-\Delta)}
    (z_1-z_2)^{(\Delta_1+\Delta_2-\Delta)}.
\end{equation}
Applying equations \eqref{1null} and \eqref{2null} to Eq \eqref{3} we can express correlation functions
$\langle W_{-1}V(z) V_1 (z_1)V_2(z_2)\rangle$ and $\langle W_{-2}V(z) V_1 (z_1)V_2(z_2)\rangle$ as
\begin{subequations}
\begin{equation}
    \langle W_{-1}V(z) V_1 (z_1)V_2
     (z_2)\rangle=-\frac{3w}{2\Delta}
    \left(\frac{\Delta+\Delta_1-\Delta_2}{(z-z_1)}+
    \frac{\Delta+\Delta_2-\Delta_1}{(z-z_2)}\right)\langle V(z) 
    V_1 (z_1)V_2 (z_2)\rangle
\end{equation}
and
\begin{multline}
   \langle W_{-2}V(z) V_1 (z_1)V_2
   (z_2)\rangle=\left[\frac{12w}{\Delta(5\Delta+1)}
   \left(\frac{(\Delta+\Delta_1-\Delta_2)(\Delta+\Delta_1-
   \Delta_2+1)}
   {(z-z_1)^2}\right.\right.+\\\left.+
   \frac{2(\Delta+\Delta_1-\Delta_2)(\Delta+\Delta_2-\Delta_1)}
   {(z-z_1)(z-z_2)}+\frac{(\Delta+\Delta_2-\Delta_1)
   (\Delta+\Delta_2-\Delta_1+1)}
   {(z-z_2)^2}\right)-\\-\left.
   \frac{6w(\Delta+1)}{\Delta(5\Delta+1)}
   \left(\frac{(2\Delta_1+\Delta-\Delta_2)}{(z-z_1)^2}+
   \frac{(2\Delta_2+\Delta-\Delta_1)}{(z-z_2)^2}-
   \frac{(\Delta_1+\Delta_2-\Delta)}{(z-z_1)(z-z_2)}\right)
   \right]\times\\\times
   \langle V(z) V_1 (z_1)V_2 (z_2)\rangle.
\end{multline}
\end{subequations}
Similar to the case of two-point correlation function we obtain five
equations, which follow from the asymptotic condition \eqref{DecreaseLawW}. The determinant of the corresponding matrix should be zero for any points $z$, $z_1$ and $z_2$. It gives the equation
\begin{equation}\label{fusion1}
12w(\Delta_1-\Delta_2)^2-3w(\Delta+1)(\Delta_1+\Delta_2)
+\Delta(5\Delta+1)(w_1+w_2)-4w\Delta(\Delta-1)=0
\end{equation}
We should take also into a account Eq \eqref{3null} and put $\langle\chi_3(z) V_1(z_1)V_2(z_2)\rangle=0$. As a result, we obtain the second algebraic equation
\begin{multline}\label{fusion2}
    32w(\Delta_1-\Delta_2)^3-12w(\Delta+1)(\Delta_1^2-\Delta_2^2)-
    w(15\Delta^2-18\Delta-1)
    (\Delta_1-\Delta_2)+\\
    +\Delta(\Delta+1)(5\Delta+1)(w_1-w_2)=0
\end{multline}
The equations \eqref{fusion1} and \eqref{fusion2} define the fusion rules in our model (one should fix parameters $(\Delta_1,w_1)$ and find admissible parameters $(\Delta_2,w_2)$) after that. If we parameterize $(\Delta_1,w_1)=(\Delta(\alpha),w(\alpha))$ and $(\Delta,w)=(\Delta(-b\omega_1),w(-b\omega_1))$, then three solutions to the equations \eqref{fusion1} and
\eqref{fusion2} are
\begin{equation}\label{Fusion}
  \Delta_2=\Delta(\alpha_1-bh_j)\qquad
  w_2=-w(\alpha_1-bh_j)\qquad
  j=1,2,3,
\end{equation}
where $\Delta(\alpha)$ and $w(\alpha)$ are defined by Eqs \eqref{delta} and \eqref{omega}.
There are analogous formulae for the other completely degenerate fields. We see, that these fusion rules coincide with those obtained in \cite{Fateev:1987zh,Fateev:1987vh,Bowcock:1992gt} and coincide with the fusion rules \eqref{Fusion_completely_degenerate-k} for the Lie algebra $\mathfrak{sl}(3)$ (see section \ref{TFT}).

Having such rather simple fusion rules \eqref{Fusion} one can hope that the four-point correlation function, which contains completely degenerate field, will satisfy differential equation of the third order. Unfortunately, this is not the case. Consider, for example, the correlation function
\begin{equation}\label{FourPoint}
   \langle V(z,\bar{z})V_{\alpha_1}(z_1,\bar{z}_1)
   V_{\alpha_2}(z_2,\bar{z}_2)V_{\alpha_3}(z_3,\bar{z}_3)\rangle.
\end{equation}
Here $V(z)$ is the degenerate field with the parameter $\alpha=-b\omega_1$. Firstly, one should notice that the number of equations in this case is not enough to write down the differential equation. Really: in this case the number of correlation functions in the  Ward identity \eqref{WWard} is nine. These nine correlation functions satisfy five projective Ward equations plus three equations \eqref{1null}, \eqref{2null} and \eqref{3null}, which arise in the case, when one of the four fields is completely degenerate. Total number of equations is eight. Hence, these equations allow us only to express all correlation functions in terms of only one correlation function, but not to write down the differential equation for this function. Therefore we need at least one more additional condition,  which connects different correlation functions in Eq \eqref{WWard} together.

Let us suppose that one of the fields $V_{\alpha_1}$, $V_{\alpha_2}$ or $V_{\alpha_3}$ in correlation function \eqref{FourPoint} is partially degenerate. For example we suppose, that quantum numbers $\Delta_3$ and $w_3$ of the field $V_{\alpha_3}$ satisfy the relation
\begin{equation}
  9w_3^2=2\Delta_3^2\left(\frac{32}{22+5c}\Bigl(\Delta_3+\frac{1}{5}\Bigr)-
  \frac{1}{5}\right),
\end{equation}
which can be written, as a condition on the vector parameter $\alpha_3$ (modulo Weyl transformation)
\begin{equation}\label{alpha3}
  \alpha_3=\varkappa\omega_2
\end{equation}
with arbitrary coefficient $\varkappa$. Corresponding field $V_{\varkappa\omega_{2}}$ satisfies the null vector condition at the first level
\begin{equation}\label{Cond_for_alpha_3}
  \left(W_{-1}-\frac{3w_3}{2\Delta_3}L_{-1}\right)V_{\varkappa\omega_2}=0.
\end{equation}
Under this assumption, correlation function \eqref{FourPoint} satisfies differential equation of the third order. In order to write it explicitly, we define function $G(x,\bar{x})$ as
\begin{equation}
  \langle V(z,\bar{z})V_{\alpha_1}(z_1,\bar{z}_1)V_{\alpha_2}(z_2,\bar{z}_2)
  V_{\varkappa\omega_2}(z_3,\bar{z}_3)\rangle\sim
  |x|^{2b(\alpha_1,h_1)}|1-x|^{\frac{2b\varkappa}{3}}
  \frac{G(x,\bar{x})}{|z-z_2|^{4\Delta}},
\end{equation}
with $x$ being the projective invariant of four points
\begin{math}
  x=\frac{z_{23}}{z_{13}}\frac{(z-z_1)}{(z-z_2)}
\end{math}
and sign ~$\sim$ ~means that we have omitted factors independent on the coordinate $z$. 
We derive from Eqs \eqref{null} and \eqref{Cond_for_alpha_3} that function $G(x,\bar{x})$ satisfies generalized Pochgamer hypergeometric differential equation of the type $(3,2)$\footnote{Of course, the same differential equation with $x$ being replaced with $\bar{x}$ is also valid.}
\begin{multline}\label{DiffEqn}
  \left[x\left(x\frac{d}{dx}+A_1\right)
  \left(x\frac{d}{dx}+A_2\right)
  \left(x\frac{d}{dx}+A_3\right)-\right.\\-\left.
  \left(x\frac{d}{dx}+B_1-1\right)
  \left(x\frac{d}{dx}+B_2-1\right)x\frac{d}{dx}\right]G(x,\bar{x})=0
\end{multline}
with
\begin{equation}
    A_k=\frac{b\varkappa}{3}-\frac{2}{3}b^2+b(\alpha_1-Q,h_1)+b(\alpha_2-Q,h_k),
\end{equation}  
and
\begin{equation}
  \begin{aligned}
    &B_1=1+b(\alpha_1-Q,e_1),\\
    &B_2=1+b(\alpha_1-Q,e_1+e_2).
  \end{aligned}
\end{equation}
Three linearly independent solutions to Eq \eqref{DiffEqn} with the diagonal monodromy around the point $x=0$ have a form 
\begin{subequations}\label{Basis1}
\begin{equation}
    G_1(x)=F\left(\genfrac{}{}{0pt}{1}{A_1\:A_2\:A_3}{B_1\:B_2}
    \biggl|x\right),
\end{equation}
\begin{equation}
    G_2(x)=x^{1-B_1}F\left(\genfrac{}{}{0pt}{1}{1-B_1+A_1\:1-
    B_1+A_2\:1-B_1+A_3}{2-B_1\:1-B_1+B_2}\biggl|x\right),
\end{equation}
and
\begin{equation}
    G_3(x)=x^{1-B_2}F\left(\genfrac{}{}{0pt}{1}{1-B_2+A_1\:1-
    B_2+A_2\:1-B_2+A_3}{1-B_2+B_1\:2-B_2}\biggl|x\right),
\end{equation}
\end{subequations}
here 
\begin{equation}
F\left(\genfrac{}{}{0pt}{1}{A_1\:A_2\:A_3}{B_1\:B_2}
    \biggl|x\right)=1+\frac{A_1A_2A_3}{B_1B_2}x+
     \frac{A_1(A_1+1)A_2(A_2+1)A_3(A_3+1)}{B_1(B_1+1)B_2(B_2+1)}\frac{x^2}{2!}+\dots
\end{equation}
is the hypergeometric function of the type $(3,2)$.

Let us take now into account the antiholomorphic part of the correlation function \eqref{FourPoint}. We wish our correlation function be invariant with respect to moving point $x$ around point $0$. The invariant combination $G(x,\bar{x})$, which defines the four-point correlation function \eqref{FourPoint} has a form	
\begin{equation}\label{Sfusion}
    G(x,\bar{x})=\sum_{j=1}^3C_{-b\omega_1,\,\alpha_1}^{\alpha_1-bh_j}
    C(\alpha_1-bh_j,\,\alpha_2,\,\varkappa\omega_2)G_j(x)G_j(\bar{x})
\end{equation}
with $C_{-b\omega_1,\,\alpha_1}^{\alpha_1-bh_j}$ being the structure constants of the operator algebra. Now we should impose the condition that this correlation function remains invariant, if we move point $x$ around points $\infty$ and $1$. Evidently, it is sufficient to provide this invariance around point $\infty$, because contour surrounding points $1$ and $\infty$ can be transformed to contour surrounding point $0$. 

There is another set of solutions to the equation \eqref{DiffEqn}, which have diagonal  monodromy around the point $x=\infty$.
\begin{subequations}\label{Basis2}
\begin{equation}
    H_1(x)=x^{-A_1}
    F\left(\genfrac{}{}{0pt}{1}{A_1\:1+A_1-B_1\:1+A_1-B_2}
    {1+A_1-A_2\:1+A_1-A_3}\biggl|\frac{1}{x}\right),
\end{equation}
\begin{equation}
    H_2(x)=x^{-A_2}
    F\left(\genfrac{}{}{0pt}{1}{1+A_2-B_1\:A_2\:1+A_2-B_2}
    {1+A_2-A_1\:1+A_2-A_3}\biggl|\frac{1}{x}\right),
\end{equation}
and
\begin{equation}
    H_3(x)=x^{-A_3}
    F\left(\genfrac{}{}{0pt}{1}{1+A_3-B_1\:1+A_3-B_2\:A_3}
    {1+A_3-A_1\:1+A_3-A_2}\biggl|\frac{1}{x}\right).
\end{equation}
\end{subequations}
Of course, these two bases \eqref{Basis1} and \eqref{Basis2} of the solutions to Eq \eqref{DiffEqn} are linearly connected. Using Mellin-Barnes representation for the generalized hypergeometric function one can obtain the relation between them. For example
\begin{multline}\label{MellinBarnes}
    \frac{\Gamma(A_1)\Gamma(A_2)\Gamma(A_3)}{\Gamma(B_1)\Gamma(B_2)}
    F\left(\genfrac{}{}{0pt}{1}{A_1\:A_2\:A_3}{B_1\:B_2}
    \biggl|x\right)=\\
    =(-x)^{-A_1}
    \frac{\Gamma(A_1)\Gamma(A_2-A_1)\Gamma(A_3-A_1)}{\Gamma(B_1-A_1)
    \Gamma(B_2-A_1)}
    F\left(\genfrac{}{}{0pt}{1}{A_1\:1+A_1-B_1\:1+A_1-B_2}
    {1+A_1-A_2\:1+A_1-A_3}\biggl|\frac{1}{x}\right)+
    \\+
    (-x)^{-A_2}
    \frac{\Gamma(A_2)\Gamma(A_1-A_2)\Gamma(A_3-A_2)}{\Gamma(B_1-A_2)
    \Gamma(B_2-A_2)}
    F\left(\genfrac{}{}{0pt}{1}{1+A_2-B_1\:A_2\:1+A_2-B_2}{1+A_2-A_1
    \:1+A_2-A_3}\biggl|\frac{1}{x}\right)+
    \\+
    (-x)^{-A_3}
    \frac{\Gamma(A_3)\Gamma(A_1-A_3)\Gamma(A_2-A_3)}{\Gamma(B_1-A_3)
    \Gamma(B_2-A_3)}
    F\left(\genfrac{}{}{0pt}{1}{1+A_3-B_1\:1+A_3-B_2\:A_3}{1+A_3-A_1
    \:1+A_3-A_2}\biggl|\frac{1}{x}\right).
\end{multline}
Our correlation function has to be also single valued at the point $x=\infty$. Hence it must be represented by the diagonal bilinear form
\begin{equation}\label{Tfusion}
    G(x,\bar{x})=\sum_{j=1}^3
    C_{-b\omega_1,\,\alpha_2}^
    {\alpha_2-bh_j}
    C(\alpha_1,\,\alpha_2-bh_j,\,\varkappa\omega_2)H_j(x)H_j(\bar{x}).
\end{equation}
The necessary conditions of the validity of the both s-channel
\eqref{Sfusion} and t-channel \eqref{Tfusion} decompositions are
\begin{equation}\label{Func-Relat-3}
  \begin{aligned}
   &\frac{C_{-b\omega_1,\,\alpha_1}^{\alpha_1-bh_1}
    C(\alpha_1-bh_1,\,\alpha_2,\,\varkappa\omega_2)}
    {C_{-b\omega_1,\,\alpha_1}^{\alpha_1-bh_2}
    C(\alpha_1-bh_2,\,\alpha_2,\,\varkappa\omega_2)}=
    \frac{\prod_{k=1}^3\gamma(A_k)\gamma(B_1-A_k)}{\gamma(B_1)
    \gamma(B_2)}
    \frac{\gamma(1-B_1+B_2)}{\gamma(B_1-1)},\\
   &\frac{C_{-b\omega_1,\,\alpha_1}^{\alpha_1-bh_1}
    C(\alpha_1-bh_1,\,\alpha_2,\,\varkappa\omega_2)}
    {C_{-b\omega_1,\,\alpha_1}^{\alpha_1-bh_3}
    C(\alpha_1-bh_3,\,\alpha_2,\,\varkappa\omega_2)}=
    \frac{\prod_{k=1}^3\gamma(A_k)\gamma(B_2-A_k)}{\gamma(B_1)
    \gamma(B_2)}
    \frac{\gamma(1-B_2+B_1)}{\gamma(B_2-1)}.
   \end{aligned}
\end{equation}
Of course, functional equations similar to \eqref{Func-Relat-3} with $\alpha_1$ being replaced by $\alpha_2$ are also valid.

One can expect, that differential equation similar to \eqref{DiffEqn} will take place in the $\mathfrak{sl}(n)$ case too\footnote{We do not give here the strict  algebraic proof of this fact for $\mathfrak{sl}(n)$ with $n>3$, but the generalization is very straightforward.}. 
The condition \eqref{alpha3} undergoes natural modification
\begin{equation}
    \alpha_3=\varkappa\omega_{n-1}.
\end{equation}
Let us consider  correlation function
\begin{equation}\label{4-point-first}
   \langle V_{-b\omega_1}(x,\bar{x})V_{\alpha_1}(0)
   V_{\alpha_2}(\infty)V_{\varkappa\omega_{n-1}}(1)\rangle=
  |x|^{2b(\alpha_1,h_1)}|1-x|^{\frac{2b\varkappa}{n}}G(x,\bar{x}).
\end{equation}
Function $G(x,\bar{x})$ satisfies  generalized Pochgamer hypergeometric differential equation of the type $(n,n-1)$ in each of the variables $x$ and $\bar{x}$
\begin{multline}\label{DiffEqnN}
  \Bigl[x\left(x\frac{d}{dx}+A_1\right)\dots
  \left(x\frac{d}{dx}+A_n\right)-\\-
  \left(x\frac{d}{dx}+B_1-1\right)
  \dots\left(x\frac{d}{dx}+B_{n-1}-1\right)
  x\frac{d}{dx}\Bigr]G(x,\bar{x})=0
\end{multline}
with
\begin{equation}\label{A_k}
  A_k=\frac{b\varkappa}{n}-\frac{(n-1)}{n}b^2+b(\alpha_1-Q,h_1)+
  b(\alpha_2-Q,h_k),
\end{equation}
and
\begin{equation}\label{B_k}
  B_k=1+b(\alpha_1-Q,e_1+\dots+e_k).
\end{equation}
The basis of the solutions to differential equation \eqref{DiffEqnN} with diagonal monodromy around the point $x=0$ has a form
\begin{subequations}\label{Basis1-n}
\begin{equation}
    G_1(x)=F\left(\genfrac{}{}{0pt}{1}{A_1\dots\:A_n}{B_1\dots B_{n-1}}
    \biggl|x\right),
\end{equation}
\begin{equation*}
    \dots
\end{equation*}
\begin{equation}
 G_{k+1}(x)=x^{1-B_k}F
 \left(\genfrac{}{}{0pt}{1}{1-B_k+A_1\dots1-B_k+A_n}{1-B_k+B_1\dots2-B_k\dots1-B_k+B_{n-1}}\biggl|x\right)
  \quad\text{for}\quad k\geq1,
\end{equation}
\end{subequations}
while the dual basis of the solutions with diagonal monodromy around the point $x=\infty$ can by represented by the functions
\begin{equation}\label{Basis2-n}
    H_k(x)=x^{-A_k}
    F\left(\genfrac{}{}{0pt}{1}{1+A_k-B_1\dots A_k\dots1+A_k-B_{n-1}}
    {1+A_k-A_1\dots1+A_k-A_{k-1}\:1+A_k-A_{k+1}\dots1+A_k-A_n}\biggl|\frac{1}{x}\right),
\end{equation}
where $F\left(\genfrac{}{}{0pt}{1}{A_1\dots\:A_n}{B_1\dots B_{n-1}}\biggl|x\right)$ is the hypergeometric function of the type $(n,n-1)$. Four-point correlation function \eqref{4-point-first} should be single valued function of the variables $x$ and $\bar{x}$. It means, that it should be represented simultaneously as
\begin{equation}\label{t-chanel-n}
 G(x,\bar{x})=\sum_{j=1}^nC_{-b\omega_1,\,\alpha_1}^{\alpha_1-bh_j}
    C(\alpha_1-bh_j,\,\alpha_2,\,\varkappa\omega_{n-1})G_j(x)G_j(\bar{x})
\end{equation}
and as
\begin{equation}\label{s-chanel-n}
 G(x,\bar{x})=\sum_{j=1}^nC_{-b\omega_1,\,\alpha_2}^{\alpha_2-bh_j}
              C(\alpha_1,\,\alpha_2-bh_j,\,\varkappa\omega_{n-1})H_j(x)H_j(\bar{x}),
\end{equation}
where functions $G_j(x)$ are given by Eqs \eqref{Basis1-n} and functions $H_j(x)$ are given by Eqs \eqref{Basis2-n}. Using formula naturally generalizing Eq \eqref{MellinBarnes} for $n>3$, we can connect two bases $G_j(x)$ and $H_j(x)$. As a result, we obtain that the condition of the validity of the both $t-$ and $s-$ channel decompositions \eqref{t-chanel-n} and \eqref{s-chanel-n} for the correlation function \eqref{4-point-first} has a form
\begin{equation}\label{FunkRelat}
   \frac{C_{-b\omega_1,\,\alpha_1}^{\alpha_1-bh_1}
    C(\alpha_1-bh_1,\,\alpha_2,\,\varkappa\omega_{n-1})}
    {C_{-b\omega_1,\,\alpha_1}^{\alpha_1-bh_k}
    C(\alpha_1-bh_k,\,\alpha_2,\,\varkappa\omega_{n-1})}=
    \frac{\prod_{j=1}^{n}\gamma(A_j)\gamma(B_{k-1}-A_j)}
    {\prod_{j=1}^{n-1}\gamma(B_j)}
    \frac{\prod_{j\neq k-1}^{n-1}\gamma(1+B_j-B_{k-1})}{\gamma(B_{k-1}-1)},
\end{equation}
where $k=2,\dots,n$. 

The structure constants $C_{-b\omega_1,\,\alpha_1}^{\alpha_1-bh_k}$ admit the free-field representation \cite{Fateev:2000ik}
\begin{equation}\label{FreeField}
C_{-b\omega_1,\,\alpha_1}^{\alpha_1-bh_k}=
 (-\mu)^{k-1}
\int\langle V_{-b\omega_1}(0)V_{\alpha_1}(1)
V_{2Q-\alpha_1+bh_k}(\infty)\prod_{i=1}^{k-1}
V_{be_i}(z_i,\bar{z}_i)\,d^2z_i\rangle_{0}.
\end{equation}
The expectation value in Eq \eqref{FreeField} is taken using the Wick rules in the theory of a free massless scalar field. This integral, as was pointed out in section \ref{TFT}, can be calculated explicitly. The answer follows from Eq \eqref{StrConst_k}
\begin{equation}\label{StrConst}
    C_{-b\omega_1,\,\alpha_1}^{\alpha_1-bh_k}=
    \left(-\frac{\pi\mu}{\gamma(-b^2)}\right)^{k-1}
    \prod_{i=1}^{k-1}
    \frac{\gamma(b(\alpha_1-Q,h_i-h_k))}
    {\gamma(1+b^2+b(\alpha_1-Q,h_i-h_k))}.
\end{equation}
Therefore, we obtain from Eqs \eqref{FunkRelat} and \eqref{StrConst} the system of $(n-1)$ functional relations. There is another dual set of functional relations with parameter $b$ being replaced with $b^{-1}$ and cosmological constant $\mu$ being replaced with dual cosmological constant $\tilde{\mu}$ defined by Eq \eqref{mu-dual}. If parameter $b^2$ real and  irrational, then the solution to the both systems of equations is unique up to a multiplicative constant, which depends only on the parameter $\varkappa$. It is easy to check  that proposed in section \ref{TFT} three-point correlation function \eqref{C} satisfies both of these systems of equations.

In conclusion of this section, we present the exact expression for the four-point correlation function \eqref{4-point-first}. This correlation function can be expressed in terms of Coulomb integral
\begin{multline}\label{4-point}
   \langle V_{-b\omega_1}(x,\bar{x})V_{\alpha_1}(0)
   V_{\alpha_2}(\infty)V_{\varkappa\omega_{n-1}}(1)\rangle=\\=
    \left(\frac{b}{\pi}\right)^{n-1}
    \left[\pi\mu\gamma(b^2)b^{2-2b^2}\right]^
    {\frac{(2Q-\alpha,\rho)}{b}}
    \frac{\left(\Upsilon(b)\right)^{n-1}\Upsilon(\varkappa)
    \prod\limits_{e>0}\Upsilon\Bigl((Q-\alpha_1,e)\Bigr)
    \Upsilon\Bigl((Q-\alpha_2,e)\Bigr)}
    {\prod\limits_{ij}\Upsilon\Bigl(\frac{\varkappa+b}{n}+
    (\alpha_1-Q,h_i)+(\alpha_2-Q,h_j)-b\delta_{ij}\Bigr)}\times\\\times
   |x|^{2b(\alpha_1,h_1)}|1-x|^{\frac{2b\varkappa}{n}}
   \int\prod_{k=1}^{n-1}d^2t_k\,|t_k|^{2(A_k-B_k)}|t_k-t_{k+1}|^{2(B_k-A_{k+1}-1)}
   |t_1-x|^{-2A_1},
\end{multline}
where $t_n\equiv1$, $\alpha=-b\omega_1+\alpha_1+\alpha_2+\varkappa\omega_{n-1}$ and parameters $A_k$ and $B_k$ are given by Eqs \eqref{A_k}--\eqref{B_k}. This expression for the correlation function can be derived by the analytical continuation to the non-integer values of numbers $s_k$ in Eq \eqref{GoulLieFormula1}\footnote{It can be also proved, that integral in Eq \eqref{4-point} satisfies holomorphic (and antiholomorphic) differential equation \eqref{DiffEqnN}.}, which permits also to find expressions for more general correlation functions (see also \cite{Fateev:2006:JETP,Fateev:2007:TMPH}). In principle, it is possible to write down explicit expressions for the correlation functions 
$$
\langle V_{-mb\omega_1}(x,\bar{x})V_{\alpha_1}(0)V_{\alpha_2}(\infty)V_{\varkappa\omega_{n-1}}(1)\rangle
$$
in terms of finite dimensional integral for $m>1$, but the result will have more tedious form. We plan to consider these and more general correlation functions in the forthcoming paper \cite{Part-Deux}.

If we consider the operator product expansion of the field $V_{-b\omega_1}$ with the field $V_{\varkappa\omega_{n-1}}$ in the correlation function \eqref{4-point}, we find that the coefficient before singularity $(1-x)^{b\varkappa/3}$ defines the three-point correlation function $C(\alpha_1,\alpha_2,\varkappa\omega_{n-1}-b\omega_1)$, which is given by the expression 
\begin{multline}\label{C-shifted}
    C(\alpha_1,\alpha_2,\varkappa\omega_{n-1}-b\omega_1)=\\=
    \left(\frac{b}{\pi}\right)^{n-1}
    \left[\pi\mu\gamma(b^2)b^{2-2b^2}\right]^
    {\frac{(2Q-\alpha,\rho)}{b}}
    \frac{\left(\Upsilon(b)\right)^{n-1}\Upsilon(\varkappa)
    \prod\limits_{e>0}\Upsilon\Bigl((Q-\alpha_1,e)\Bigr)
    \Upsilon\Bigl((Q-\alpha_2,e)\Bigr)}
    {\prod\limits_{ij}\Upsilon\Bigl(\frac{\varkappa+b}{n}+
    (\alpha_1-Q,h_i)+(\alpha_2-Q,h_j)+b\delta_{ij}\Bigr)}\times\\\times
   \int\prod_{k=1}^{n-1}d^2t_k\,|t_k|^{2(A_k-B_k)}|t_k-t_{k+1}|^{2(B_k-A_{k+1}-1)}
   |t_1-1|^{-2A_1}.
\end{multline}
where $t_n\equiv1$. Integral in Eq \eqref{C-shifted} can be calculated in many different ways. The simplest one is to combine Eqs \eqref{4-point-first}, \eqref{4-point} and \eqref{t-chanel-n} and express it in terms of hypergeometric functions of the type $(n,n-1)$. As a result we obtain that
\begin{multline}\label{C-shifted-1}
   \int\prod_{k=1}^{n-1}d^2t_k\,|t_k|^{2(A_k-B_k)}|t_k-t_{k+1}|^{2(B_k-A_{k+1}-1)}
   |t_1-1|^{-2A_1}=
   \pi^{n-1}\frac{\prod_{j=1}^{n-1}\gamma(B_j-A_{j+1})}
   {\gamma(A_1)}\times\\\times
   \Biggl[\mathfrak{G}\left(\genfrac{}{}{0pt}{1}{A_1\:\:\dots\:A_n}{B_1\:\:\dots\:B_{n-1}}
    \right)
   +\sum_{k=1}^{n-1}
    \mathfrak{G}\left(\genfrac{}{}{0pt}{1}{1-B_k+A_1\:\dots\:1-B_k+A_n}
      {1-B_k+B_1\:\dots,2-B_k,\dots\:1-B_k+B_{n-1}}
    \right)\Biggr],
\end{multline}
where
\begin{equation}
\mathfrak{G}\left(\genfrac{}{}{0pt}{1}{A_1\:\:\dots\:A_n}{B_1\:\:\dots\:B_{n-1}}
    \right)\overset{\text{def}}{=}\frac{\gamma(A_1)\dots\gamma(A_n)}{\gamma(B_1)\dots\gamma(B_{n-1})}
     F\left(\genfrac{}{}{0pt}{1}{A_1\:\:\dots\:A_n}{B_1\:\:\dots\:B_{n-1}}\bigl|1\right)^2.
\end{equation}

As we see from the above, the four-point correlation function in $\mathfrak{sl}(n)$ TFT, which contains completely degenerate field, satisfies differential equation, only if at least one of the other fields is special. Namely, if field $V_{\alpha_3}$ is degenerate at the first level (parameter $\alpha_3$ takes the special value $\alpha_3=\varkappa\omega_{n-1}$), then the four-point correlation function satisfies Fuchsian differential equation of the order $n$, which can be reduced to the generalized Pochgamer differential equation \eqref{DiffEqnN} and, hence, it can be represented by the Coulomb integral \eqref{4-point}. Without such a condition, the four-point correlation function seems to be more complicated object. One can prove, that for $n>2$ it can not be a solution to the ordinary Fuchsian differential equation of the order $n$ \cite{Fateev:2005gs}. However, if the field $V_{\alpha_3}$ is degenerate (but not completely degenerate) at the higher levels $m_k+1$, $k=1,\dots,n-2$, i.~e. parameter $\alpha_3$ takes the special values
\begin{equation}
   \alpha_3=\varkappa\omega_{n-1}-\sum_{k=1}^{n-2}m_kb\omega_k
\end{equation}
with non-negative integers $m_k$, then the four-point correlation function can be represented by the Coulomb integral of the finite order \footnote{We will show it for the case of $\mathfrak{sl}(3)$ TFT in Ref \cite{Part-Deux}.}. 
\section{Classical limit (heavy exponential fields)}\label{CL-heavy}
In this section we consider the semi-classical limit $b\rightarrow 0$ of the conformal TFT. 
Let us define classical field as
\begin{equation}
    \phi=b\varphi.
\end{equation}
Its dynamics is described by the classical action
\begin{equation}\label{Sclass}
    S_{class}=\frac{1}{8\pi b^2}\int\left[
    (\partial\phi)^2+8\pi\mu b^2\sum_{k=1}^{n-1}e^{(e_k,\phi)}\right].
\end{equation}
In this limit the leading asymptotic of the correlation functions (saddle point asymptotic) is governed by the classical action calculated on some specific solution to the equations of motion, which follow from the action \eqref{Sclass}.

We will consider here the case of $\mathfrak{sl}(3)$ TFT, as an example ($\mathfrak{sl}(2)$ case corresponding to Liouville field theory was considered in \cite{Zamolodchikov:1995aa}), which is already non-trivial. The main asymptotic  at $b\rightarrow0$ of the correlation functions 
$\langle V_{\alpha_1}(z_1,\bar{z}_1)\dots V_{\alpha_N}(z_N,\bar{z}_N)\rangle$ of heavy operators with parameters
\begin{equation}
    \alpha_k=\frac{\eta_k}{b}
\end{equation}
is given by the  regularized action\footnote{The divergences arise from the vicinity of sources corresponding to the insertion of the operators $V_{\alpha}$. To obtain finite answer one should re-normalize them. See \cite{Zamolodchikov:1995aa,Seiberg:1990eb} for the regularization prescription.}
\begin{equation}
    \langle V_{\alpha_1}(z_1,\bar{z}_1)\dots V_{\alpha_N}(z_N,\bar{z}_N)\rangle\sim
    \exp\Bigl(-S_{class}^{reg}\left[
    \phi(\eta_1\dots\eta_N\vert z_1,\bar{z}_1\dots z_N,\bar{z}_N)\right]\Bigr),
\end{equation}
here $\phi(\eta_1\dots\eta_N\vert z_1,\bar{z}_1\dots z_N,\bar{z}_N)$ -- real single-valued solution to the Toda equation 
\begin{subequations}\label{Toda3}
  \begin{align}\label{Toda31}
     &\partial\bar{\partial}\phi=\pi\mu b^2\left(
     e_1e^{(e_1,\phi)}+e_2e^{(e_2,\phi)}\right)\\
  \intertext{with the asymptotic conditions}\label{AssOfFields1} 
     &\phi=-4\rho\log |z|+\dots\quad\text{at}\quad |z|
     \rightarrow\infty,\\
     &\phi=-2\eta_j\log |z-z_j|+X_j+\dots\quad\text{at}
     \quad z\rightarrow z_j.\label{AssOfFields2}
   \end{align}
\end{subequations}
In Eq \eqref{AssOfFields2} $X_j$ is a $z$ independent term.
The solution to the boundary problem \eqref{Toda3} with positive cosmological constant $\mu$ exists if
\begin{equation}\label{condition_for_class_solution}
    \sum_{i=1}^N(\eta_i,\omega_k)-2>0\qquad k=1,2.
\end{equation}
The regularized action  $S_{class}^{reg}$ on this solution can be calculated as follows \cite{Zamolodchikov:1995aa}. By definition of the classical regularized action its differential is related with parameters $X_j$ defined by Eq \eqref{AssOfFields2} in a simple way
\begin{equation}\label{trick}
    dS_{class}^{reg}=-\sum_{j=1}^N\left(X_j,d\eta_j\right).
\end{equation}
The constant of integration in Eq \eqref{trick} can be fixed by the condition\footnote{In quantum case this condition means, that correlation function $\langle V_{\alpha_1}(z_1,\bar{z}_1)\dots V_{\alpha_N}(z_N,\bar{z}_N)$ is trivial in the case, then $\sum\alpha_k=2Q$. Namely, correlation function has a multiple pole under this condition with residue expressed in terms of free field correlation function without screening fields (see Eq \eqref{GoulLieFormula1}).}
\begin{equation}\label{trick-condition}
    S_{class}^{reg}\Bigl\vert_{\sum_{i}(\eta_i,\omega_k)=2}=
    \sum_{i<j}^N\left(\eta_i,\eta_j\right)\log\vert z_i-z_j\vert^2.
\end{equation}
In the case of $\mathfrak{sl}(3)$ TFT it is convenient to introduce the projection of the field $\phi$ on the fundamental weights $\omega_k$, $k=1,2$:
\begin{equation}
 \Phi_k=(\phi,\omega_k).
\end{equation}
In terms of fields $\Phi_k$ equation \eqref{Toda31} has a form
\begin{subequations}\label{TodaEq-normal}
\begin{align}\label{TodaEq-normal-1}
  &\partial\bar{\partial}\Phi_1=\pi\mu b^2e^{2\Phi_1-\Phi_2},\\\label{TodaEq-normal-2}
  &\partial\bar{\partial}\Phi_2=\pi\mu b^2e^{2\Phi_2-\Phi_1}.
\end{align}
\end{subequations}

General solution to the system of equations \eqref{TodaEq-normal} can be obtained by introducing
the holomorphic currents
\begin{equation}\label{Tclass}
    \mathsf{T}=(\partial\Phi_1)^2+(\partial\Phi_2)^2-
    \partial\Phi_1\partial\Phi_2-\partial^2\Phi_1-\partial^2\Phi_2
\end{equation}
and
\begin{multline}\label{Wclass}
    \mathsf{W}=\left(\partial\Phi_1(\partial\Phi_2)^2+
    \partial\Phi_1\partial^2\Phi_1
    -\frac{1}{2}\partial\Phi_1\partial^2\Phi_2-
    \frac{1}{2}\partial^3\Phi_1\right)
    -\\-
    \left(\partial\Phi_2(\partial\Phi_1)^2+
    \partial\Phi_2\partial^2\Phi_2
    -\frac{1}{2}\partial\Phi_2\partial^2\Phi_1-
    \frac{1}{2}\partial^3\Phi_2\right).
\end{multline}
Using Eq \eqref{TodaEq-normal}, one can easily verify that $\bar{\partial}\mathsf{T}=\bar{\partial}\mathsf{W}=0$. In a similar way, if we change $\partial\rightarrow\bar{\partial}$ in \eqref{Tclass} and \eqref{Wclass}, we obtain anti-holomorphic currents $\bar{\mathsf{T}}$ and $\bar{\mathsf{W}}$. It follows from the explicit form of the currents $\mathsf{T}$ and $\mathsf{W}$, that field $e^{-\Phi_1}$ satisfies both holomorphic and anti-holomorphic linear differential equations of the third order
\begin{subequations}\label{HolomorphicEquation}
\begin{align}\label{HolomorphicEquation1}
     &\left(-\partial^3+\frac{1}{2}
     \partial\mathsf{T}+\mathsf{T}\partial+
     \mathsf{W}\right)e^{-\Phi_1}=0,\\
     \label{HolomorphicEquation2}
     &\left(-\bar{\partial}^3+\frac{1}{2}
     \bar{\partial}\bar{\mathsf{T}}+
     \bar{\mathsf{T}}\bar{\partial}+
     \bar{\mathsf{W}}\right)e^{-\Phi_1}=0\,.
\end{align}
\end{subequations}
Similar equations for $e^{-\Phi_2}$ with changed sign before $\mathsf{W}$ and $\bar{\mathsf{W}}$ are also valid\footnote{It is evident because current $\mathsf{T}(z)$ is symmetric and current $\mathsf{W}(z)$ is antisymmetric under the substitution $1\leftrightarrow2$.}. Differential equations \eqref{HolomorphicEquation} will play an important role in the following.

From the other hand, equations \eqref{HolomorphicEquation1} and \eqref{HolomorphicEquation2}, being viewed as a  system of linear holomorphic and anti-holomorphic differential equations with arbitrary functions $\mathsf{T}(z)$, $\bar{\mathsf{T}}(\bar{z})$, $\mathsf{W}(z)$ and $\bar{\mathsf{W}}(\bar{z})$ can be used to solve the system \eqref{TodaEq-normal}. Namely, let $\Psi_k=\Psi_k(z)$ are three linearly independent solutions to Eq \eqref{HolomorphicEquation1} and $\bar{\Psi}_k=\bar{\Psi}_k(\bar{z})$ are three linearly independent solutions to Eq \eqref{HolomorphicEquation2}\footnote{Generally speaking functions $\Psi_k$ and $\bar{\Psi}_k$ do not complex conjugated to each other.}. Then we can express the field $e^{-\Phi_1}$, as a bilinear combination
\begin{equation}
   e^{-\Phi_1}=\sum_{k=1}^3\Psi_k\bar{\Psi}_k.
\end{equation}
After that we find the field $e^{-\Phi_2}$ from the equation \eqref{TodaEq-normal-1}
\begin{equation}
  e^{-\Phi_2}=-(\pi\mu b^2)^{-1}\sum_{i<j=1}^3(\Psi_i\partial\Psi_j-\Psi_j\partial\Psi_i)
  (\bar{\Psi}_i\bar{\partial}\bar{\Psi}_j-\bar{\Psi}_j\bar{\partial}\bar{\Psi}_i).
\end{equation}
Second equation \eqref{TodaEq-normal-2} is satisfied only if
\begin{equation}\label{Wronskian}
  \mathbb{W}\left[\Psi_1,\Psi_2,\Psi_3\right]
  \mathbb{W}\left[\bar{\Psi}_1,\bar{\Psi}_2,\bar{\Psi}_3\right]=
  -\left(\pi\mu b^2\right)^3,
\end{equation}
here $\mathbb{W}\left[\Psi_1,\Psi_2,\Psi_3\right]$ is Wronskian. Henceworth, the solution to the system \eqref{TodaEq-normal} can be build up from the solutions of any pair of holomorphic and anti-holomorphic linear differential equations of the third order with the condition \eqref{Wronskian}.

In our case, we should solve Eq \eqref{Toda31} with boundary conditions \eqref{AssOfFields1} and \eqref{AssOfFields2}. It puts the limitations on the possible form of the currents $\mathsf{T}(z)$ and $\mathsf{W}(z)$. As a consequence of Eq \eqref{AssOfFields1}, the
currents $\mathsf{T}(z)$ and $\mathsf{W}(z)$ have asymptotic at infinity
\begin{equation}\label{AssOfCurrents}
    \mathsf{T}(z)\sim\frac{1}{z^4}\quad
     \mathsf{W}(z)\sim\frac{1}{z^6}\quad\text{at}
    \quad z\rightarrow\infty
\end{equation}
and  due to \eqref{AssOfFields2} are in fact the rational functions
\begin{equation}\label{Currents_rational}
  \begin{aligned}
    &\mathsf{T}(z)=\sum_{k=1}^N\left(\frac{\delta_k}{(z-z_k)^2}+
    \frac{C_k}{(z-z_k)}\right),\\
    &\mathsf{W}(z)=\sum_{k=1}^N\left(\frac{\mathsf{w}_k}{(z-z_k)^3}+
    \frac{D_k}{(z-z_k)^2}+
    \frac{E_k}{(z-z_k)}\right),
  \end{aligned}
\end{equation}
here parameters $\delta_k$ and $\mathsf{w}_k$ are expressed in terms of vector parameters $\eta_k$ as
\begin{equation}\label{TandW}
\begin{aligned}
    &\delta_k=\frac{(\eta_k,\eta_k)}{2}-(\eta_k,\rho)\\
    &\mathsf{w}_k=\bigl((\eta_k,\omega_1)-(\eta_k,\omega_2)\bigr)
    \bigl((\eta_k,\omega_1)-1\bigr)\bigl((\eta_k,\omega_2)-1\bigr)
\end{aligned}
\end{equation}
and coincide up to a sign with semiclassical limit of the quantum numbers \eqref{delta} and \eqref{omega}.
The parameters $C_k$, $D_k$ and $E_k$ are not defined  from the main asymptotic  \eqref{AssOfFields2} at $z\rightarrow z_k$, but contain information about next subleading terms. In fact, they are not  linearly independent, but satisfy linear algebraic relations, which follow from the asymptotic \eqref{AssOfCurrents} (analog of Ward identities \eqref{LWard} and \eqref{WWard} in quantum case).

First interesting case is the case of three singular points, which corresponds to the semiclassical limit of the three-point correlation function. Let us consider it in more details. In this case the number of equations, which follow from the asymptotic of the current $\mathsf{T}(z)$, is enough to find parameters $C_k$. Really, we have three parameters $C_1$, $C_2$ and $C_3$ and three conditions, which follow from the asymptotic of the current $\mathsf{T}(z)$ at infinity. Unfortunately, this is not true for the asymptotic of the current $\mathsf{W}$. In this case, we have six parameters $D_k$ and $E_k$ and only five equations, which appear from the asymptotic $\mathsf{W}(z)\sim\frac{1}{z^6}$. Therefore one parameter remains free. Evidently, it corresponds to the possibility to add to the current $\mathsf{W}(z)$ the term
\begin{equation*}
 \frac{1}{(z-z_1)^2(z-z_2)^2(z-z_3)^2}
\end{equation*}
with arbitrary coefficient. In order to emphasize this one-parameter freedom, let us fix first non-vanishing term of the asymptotic of the current $\mathsf{W}(z)$ at infinity as
\begin{multline}\label{Lambda-definition}
    \mathsf{W}(z)=\frac{1}{2z^6}\Bigl[\mathsf{w}_1z_{12}z_{13}(z_{12}+
    z_{13})+\mathsf{w}_2z_{21}z_{23}(z_{21}+z_{23})+\\+
    \mathsf{w}_3z_{31}z_{32}(z_{31}+z_{32})+
    2\Lambda z_{12}z_{13}z_{23}\Bigr]+
    O\left(\frac{1}{z^7}\right).
\end{multline}
The parameter $\Lambda$, which we call accessory parameter, is not known a priori. Here we arrive at the main difference with $\mathfrak{sl}(2)$ case, where the accessory parameters do not appear in the case of three singular points \cite{Zograf}. This difference explains at the classical level why the three-point correlation function is much more complicated object in higher Toda systems. As we will show below, the  parameter $\Lambda$ can be found, in principle, from rather different arguments, which resemble the conformal bootstrap program. 

Using projective invariance of Eqs \eqref{HolomorphicEquation}\footnote{One can show, that differential equation
\begin{equation*}
\left(-\partial^3+\frac{1}{2}\partial\mathsf{T}+\mathsf{T}\partial+\mathsf{W}\right)\Psi=0
\end{equation*}
is invariant under the substitution
$z\rightarrow w(z)$, $\Psi(z)\rightarrow \left(\frac{dw}{dz}\right)^{-1}\Psi(w)$,
$\mathsf{T}(z)\rightarrow\left(\frac{dw}{dz}\right)^{2}\mathsf{T}(w)-2\{w,z\}$ and
$\mathsf{W}(z)\rightarrow \left(\frac{dw}{dz}\right)^{3}\mathsf{W}(w)$,
where $\{w,z\}$ is Schwartz derivative.} one can rewrite them through the invariants of four points 
\begin{equation*}
x=\frac{(z-z_1)(z_2-z_3)}{(z-z_3)(z_2-z_1)}\;\;\;\;\;\text{and}\;\;\;\;\;
\bar{x}=\frac{(\bar{z}-\bar{z}_1)(\bar{z}_2-\bar{z}_3)}{(\bar{z}-\bar{z}_3)(\bar{z}_2-\bar{z}_1)}.
\end{equation*}
For example, Eq \eqref{HolomorphicEquation1} will have a form 
\begin{equation}\label{Fuchsian-3}
  \left(-\partial_x^3+\frac{1}{2}\partial_x\mathsf{T}(x)+\mathsf{T}(x)\partial_x+\mathsf{W}(x)\right)
  \Psi(x)=0
\end{equation}
with
\begin{equation}
  \begin{aligned}
    &\mathsf{T}(x)=\frac{\delta_1}{x^2}+\frac{\delta_2}{(x-1)^2}+
   \frac{\delta_3-\delta_1-\delta_2}{x(x-1)},\\
    &\mathsf{W}(x)=\frac{\mathsf{w}_1}{x^3}+\frac{\mathsf{w}_2}{(x-1)^3}
    +\frac{1}{2}(\mathsf{w}_1+\mathsf{w}_2+\mathsf{w}_3)\Bigl(\frac{1}{x^2}-\frac{1}{(x-1)^2}\Bigr)+
    \frac{(\mathsf{w}_1-\mathsf{w}_2+\Lambda)}{x^2(x-1)^2}.
  \end{aligned}
\end{equation}
Equation \eqref{Fuchsian-3} is the most general Fuchsian differential equation of the third order with three singular points $0$, $1$ and $\infty$ modulo "gauge" transformation $\Psi(x)\rightarrow x^{\alpha}(x-1)^{\beta}\Psi(x)$. The "gauge" is fixed by the condition, that term with second derivative
$\partial^2_x\Psi(x)$ is absent  in Eq \eqref{Fuchsian-3}.

In order to solve the problem \eqref{Toda3}, one should find real single valued solution to Eqs \eqref{HolomorphicEquation}. The last requirement is not trivial, because the general solution to Eqs
\eqref{HolomorphicEquation} does not satisfy this property. Let $\psi_k$ be the basis of the solutions to Eq \eqref{HolomorphicEquation1} with diagonal monodromy around point $z=z_1$
\begin{equation}
    \psi_k=(z-z_1)^{1+(\eta_1-\rho,h_k)}\left(1+O(z-z_1)\right)
    \qquad k=1,2,3
\end{equation}
\begin{figure}
\psfrag{z1}{$z_1$}\psfrag{z2}{$z_2$}\psfrag{z3}{$z_3$}
\psfrag{c1}{$C_1$}\psfrag{c2}{$C_2$}\psfrag{c3}{$C_3$}
  \begin{center}
   \epsfig{figure=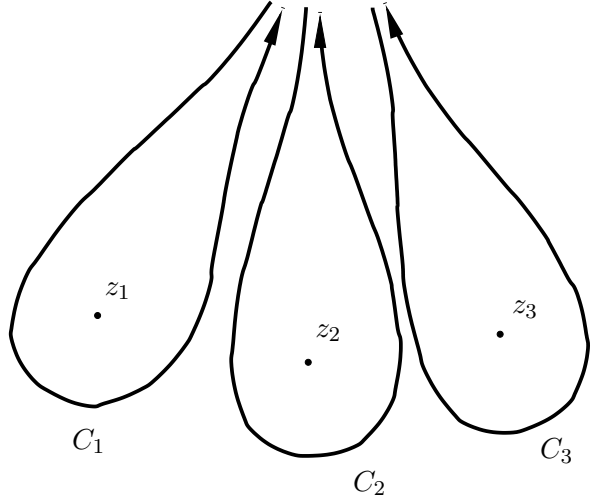,width=.514\textwidth}
   \caption{Basic monodromy contours for the equation  
   \protect\eqref{HolomorphicEquation} in the case of three singular points.}
   \label{fig:contur}
  \end{center}
\end{figure} 
If we write a diagonal bilinear combination
\begin{equation}\label{AssosRelat1}
  e^{-\Phi_1}=\lambda_1|\psi_1|^2+\lambda_2|\psi_2|^2+
  \lambda_3|\psi_3|^2,
\end{equation}
such a solution is evidently invariant if we move point $z$ around point $z_1$ (contour $C_1$ on figure \ref{fig:contur}). But we need also such an invariance around points $z_2$ and $z_3$ (contour $C_2$ and $C_3$ on figure \ref{fig:contur} respectively). Let $\chi_k$ be the basis of the solutions to Eq \eqref{HolomorphicEquation1} with diagonal monodromy around point $z=z_2$
\begin{equation}
    \chi_k=(z-z_2)^{1+(\eta_2-\rho,h_k)}\left(1+O(z-z_2)\right)
    \qquad k=1,2,3.
\end{equation}
The following formula also should be valid
\begin{equation}\label{AssosRelat2}
  e^{-\Phi_1}=\tilde{\lambda}_1|\chi_1|^2+
  \tilde{\lambda}_2|\chi_2|^2+\tilde{\lambda}_3|\chi_3|^2
\end{equation}
with some other constants $\tilde{\lambda}_k$. If the solution $e^{-\Phi_1}$ can be represented simultaneously as \eqref{AssosRelat1} and as \eqref{AssosRelat2} it becomes single-valued on a total sphere, because the contour surrounding point $z_3$ can be transformed to the contour surrounding points $z_1$ and $z_2$, as guaranteed by the condition \eqref{AssOfCurrents}. As  functions $\psi_k$ and $\chi_k$ satisfy the same differential equation, they are linearly connected
\begin{equation}\label{Assos}
  \psi_i=M_{ij}\chi_j.
\end{equation}
Entries of the matrix $M_{ij}$ are believed to be meromorphic functions of the parameters $\delta_k$, $\mathsf{w}_k$ and the accessory parameter $\Lambda$ (for the real values of the parameters $\delta_k$, $\mathsf{w}_k$ and $\Lambda$ matrix $M_{ij}$ is real). If we substitute relation \eqref{Assos} into Eq \eqref{AssosRelat1}, we obtain unwanted  cross terms like $\chi_1\bar{\chi}_2$  destroying the property \eqref{AssosRelat2}, which guarantees that solution is single-valued. So, one should set all coefficients before such terms equal to zero. As a result, we arrive to the system of equations
\begin{equation}
  \begin{pmatrix}
     M_{11}M_{12} & M_{11}M_{13} & M_{12}M_{13} \\
     M_{21}M_{22} & M_{21}M_{23} & M_{22}M_{23} \\
     M_{31}M_{32} & M_{31}M_{33} & M_{32}M_{33} 
  \end{pmatrix}
  \begin{pmatrix}
     \lambda_1\\
     \lambda_2 \\
     \lambda_3 
  \end{pmatrix}=0.
\end{equation} 
The determinant of the corresponding matrix should be zero
\begin{equation}\label{det}
   \det \begin{pmatrix}
     M_{11}M_{12} & M_{11}M_{13} & M_{12}M_{13} \\
     M_{21}M_{22} & M_{21}M_{23} & M_{22}M_{23} \\
     M_{31}M_{32} & M_{31}M_{33} & M_{32}M_{33} 
  \end{pmatrix}=0. 
\end{equation}
The condition \eqref{det} can be viewed as an equation on accessory parameter $\Lambda$. Each accessory parameter $\Lambda$, which solves equation \eqref{det} gives single-valued solution to the boundary Toda problem \eqref{Toda3}.

Let us try to find a solution to this equation in a special situation, which corresponds to the classical limit of three-point correlation function \eqref{C}. Namely, we suppose that
\begin{equation}\label{ClassCond}
   \eta_3=\kappa\omega_2.
\end{equation}
In this case one can guess accessory parameter $\Lambda$, which solves equation \eqref{det}:
\begin{equation}\label{Lambda}
   \Lambda=\left(\frac{\kappa}{3}-\frac{1}{2}\right)
   \left(\delta_1-\delta_2\right)-\frac{1}{2}\left(\mathsf{w}_1-\mathsf{w}_2\right)
\end{equation}
from the  rather simple reasoning. The logic is the following: if the condition \eqref{ClassCond} is satisfied, the equations \eqref{HolomorphicEquation} have the same behavior near the singular points as a hypergeometric equation of the type $(3,2)$, but do not coincide with it if parameter $\Lambda$ is general. So, we select such special value of the parameter $\Lambda$ defined by Eq \eqref{Lambda}, that these equations are identical. One can easily check that the equation \eqref{det} is satisfied in this case. The solution to the boundary problem \eqref{Toda3} can be obtained in this case in a simple way: expression for the field $e^{-\Phi_1}$ can be derived by semiclassical limit of four-point correlation function \eqref{4-point} for $n=3$, while expression for the field  $e^{-\Phi_2}$ can be obtained from Eq \eqref{TodaEq-normal-1}. Both of them can be expressed in terms of Coulomb integrals (for simplicity we set $z_1=0$, $z_2=\infty$, $z_3=1$ and $z=x$)
\begin{equation}\label{Coulomb-for-hypergeom}
\begin{aligned}
  &e^{-\Phi_1}=\mathfrak{C}\;
    |x|^{2(\eta_1,\omega_1)}|x-1|^{\frac{2\kappa}{3}}
    \int d^2t\,d^2y \;
    |t-x|^{2\mathfrak{b}_1}|t-y|^{2\mathfrak{b}_2}|y-1|^{2\mathfrak{b}_3}
     \;|t|^{2\mathfrak{a}_1}|y|^{2\mathfrak{a}_2},\\
  &e^{-\Phi_2}=\tilde{\mathfrak{C}}\;
    |x|^{2(\eta_1,\omega_2)}|x-1|^{2-\frac{2\kappa}{3}}
    \int d^2t\,d^2y\;
    |t-x|^{2\tilde{\mathfrak{b}}_1}|t-y|^{2\tilde{\mathfrak{b}_2}}|y-1|^{2\tilde{\mathfrak{b}}_3}
    \;|t|^{2\tilde{\mathfrak{a}}_1}|y|^{2\tilde{\mathfrak{a}}_2},
\end{aligned}
\end{equation}
with
\begin{equation}
\begin{aligned}
&\mathfrak{a}_k=-1+\frac{\kappa}{3}+(\eta_1-\rho,h_{k+1})+(\eta_2-\rho,h_{k}),\;
&&\mathfrak{b}_k=-\frac{\kappa}{3}-(\eta_1-\rho,h_{k})-(\eta_2-\rho,h_{k}),\\
&\tilde{\mathfrak{a}}_k=-\frac{\kappa}{3}+(\eta_1-\rho,h_{k+1}^{*})+(\eta_2-\rho,h_k^{*}),\;
&&\tilde{\mathfrak{b}}_k=-1+\frac{\kappa}{3}-(\eta_1-\rho,h_k^{*})-(\eta_2-\rho,h_k^{*}),
\end{aligned}
\end{equation}
where $h_k^{*}=-h_{4-k}$ and
\begin{equation}
\begin{aligned}
  &\mathfrak{C}=\frac{\mu b^2}{\pi}\frac
  {\prod_{k=1}^3\gamma(\frac{\kappa}{3}+(\eta_1-\rho,h_k)+(\eta_2-\rho,h_k))}
  {\prod_{i=1}^3\prod_{j=1}^3\left[\gamma(\frac{\kappa}{3}+(\eta_1-\rho,h_i)+(\eta_2-\rho,h_j))\right]
  ^{\frac{1}{3}}},\\
  &\tilde{\mathfrak{C}}=\frac{\mu b^2}{\pi}\frac
  {\prod_{i=1}^3\prod_{j=1}^3\left[\gamma(\frac{\kappa}{3}+(\eta_1-\rho,h_i)+(\eta_2-\rho,h_j))\right]
  ^{\frac{1}{3}}}
  {\prod_{k=1}^3\gamma(\frac{\kappa}{3}+(\eta_1-\rho,h_k)+(\eta_2-\rho,h_k))}.
\end{aligned}
\end{equation}
This solution can be also written through the hypergeometric function of the type $(3,2)$.
The regularized classical action on this solution $S_{class}^{reg}$ can be easily found using Eqs \eqref{trick} and \eqref{trick-condition} and has a form
\begin{multline}\label{ClassAssym}
       S_{class}^{reg}=\left((\eta_1+\eta_2,\rho)+\kappa-4\right)
       \log(\pi\mu b^2)+F(\kappa)+
       \sum_{e>0}F((\rho-\eta_1,e))+
       \sum_{e>0}F((\rho-\eta_2,e))-\\-
       \sum_{ij}F\left(\frac{\kappa}{3}+(\eta_1-\rho,h_i)
       +(\eta_2-\rho,h_i)\right)-F^2(0)
\end{multline}
with 
\begin{equation}
  F(x)=\int_{\frac{1}{2}}^{x}\log\gamma(t)dt.
\end{equation}
We see that the classical limit of the proposed three-point correlation function \eqref{C} is in complete agreement with expression \eqref{ClassAssym}. 

We have found a solution to the boundary Toda problem \eqref{Toda3} and an explicit expression for the accessory parameter $\Lambda$ in the case of five-parametric family (vector parameter $\eta_3$ restricted by the condition $\eta_3=\kappa\omega_2$). In general case boundary problem \eqref{Toda3} is rather complicated, because it corresponds to the most general differential equation of the third order with three singular points of the Fuchsian type (by projective invariance this equation can be transformed to equation \eqref{Fuchsian-3}). It is interesting to notice, that to the same type belongs differential equation for the four-point correlation function $\langle V_{-b}(z)V_{\alpha_1}(z_1)V_{\alpha_2}(z_2)V_{\alpha_3}(z_3)\rangle$ in the Liouville field theory \cite{Belavin:1984vu} ($\mathfrak{sl}(2)$ TFT)\footnote{Here we use standart for the Liouville field theory (LFT) notations, which differ from those used in this paper. Namely central charge in LFT equals $c_L=1+6(b+b^{-1})^2$ and exponential fields $V_{\alpha}$ have conformal dimensions $\Delta_L(\alpha)=\alpha(b+b^{-1}-\alpha)$. One should emphasize, that parameter $b$ here is formal parameter, which does not goes to zero.}. It can be transformed to the differential equation \eqref{Fuchsian-3} with parameters (one should take into account projective invariance)
\begin{equation}\label{DiffEq-Phi13}
   \delta_k=-\frac{2}{3}b^2(6\Delta_L(\alpha_k)-3-2b^2),\;\;
   \mathsf{w}_k=-\frac{2}{27}b^2(36b^2\Delta_L(\alpha_k)-8b^4-18b^2-9),\;\;
\Lambda=0.
\end{equation}
If we introduce auxiliary parameter $g=-b^2$, then the numbers $\delta_k$ and $\mathsf{w}_k$ are subject the condition
\begin{equation}
  \delta_k-\frac{\mathsf{3w}_k}{2g}=\frac{4g^2}{9}-1,
\end{equation}
which can be parameterized in terms of vector parameter $\eta_k$, which enter in Eq \eqref{TandW}, as
\begin{equation}\label{eta_for_Liouville}
  \eta_k=\lambda_k\omega_1+(\lambda_k-2g)\omega_2\;\;\;\;k=1,2,3,
\end{equation}
where $\omega_1$ and $\omega_2$ are the fundamental weights of the Lie algebra $\mathfrak{sl}(3)$ and $\lambda_k$ are auxiliary scalar parameters. Simultaneous single valued solution to equation \eqref{Fuchsian-3} and to corresponding antiholomorphic equation can be written in terms of Coulomb integral \cite{Fateev:2006:JETP,Fateev:2007:TMPH} (this solution coinsides up to multiplicative constant with the field $e^{-\Phi_1}$)
\begin{equation}\label{Coulomb-for-Phi13}
   e^{-\Phi_1}=\mathfrak{D}
   |x|^{2\lambda_1-4g/3}|x-1|^{2\lambda_2-4g/3}
   \int \prod_{k=1}^2|t_k|^{2A}|t_k-1|^{2B}|t_k-x|^{2C}\mathcal{D}_2^{2g}(t)\;d^2t_1\,d^2t_2,
\end{equation}
where $\mathcal{D}_2(t)=|t_1-t_2|^2$ and
\begin{equation}
A=-\lambda_1-C,\;\;B=-\lambda_2-C,\;\;C=1+g-\frac{1}{2}(\lambda_1+\lambda_2+\lambda_3).
\end{equation}
The dual solution to the differential equation \eqref{Fuchsian-3} with changed sign before current $W(x)$ (this solution corresponds to the field $e^{-\Phi_2}$) has a form
\begin{equation}\label{Coulomb-for-Phi13-1}
   e^{-\Phi_2}=\mathfrak{D}'
   |x|^{2\lambda_1-8g/3}|x-1|^{2\lambda_2-8g/3}
   \int \prod_{k=1}^2|t_k|^{2A'}|t_k-1|^{2B'}|t_k-x|^{2C'}\mathcal{D}_2^{2g'}(t)\;d^2t_1\,d^2t_2
\end{equation}
with
\begin{equation*}
  A'=A+g,\;\;\;\;B'=B+g,\;\;\;\;C'=C+g,\;\;\;\;g'=-g.
\end{equation*}
Functions $e^{-\Phi_1}$ and $e^{-\Phi_2}$ given by Eqs \eqref{Coulomb-for-Phi13} and
\eqref{Coulomb-for-Phi13-1} define modulo numerical factors the solution to the Toda boundary problem \eqref{Toda3} with three singular points $0$, $1$ and $\infty$ and parameters $\eta_k$ given by Eq \eqref{eta_for_Liouville}. We do not give here explicit expressions for the numerical constants $\mathfrak{D}$ and $\mathfrak{D}'$ before integrals \eqref{Coulomb-for-Phi13} and \eqref{Coulomb-for-Phi13-1} and for the action calculated on this solution, because in this paper we do not suppose to quantize it. 

The results of this section show, that the boundary problem \eqref{Toda3} for the $\mathfrak{sl}(3)$ Toda equation is much more complicated than the corresponding boundary problem for Liouville equation. Even in the case of three singular points one should deal with accessory parameters. Boundary problem \eqref{Toda3} can be reduced to the problem to finding single valued solution to the holomorphic and antiholomorphic Fuchsian differential equations of the third order with given behavior near the singular points \footnote{The most general such equation with three singular points $0$, $1$ and $\infty$ is given by Eq \eqref{Fuchsian-3}.}. We have shown that it can be reduced to the problem of finding values of accessory parameter $\Lambda$ which solves equation \eqref{det}. We suppose, that in the domain \eqref{condition_for_class_solution} the solution to this equation is unique. 

An interesting question how to find parameter $\Lambda$. We have found it in two different cases. First case is when one of the parameters $\eta_k$ is proportional to the fundamental weight (for example $\eta_3=\kappa\omega_2$). In this case accessory parameter $\Lambda$ is given by Eq \eqref{Lambda}. Another interesting case, which corresponds to the differential equation for the quantum field $V_{-b}$ in Liouville field theory gives the value of the parameter $\Lambda=0$. One has to notice, that in both cases the solution to Eq \eqref{Fuchsian-3} is given in terms of Coulomb integrals over a plane (Eqs \eqref{Coulomb-for-hypergeom} and \eqref{Coulomb-for-Phi13} respectively), so these solutions are evidently single valued. An important problem remains unsolved: how to find parameter $\Lambda$ in general case? It is difficult to expect, that solution to Eq \eqref{Fuchsian-3} in general case can be expressed in terms of finite dimensional integral. Because of that we do not have a efficient procedure to find matrix $M_{ij}$ defined by Eq \eqref{Assos}. We suppose to develop the effective numerical method to solve this problem in a future publication.
\section{Classical limit (light exponential fields)}\label{CL-light}
In this section we consider the semiclassical limit of $\mathfrak{sl}(3)$ TFT in the opposite case of light exponential fields $V_{\alpha_k}$ with parameters 
\begin{equation}
  \alpha_k=b\eta_k.
\end{equation}
The solution to the Toda equation \eqref{Toda31} with positive cosmological constant $\mu$ in this case does not exist, because the condition \eqref{condition_for_class_solution} does not satisfied. In order to have a solution, it is useful to perform analytical continuation $\mu\rightarrow-\mu$\footnote{Alternatively, one can consider the correlation functions with the fixed "area". See Ref \cite{Zamolodchikov:1995aa} for details.}. The leading asymptotic behavior of correlation functions at $b\rightarrow0$ is now governed by the solution to the Toda equation with the opposite sign in the r.~h.~s.
\begin{equation}\label{Toda3-}
     \partial\bar{\partial}\phi=-\pi\mu b^2\left(e_1e^{(e_1,\phi)}
     +e_2e^{(e_2,\phi)}\right).
\end{equation}
It is evident, that light exponential fields $V_{b\eta_k}$ do not affect on dynamics, it means, that in this case one has to set
\begin{equation}
   \mathsf{T}=\mathsf{W}=\bar{\mathsf{T}}=\bar{\mathsf{W}}=0
\end{equation}
in Eqs \eqref{HolomorphicEquation}. General solution to Eq \eqref{Toda3-} expressed in terms of solutions to Eqs \eqref{HolomorphicEquation}, which in this case are the polynomials of degree $2$. It is convenient to parameterize these polynomials by nine complex parameters $a_k$, $b_k$ and $c_k$ ($k=1,2,3$) as follows
\begin{equation}\label{Polynom}
  p_k=a_k+b_kz+\frac{c_k}{2}z^2.
\end{equation}
The solution to the Toda equation \eqref{Toda3-} is given by
\begin{equation}\label{Phi-unitary}
     \phi^0(z,\bar{z})=-\rho\log(\pi\mu b^2)+e_1\Phi_1^0(z,\bar{z})+e_2\Phi_2^0(z,\bar{z})  
\end{equation}
with $\rho$ being the Weyl vector and
\begin{equation}\label{Phi-unitary-1}
     \Phi_1^0(z,\bar{z})=-\log
     \left(\vert p_1\vert^2+\vert p_2\vert^2+\vert p_3\vert^2\right),\;\;\;\;
     \Phi_2^0(z,\bar{z})=-\log\left(\vert\tilde{p}_1\vert^2+\vert\tilde{p}_2\vert^2+\vert  
     \tilde{p}_3\vert^2\right),
\end{equation}
where the dual polynomials $\tilde{p}_i$ are defined as
\begin{equation}\label{Polynom-dual}
   \tilde{p}_1=p_2p_3'-p_3p_2',\qquad
   \tilde{p}_2=p_1p_3'-p_3p_1',\qquad
   \tilde{p}_3=p_1p_2'-p_2p_1'.
\end{equation}
Due to Eq \eqref{Wronskian} nine complex parameters $a_k$, $b_k$ and $c_k$ are subject to the $SL(3,C)$ constraint
\begin{equation}\label{SL(3)-constraint}
   \det \begin{pmatrix}
     a_1 & b_1 & c_1 \\
     a_2 & b_2 & c_2 \\
     a_3 & b_3 & c_3 
  \end{pmatrix}=1. 
\end{equation}
The difference with the semiclassical limit of the correlation function of ~"heavy" exponential fields considered in section \ref{CL-heavy} is that the saddle point now is not unique. The action is minimazed on any function $\phi^0(z,\bar{z})$ defined by Eq \eqref{Phi-unitary}. So, in order to obtain the semiclassical expression for the correlation function of the light operators $V_{b\eta_k}$ one should integrate over the space of all polynomials \eqref{Polynom} restricted by the condition \eqref{SL(3)-constraint}. Namely, the semiclassical limit of the $N$-point correlation function is given by the integral
\begin{equation}\label{ClassInt}
      \frac{1}{Z_0}\langle V_{b\eta_1}(z_1,\bar{z}_1)
      \dots V_{b\eta_N}(z_N,\bar{z}_N)\rangle\underset{b\rightarrow0}{\longrightarrow}
      \int\prod_{k=1}^N e^{(\eta_k,\phi^0(z_k,\bar{z}_k))}d\Omega(a_k,b_k,c_k),
\end{equation}
where $Z_0$ is TFT partition function and $d\Omega(a_k,b_k,c_k)$ is the $SL(3,C)$ invariant measure. 

It is evident from Eq \eqref{Phi-unitary-1}, that fields $\Phi_1^0(z,\bar{z})$, $\Phi_2^0(z,\bar{z})$ and  the integral \eqref{ClassInt} have $SU(3)$ invariance. Hence, the integral \eqref{ClassInt} is just proportional to the volume of $SU(3)$. We use Iwasawa decomposition for the $SL(3,C)$ to fix the gauge. Namely, each $SL(3,C)$ matrix can be represented, as a product of $SU(3)$ matrix and uppertriangle matrix with unit determinant:
\begin{equation}\label{Iwasawa}
   SL(3,C)=SU(3)\times
   \begin{pmatrix}
       \varrho &a   &b\\
       0      &\nu &c\\
       0       &0   &\tau 
   \end{pmatrix}
\end{equation}
here $a$, $b$ and $c$ are the complex numbers and $\varrho$, $\nu$ and $\tau$ are the real numbers with the  condition $\varrho\nu\tau=1$. In this gauge polynomials $p_k$ and $\tilde{p}_k$ defined by Eqs \eqref{Polynom} and \eqref{Polynom-dual} will have the following form
\begin{equation}\label{Polynom-fixed}
 \begin{aligned}
  &p_1=\frac{\varrho z^2}{2}+az+b,&\;\;\;\;&p_2=\nu z+c,&\;\;\;\;&p_3=\tau,\\
  &\tilde{p}_1=\nu\tau,&\;\;\;&\tilde{p}_2=\tau(\varrho z+a),&\;\;\;
  &\tilde{p}_3=\frac{\varrho\nu z^2}{2}+c\varrho z+ac-b\nu.
 \end{aligned}
\end{equation}
The measure will transform to
\begin{equation}
   d\Omega(a_k,b_k,c_k)=d\Theta\;d^2a\;d^2b\;d^2c\;
   \varrho^3d\varrho\;\nu d\nu,
\end{equation}
where $d\Theta$ is $SU(3)$ invariant measure. 

In the case of three-point correlation function it is convenient to introduce the notations 
\begin{equation}\label{eta->lambda}
   (\eta_1,e_j)=\lambda_j;\;\;
   (\eta_2,e_j)=\kappa_j;\;\;
   (\eta_3,e_j)=\sigma_j;\;\;j=1,2.
\end{equation}
The semiclassical limit of three-point correlation function has a form
\begin{multline}\label{ClassInt-3point-def}
      \frac{1}{Z_0}\langle V_{b\eta_1}(z_1,\bar{z}_1)V_{b\eta_2}(z_2,\bar{z}_2)
       V_{b\eta_3}(z_3,\bar{z}_3)\rangle\underset{b\rightarrow0}{\longrightarrow}
      \left(\pi\mu b^2\right)^{-\lambda_1-\lambda_2-\kappa_1-\kappa_2-\sigma_1-\sigma_2}
      \times\\\times
      J(\lambda_1,\lambda_2;\kappa_1,\kappa_2;\sigma_1,\sigma_2|z_1,\bar{z}_1,z_2,\bar{z}_2,z_3,\bar{z}_3),
\end{multline}
with
\begin{multline}\label{ClassInt-3point}
       J(\lambda_1,\lambda_2;\kappa_1,\kappa_2;\sigma_1,\sigma_2|z_1,\bar{z}_1,z_2,\bar{z}_2,z_3,\bar{z}_3)
      =\\=\int \exp\left(\sum_{k=1}^2\left(
      \lambda_k\Phi^0_k(z_1,\bar{z}_1)+\kappa_k\Phi^0_k(z_2,\bar{z}_2)+
      \sigma_k\Phi^0_k(z_3,\bar{z}_3)\right)\right)
      d^2a\;d^2b\;d^2c\;\varrho^3d\varrho\;\nu d\nu,
\end{multline}
where functions $\Phi^0_k(z)$ are given by Eq \eqref{Phi-unitary-1} with polynomials $p_k$  and $\tilde{p}_k$ defined by Eqs \eqref{Polynom-fixed}. The coordinate dependence of the integral \eqref{ClassInt-3point} is fixed by the projective invariance and we can set $z_1=0$, $z_2=1$ and $z_3=\infty$. More exactly we consider the limit of the integral \eqref{ClassInt-3point}
\begin{equation}\label{ClassInt-3}
   \mathfrak{J}(\lambda_1,\lambda_2;\kappa_1,\kappa_2;\sigma_1,\sigma_2)=
   \lim_{z_3\rightarrow\infty}
   |z_3|^{4(\sigma_1+\sigma_2)}J(\lambda_1,\lambda_2;\kappa_1,\kappa_2;\sigma_1,\sigma_2|0,1,z_3).
\end{equation}
Function $\mathfrak{J}(\lambda_1,\lambda_2;\kappa_1,\kappa_2;\sigma_1,\sigma_2)$, which defines the semiclassical limit of three-point correlation function, will be the main object of this section.
For convenience, it is better to renormalize parameters $a,b,c,\varrho$ and $\nu$
\begin{equation}
  \begin{gathered}
      a\rightarrow\nu a;\;\;b\rightarrow\tau b;\;\;c\rightarrow\tau c,\\
      \varrho\rightarrow\tau\varrho;\;\;\nu\rightarrow\tau\nu
  \end{gathered}
\end{equation}
and take into account the condition $\tau\nu\varrho=1$. Then function $\mathfrak{J}(\lambda_1,\lambda_2;\kappa_1,\kappa_2;\sigma_1,\sigma_2)$ can be represented by the eight-dimensional integral
\begin{equation}\label{Iwasawa1}
  \mathfrak{J}(\lambda_1,\lambda_2;\kappa_1,\kappa_2;\sigma_1,\sigma_2)=4^{\sigma_1+\sigma_2}\,
  \int\frac{\varrho^{2\delta}\nu^{2\Delta}}
  {Z_1^{\lambda_1}Z_2^{\lambda_2}Z_3^{\kappa_1}Z_4^{\kappa_2}}\;\frac{d\varrho}{\varrho}
  \frac{d\nu}{\nu}\;d^2a\,d^2b\,d^2c
\end{equation}
with
\addtocounter{equation}{-1}
\begin{subequations}
\begin{equation}\label{Zk}
   \begin{aligned}
    &Z_1=1+|b|^2+|c|^2,\qquad&&Z_3=1+|c+\nu|^2+\Bigl|b+\nu a+\frac{\varrho}{2}\Bigr|^2,\\
    &Z_2=1+|a|^2+|ac-b|^2,\qquad&&Z_4=1+\Bigl|a+\frac{\varrho}{\nu}\Bigr|^2+
     \Bigl|\frac{\varrho}{2}+\frac{c\varrho}{\nu}+ac-b\Bigr|^2
   \end{aligned}
\end{equation}
and
\begin{equation}\label{delta-Delta}
  \begin{aligned}
    &\delta=\frac{1}{3}\left(\lambda_1+2\lambda_2+\kappa_1+2\kappa_2-2\sigma_1-\sigma_2\right),\\
    &\Delta=\frac{1}{3}\left(\lambda_1-\lambda_2+\kappa_1-\kappa_2+\sigma_1-\sigma_2\right).  
  \end{aligned}
\end{equation}
\end{subequations}
After non-trivial transformations (see appendix \ref{Eval_Iwasawa} for details), integral \eqref{Iwasawa1} can be reduced to the three dimensional Barnes like integral
\begin{multline}\label{3_Fold_Int}
  \mathfrak{J}(\lambda_1,\lambda_2;\kappa_1,\kappa_2;\sigma_1,\sigma_2)=
  4^{\lambda_1+\kappa_1+\sigma_1-\Delta}
  \times\\\times
  \frac{\Gamma\bigl(\lambda_1+\kappa_1+\sigma_1-\Delta-2\bigr)
   \Gamma\bigl(\lambda_2+\kappa_2+\sigma_2+\Delta-2\bigr)}
  {\Gamma(\lambda_1)\Gamma(\lambda_2)\Gamma(\lambda_1+\lambda_2-1)
  \Gamma(\kappa_1)\Gamma(\kappa_2)\Gamma(\kappa_1+\kappa_2-1)
  \Gamma(\sigma_1)\Gamma(\sigma_2)\Gamma(\sigma_1+\sigma_2-1)}
  \times\\\times\frac{1}{(2\pi i)^3}\int\,du\,ds\,dy\,\,4^{-u}\,
  \Gamma(y)\Gamma(s)\Gamma(u)\Gamma(\sigma_1-u-s)\Gamma(\lambda_2+\Delta-u-s)
  \Gamma(\sigma_1-\kappa_2-\Delta+y)\cdot\\
  \Gamma(\sigma_1-\kappa_2+\lambda_1-\Delta-u-s)\Gamma(\kappa_1+\kappa_2-1-y)
  \Gamma(\kappa_1+\lambda_1-\Delta-1-y)\Gamma(\kappa_2+\sigma_2+\Delta-1-y)\cdot\\
  \Gamma(\lambda_2-\kappa_1+\Delta+y)
  \frac{\Gamma(\kappa_1-\lambda_2-\Delta+s)\Gamma(\kappa_2+\Delta-\sigma_1+s)
  \Gamma(u-\Delta)}{\Gamma(s+y)\Gamma(\lambda_1+\kappa_1+\sigma_1-\Delta-1-u-s-y)},
\end{multline}
where the integral over variables $u$, $s$ and $y$ goes along imaginary axis. Integral \eqref{3_Fold_Int} is convergent in the domain
\begin{equation}
   \begin{aligned}
     &1+(\eta_1-\rho,h_i)+(\eta_2-\rho,h_j)+(\eta_3-\rho,h_k)>0\;\;\;
     \text{if}\;\;\;(h_i+h_j+h_k,\rho)>-1,\\
     &1+(\eta_1-\rho,h_i)+(\eta_2-\rho,h_j)+(\eta_3-\rho,h_k)<0
     \;\;\;\text{if}\;\;\;(h_i+h_j+h_k,\rho)\leq-1,
   \end{aligned}
\end{equation}
where vector parameters $\eta_1$, $\eta_2$ and $\eta_3$ are related with parameters $\lambda_k$, $\kappa_k$ and $\sigma_k$ by Eq \eqref{eta->lambda}. By definition, the integral \eqref{ClassInt} and hence the integral \eqref{3_Fold_Int} should be symmetric with respect to substitution $\lambda\rightarrow\kappa\rightarrow\sigma$ and also with respect to substitution $1\rightarrow2$, but these symmetries are not evident from it explicit form. We were not able to represent this integral in terms of the finite sum of the known functions. Another alternative expression for the function $\mathfrak{J}(\lambda_1,\lambda_2;\kappa_1,\kappa_2;\sigma_1,\sigma_2)$ in terms of three-dimensional integral of Tricomi functions is given in the appendix \ref{Eval_Iwasawa}.
We see, that the integral \eqref{3_Fold_Int}, which is semiclassical limit of the three-point correlation function, is already rather nontrivial object. Henceworth, it is difficult to expect that quantum expression will have  a simple form.

Integral in Eq \eqref{3_Fold_Int} simplifies drastically if one of the parameters $\lambda_k$, $\kappa_k$ or $\sigma_k$ equals to zero. For example, let us consider the limit  $\sigma_1\rightarrow0$. We see, that due to the factor $\Gamma(s)\Gamma(u)\Gamma(\sigma_1-u-s)$ in the integrand in \eqref{3_Fold_Int} in this case integral \eqref{3_Fold_Int} develops a pole, when we pinch points $u$ and $s$ near the point $u=s=0$. This pole cancelles with a zero coming from function $\Gamma(\sigma_1)^{-1}$ in the prefactor of \eqref{3_Fold_Int} and remaining integral over the variable $y$ can be performed using the first Barnes lemma \eqref{Barnes1}. As a result we obtain a simple expression for the function \eqref{3_Fold_Int} 
\begin{multline}\label{Semiclass}
   \mathfrak{J}(\lambda_1,\lambda_2;\kappa_1,\kappa_2;0,\sigma_2)=
    4^{\frac{\sigma_2}{3}+(\eta_1+\eta_2,\rho)}\times\\\times
   \frac{\prod_{ij}\Theta_{ij}}
   {\Gamma(\sigma_2)\prod_{e>0}\Gamma(1+(\eta_1-\rho,e))
   \Gamma(1+(\eta_2-\rho,e))},
\end{multline}
where vectors $\eta_k$ are given by Eq \eqref{eta->lambda} and
\begin{equation*}
   \Theta_{ij}=\Biggl\{
   \begin{array}{ll}
       \Gamma(\frac{\sigma_2}{3}+(\eta_1-\rho,h_i)+
    (\eta_2-\rho,h_j))&\text{if}\;\;(h_i+h_j,\rho)>-1\\
       \Gamma(1-\frac{\sigma_2}{3}-(\eta_1-\rho,h_i)-
    (\eta_2-\rho,h_j))&\text{if}\;\;(h_i+h_j,\rho)\leqslant-1 
   \end{array}
\end{equation*}
The result \eqref{Semiclass} agrees with the corresponding limit of the three-point correlation function \eqref{C}, where $\alpha_1=b\eta_1$, $\alpha_2=b\eta_2$ and $\varkappa=b\sigma_2$. If all vectors $\eta_k$ are general, the integral \eqref{3_Fold_Int} is much more involved object and we plan to study its analytical properties in a future publication.

Another interesting point, where the integral \eqref{3_Fold_Int} can be calculated exactly, is defined by the condition $\sigma_1=-m$ with integer $m$, i.~e. $\eta_3=\sigma_2\omega_2-m\omega_1$\footnote{In quantum case field $V_{\alpha}$ with parameter $\alpha=\sigma_2\omega_2-mb\omega_1$ is partially degenerate (it has a null-vector at the level $m+1$). The three-point correlation function with such a field can be expressed in terms of finite dimensional Coulomb integral, as it will be shown in Ref \cite{Part-Deux}, and generalizes the answer \eqref{C-shifted} for the case of Lie algebra $\mathfrak{sl}(3)$.}. In this case function \eqref{3_Fold_Int} can be given by triple finite sum, which in general contains 
\begin{equation}\label{N(m)}
  N(m)=\frac{(m+1)(m+2)(m+3)}{6}
\end{equation}
terms and has a form
\begin{multline}\label{3_Fold_Int-m}
  \mathfrak{J}(\lambda_1,\lambda_2;\kappa_1,\kappa_2;-m,\sigma_2)= 
  4^{\lambda_1+\kappa_1-\Delta}\Gamma(\lambda_1+\kappa_1-2-m-\Delta)\Gamma(-m-\Delta)
  \Gamma(\lambda_1-\kappa_2-m-\Delta)\\\Gamma(\kappa_1-\lambda_2-m-\Delta)
  \Gamma(\lambda_2+\kappa_2+\Delta-1)\times\\\times
  \frac{\Gamma(\lambda_1-(m+1)-\Delta)\Gamma(\kappa_1-(m+1)-\Delta)
  \Gamma(\lambda_2+\Delta)\Gamma(\kappa_2+\Delta)}
 {\Gamma(\lambda_1)\Gamma(\lambda_2)\Gamma(\kappa_1)\Gamma(\kappa_2)\Gamma(\sigma_2)
 \Gamma(\lambda_1+\lambda_2-1)\Gamma(\kappa_1+\kappa_2-1)}\\
 \sum_{s_1,s_2,s_3=0}^m\frac{(-4)^{-s_1-s_2-s_3}(-m)_{s_1+s_2+s_3}(-\Delta-m)_{s_1+s_2+s_3}
  (1-\Delta-m-\sigma_2+s_1+s_2+s_3)_{s_1}}
  {s_1!s_2!s_3!}\\\times
  (\lambda_1-\kappa_2-m-\Delta)_{m-s_1-s_3}(\kappa_1-\lambda_2-m-\Delta)_{m-s_1-s_2}
  (\lambda_1-(m+1)-\Delta)_{s_3}(\kappa_1-(m+1)-\Delta)_{s_2}\\
  (\lambda_2+\Delta)_{m-s_3}(\kappa_2+\Delta)_{m-s_2}(\sigma_2-(m+1))_{s_2+s_3}
  (3-\Delta-\sigma_2-\lambda_2-\kappa_2)_{m-s_2-s_3}
\end{multline}
where
\begin{equation*}
  (x)_k=x(x+1)\dots(x+k-1)
\end{equation*}
and $\Delta$ is defined by Eq \eqref{delta-Delta} with $\sigma_1=-m$.

\begin{figure}
  \begin{center}
   \epsfig{figure=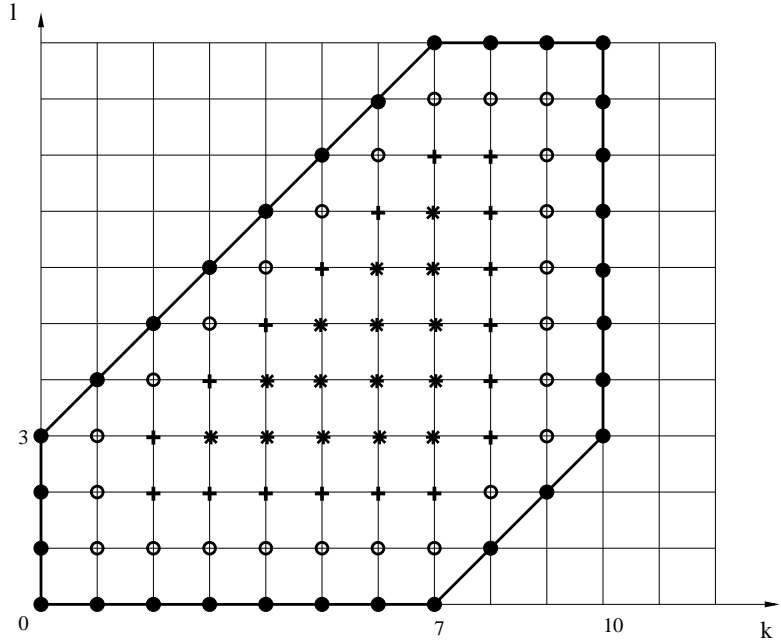,width=.68\textwidth}
   \caption{Fusion rules for the completely degenerate field specified by the highest weight  
   $\Omega=m\omega_1+n\omega_2$ with $(m,n)=(7,3)$. Each point corresponds to
    admissible pair $(k,l)$ in Eq \eqref{Classical_Fusion_Rules}. Points on the boundary correspond to the 
    weights $h_{mn}^{kl}$ with multiplicity one, points on the next layer boundary (which are shown by 
    circles) correspond to the weights $h_{mn}^{kl}$ with multiplicity two, points, shown by crosses, 
    correspond to multiplicity three and points, shown by stars (hexagon is degenerate into triangle), 
    correspond to multiplicity four. Total number of points (including multiplicities) coincides with 
    dimension of the representation $(7,3)$: $(m+1)(n+1)(m+n+2)/2=192$.}
   \label{fig:fusion}
  \end{center}
\end{figure}

It is interesting to consider Eq \eqref{3_Fold_Int-m} in the limit $\sigma_2\rightarrow-n$ (corresponding quantum field will have parameter $\alpha=-mb\omega_1-nb\omega_2$ and will be completely degenerate). If we substitute $\sigma_2=-n$ in Eq \eqref{3_Fold_Int-m}, then for general parameters $\lambda_k$ and $\kappa_k$ function $\mathfrak{J}(\lambda_1,\lambda_2;\kappa_1,\kappa_2;-m,-n)$ will be zero due to the factor $1/\Gamma(\sigma_2)$. This represents the fact, that the quantum three-point correlation function $C(-mb\omega_1-nb\omega_2,\alpha_1,\alpha_2)$ (where $\alpha_k=b\eta_k$) with completely degenerate field and two arbitrary fields equals to zero. However, if one tunes the parameter $\alpha_2$ for the fixed value of the parameter $\alpha_1$ in a special way (there are only finite number of such possibilities), then this correlation function will be infinite, namely, it will have a double pole in this limit.  In this case, as was explained in section \ref{TFT}, it is reasonable to study the structure constants of OPE, which are defined as the main residues of three-point correlation function. At semiclassical level it means, that  one should fix parameters $\lambda_1$ and $\lambda_2$ and for given value of the parameter $\sigma_1=-m$  find such values of the parameters $\kappa_1$ and $\kappa_2$, that function \eqref{3_Fold_Int-m} has a double pole in the limit $\sigma_2\rightarrow-n$. Namely, if parameters  $\kappa_1$ and $\kappa_2$ approach to the values
\begin{equation}\label{kappas}
  \kappa_1=\lambda_2-n+2l-k;\;\;\kappa_2=\lambda_1-m+2k-l,
\end{equation}
i.~e. $\eta_2\rightarrow\eta_1^{*}-n\omega_1-m\omega_2+le_1+ke_2$, 
then the double pole appears for integer $k$ and $l$ restricted by the conditions
\begin{equation}\label{zone-1}
  k\geq0;\;\;l\leq m+n;\;\;l\geq k-m.
\end{equation}
From the quantum point of view, it means, that correlation function
\begin{equation}
  C(-mb\omega_1-(nb+\epsilon)\omega_2,\alpha_1,\alpha_1^{*}-nb\omega_1-mb\omega_2+lbe_1+kbe_2)
  \sim\frac{1}{\epsilon^2}
\end{equation}
becomes infinite in the limit $\epsilon\rightarrow0$. It follows from the fact, that this correlation function coinsides up to Weyl transformation with correlation function
\begin{multline}\label{C-light} 
  C(-mb\omega_1-(nb+\epsilon)\omega_2,\alpha_1,\alpha^{*}_1-b\tilde{h}_{mn}^{kl})=\\=
  R^{-1}(\alpha_1-bh_{mn}^{kl})
  C(-mb\omega_1-(nb+\epsilon)\omega_2,\alpha_1,2Q-\alpha_1+bh_{mn}^{kl}),
\end{multline}
where
\begin{equation}\label{h-mn-kl}
  h_{mn}^{kl}=m\omega_1+n\omega_2-ke_1-le_2,\quad
  \tilde{h}_{mn}^{kl}=m\omega_2+n\omega_1-ke_2-le_1
\end{equation}
and $R(\alpha)$ is the maximal reflection amplitude defined by Eq \eqref{MaxRefl}. Correlation function in the r.~h.~s. of Eq \eqref{C-light} has a double pole, because the sum of all parameters satisfies the screening condition in the limit $\epsilon\rightarrow0$
\begin{equation}
    -mb\omega_1-nb\omega_2+\alpha_1+2Q-\alpha_1+bh_{mn}^{kl}=2Q-kbe_1-lbe_2.
\end{equation}
The main residue in this pole should be associated with structure constant
\begin{equation}\label{Structure_Constant_light}
   C(-mb\omega_1-(nb+\epsilon)\omega_2,\alpha_1,2Q-\alpha_1+bh_{mn}^{kl})=
   \frac{1}{\epsilon^2}C_{-mb\omega_1-nb\omega_2,\alpha_1}^{\alpha_1-bh_{mn}^{kl}}.
\end{equation}
So, in this case it is reasonable to consider the semiclassical limit of this structure constant. Maximal reflection amplitude $R(b\eta)$, which due to Eq \eqref{MaxRefl-def} coinsides with inverse two-point correlation function, has the following semiclassical limit (here $\eta=\kappa_1\omega_1+\kappa_2\omega_2$)
\begin{equation}\label{R-semiclass}
   Z_0R(b\eta)\underset{b\rightarrow0}{\longrightarrow}
   \left(\frac{\pi\mu b^2}{2}\right)^{2\kappa_1+2\kappa_2}
   (\kappa_1-1)(\kappa_2-1)(\kappa_1+\kappa_2-2). 
\end{equation}
Semiclassical limit of the structure constant \eqref{Structure_Constant_light} can be obtained by multiplying Eqs \eqref{3_Fold_Int-m} and \eqref{R-semiclass}, substituting \eqref{kappas}, taking the limit $\sigma_2\rightarrow-n$ and finding the main residue in this limit (one should take also into account the factor $(\pi\mu b^2)^{-\lambda_1-\lambda_2-\kappa_1-\kappa_2-\sigma_1-\sigma_2}$ in Eq \eqref{ClassInt-3point-def}, which relates function \eqref{Semiclass} with the semiclassical limit of the three-point correlation function). We see from Eqs \eqref{ClassInt-3point-def}, \eqref{C-light} and \eqref{R-semiclass}, that the structure constant defined by Eq \eqref{Structure_Constant_light} has a smooth semiclassical limit independent on partition function $Z_0$. If we parameterize $\alpha=b\lambda_1\omega_1+b\lambda_2\omega_2$, then the semiclassical limit of the structure constant $C_{-mb\omega_1-nb\omega_2,\alpha}^{\alpha-bh_{mn}^{kl}}$ defined by the relation
\begin{equation}
  C_{-mb\omega_1-nb\omega_2,\alpha}^{\alpha-bh_{mn}^{kl}}
  \underset{b\rightarrow0}{\longrightarrow}\mathbb{C}_{mn}^{kl}(\lambda_1,\lambda_2)
\end{equation}
can be written in a form
\begin{multline}\label{Classical_Fusion_Rules}
   \mathbb{C}_{mn}^{kl}(\lambda_1,\lambda_2)=\left(\frac{\pi\mu b^2}{2}\right)^{k+l}
   \frac{m!\,n!}
   {\Gamma(\lambda_1)\Gamma(\lambda_2)\Gamma(\lambda_1+\lambda_2-1)}\times\\\times
   \frac{\Sigma_{m,n}^{k,l}(\lambda_1,\lambda_2)}
   {\Gamma(\lambda_1-m+2k-l-1)\Gamma(\lambda_2-n+2l-k-1)\Gamma(\lambda_1+\lambda_2-m-n+k+l-2)}
\end{multline}
with function $\Sigma_{m,n}^{k,l}(\lambda_1,\lambda_2)$ defined as
\begin{multline}\label{Sigma_mn^kl}
   \Sigma_{m,n}^{k,l}(\lambda_1,\lambda_2)=
   \frac{(-1)^k2^{l-k}4^m}{m!\,k!\,(m+l-k)!\,(m+n-l)!}
   \Gamma(\lambda_1-m+k)\Gamma(\lambda_2+l-k)\times\\\times
   \Gamma(\lambda_1+\lambda_2-m+k-1)
   \Gamma(\lambda_1-m+k-l-1)\Gamma(\lambda_2-m-n+l-1)\Gamma(\lambda_1+\lambda_2-m-n+l-2)
   \times\\\times
   \sum_{s_1,s_2,s_3=0}^m\biggl[\frac{(-4)^{-s_1-s_2-s_3}}{s_1!s_2!s_3!}\,
   (1-l+k-m+n+s_1+s_2+s_3)_{s_1}\,(k-l-m)_{s_1+s_2+s_3}\,(-k)_{m-s_1-s_3}\times\\\times
   (-m)_{s_1+s_2+s_3}
   (l-m-n)_{m-s_1-s_2}\,(-1-m-n)_{s_2+s_3}\,(\lambda_1+k-m-l-1)_{s_3}\,(\lambda_2+l-m-n-1)_{s_2}\,
   \times\\\times
   (\lambda_1-m+k)_{m-s_2}\,(\lambda_2+l-k)_{m-s_3}\,
   (3-k+m+n-\lambda_1-\lambda_2)_{m-s_2-s_3}\biggr].
\end{multline}
From the explicit form of the function $\Sigma_{m,n}^{k,l}(\lambda_1,\lambda_2)$ it can be shown, that
it is non-zero only if numbers $k$ and $l$, besides inequalities \eqref{zone-1}, are restricted by the conditions\footnote{These conditions follow also evidently from the functional relations \eqref{Sigma-Relat} (see below).}
\begin{equation}\label{zone-2}
   l\geq0,\quad k\leq m+n,\quad k\geq l-n.
\end{equation}
It is convenient to picture admissible pairs $(k,l)$ (they are defined by the conditions \eqref{zone-1} and \eqref{zone-2}), as a set of points on a plane. Namely, function \eqref{Classical_Fusion_Rules} is non-zero only if points $(k,l)$ lay inside of a hexagon
\begin{equation}\label{Cond-nonzeros}
   k\geq0;\;\;l\geq0;\;\;\;\;\;\;m+n\geq k;\;\;m+n\geq l;\;\;\;\;\;\;
   n+k\geq l;\;\;m+l\geq k.
\end{equation}
An example of fusion rules for the completely degenerate field specified by the highest weight $\Omega=m\omega_1+n\omega_2$ with $(m,n)=(7,3)$ is shown on a fig. \ref{fig:fusion}. 
We note also, that these fusion rules coinside exactly with quantum fusion rules, which were defined in section \ref{TFT}.

The sum in \eqref{Sigma_mn^kl} can be reduced to a simple product for all points $(k,l)$, which have multiplicity one (i.~e. for points on the boundary of the hexagon). It follows from the fact, that function $\Sigma_{mn}^{kl}(\lambda_1,\lambda_2)$ satisfies remarkable functional relations\footnote{We suppose to give a proof of quantum version of relations \eqref{Sigma-Relat} in Ref \cite{Part-Deux}.}. Namely,
\begin{multline}\label{Sigma-Relat}
  \Sigma_{m,n}^{k,l}(\lambda_1,\lambda_2)=\Sigma_{n,m}^{l,k}(\lambda_2,\lambda_1)=\\=
  (-1)^{m-k}\Sigma_{k,m+n-k}^{m,m+l-k}(\lambda_1+k-m,\lambda_2)
  =\Sigma_{m+l-k,n+k-l}^{l,k}(\lambda_1-l+k,\lambda_2-k+l).
\end{multline}
In quantum case the structure constant, which corresponds to the weight $h_{mn}^{kl}$ with multiplicity 1, as was noticed in section \ref{TFT}, can be represented as a product of $\gamma-$functions.

Using functional relations \eqref{Sigma-Relat} it can be shown, that the number of terms in the sum \eqref{Sigma_mn^kl} can be reduced to the minimum of numbers
\begin{equation}
  N(m),\;N(n),\;N(k),\;N(l),\;N(m+n-k),\;N(m+n-l),\;N(m+l-k),\;N(n+k-l),
\end{equation}
where $N(m)$ is given by Eq \eqref{N(m)}. It is evident from the figure \ref{fig:fusion} that this number 
depends only on the multiplicity $\mathfrak{h}$ of the corresponding weight $h_{mn}^{kl}$ of the representation with highest weight $m\omega_1+n\omega_2$. In quantum case expression for the structure constant \eqref{C-light} for completely degenerate field can be reduced to the $4(\mathfrak{h}-1)$-dimensional Coulomb integral  \cite{Part-Deux}. For $\mathfrak{h}=1$ this correlation function can be given, as a product of $\gamma$-functions. For $\mathfrak{h}=2$ it can be expressed in terms of hypergeometric function of the type $(3,2)$ at unity, while for $\mathfrak{h}>2$ it has more complicated structure.  This fact clarifies the statement, which was done in section \ref{TFT}, that the complexity of the structure constant depends drastically on the multiplicity of the corresponding weight. 

We note also, that due to Eq \eqref{Structure_Constant_degenerate} structure constant \eqref{Structure_Constant_light} coinsides up to multiplicative factor with Coulomb integral 
\begin{equation}
   C_{-mb\omega_1-nb\omega_2,\alpha}^{\alpha-bh_{mn}^{kl}}=
   (-\pi\mu)^{k+l}I_{k,l}
   (-mb\omega_1-nb\omega_2,\alpha,2Q-\alpha+mb\omega_1+nb\omega_2-kbe_1-lbe_2),
\end{equation}
where integral $I_{k,l}(-mb\omega_1-nb\omega_2,\alpha,2Q-\alpha+mb\omega_1+nb\omega_2-kbe_1-lbe_2)$ is defined by Eq \eqref{I} for the case $n=3$, which is studied in details in appendix \ref{SL(3)-Int-def}. Using the notations of the appendix \ref{SL(3)-Int-def} this integral equals
\begin{multline}
  I_{k,l}(-mb\omega_1-nb\omega_2,\alpha,2Q-\alpha+mb\omega_1+nb\omega_2-kbe_1-lbe_2)=\\=
  \frac{k!l!}{\pi^{k+l}}\,\mathcal{I}_{kl}(-b^2\lambda_1,-b^2\lambda_2,-mb^2,-nb^2),
\end{multline}
where the integral $\mathcal{I}_{kl}(-b^2\lambda_1,-b^2\lambda_2,-mb^2,-nb^2)$ is given by Eq \eqref{SL(3)-Int}.
Using  asymptotic \eqref{SL(3)-Int-ass1} for this integral, we can obtain additional representation for the semiclassical limit of the structure constant \eqref{Structure_Constant_light}.

Let us say few words about the semiclassical limit of general $\mathfrak{sl}(n)$ TFT. This limit is governed by the classical equation\footnote{In Eq \eqref{sl(n)toda} we have set for shortness $\pi\mu b^2=1$.}
\begin{equation}\label{sl(n)toda}
   \partial\bar{\partial}\phi+\sum_{k=1}^{n-1}e_ke^{(e_k,\phi)}=0.
\end{equation}
One can show, that as consequence of Eq \eqref{sl(n)toda}, field $e^{-(\omega_1,\phi)}$ satisfies holomorphic and antiholomorphic differential equations of the order $n$
\begin{equation}\label{sl(n)holomorphic}
  \begin{aligned}
    &(-\partial^n+\mathsf{T}\partial^{n-2}+\dots)e^{-(\omega_1,\phi)}=0\\
    &(-\bar{\partial}^n+\bar{\mathsf{T}}\bar{\partial}^{n-2}+\dots)e^{-(\omega_1,\phi)}=0
  \end{aligned}
\end{equation}
where by $\dots$ denoted terms with derivatives of the smaller order. As a consequence of Eq \eqref{sl(n)holomorphic} field $e^{-\Phi_1}$ can be presented as a bilinear combination of the holomorphic and antiholomorphic solutions to \eqref{sl(n)holomorphic}
\begin{equation}
   e^{-(\omega_1,\phi)}=\sum_{k=1}^n\Psi_k\bar{\Psi}_k.
\end{equation}
It follows from Eq \eqref{sl(n)toda} that fields $e^{-(\omega_k,\phi)}$ with $k\neq 1$ have a form
\begin{equation}\label{Small_Wron}
   e^{-(\omega_k,\phi)}=
   \hspace*{-15pt}\sum_{j_1<j_2<\dots<j_k}^{n}
   \begin{vmatrix}
    \Psi_{j_1}  & \Psi_{j_2}  & \dots & \Psi_{j_k}\\
    \partial\Psi_{j_1} & \partial\Psi_{j_2} & \dots & \partial\Psi_{j_k}\\
    \dots&\hdotsfor{2}&\dots\\
    \dots&\hdotsfor{2}&\dots\\
    \partial^{k-1}\Psi_{j_1} & \partial^{k-1}\Psi_{j_2} & \dots & \partial^{k-1}\Psi_{j_k} 
   \end{vmatrix}
   \begin{vmatrix}
    \bar{\Psi}_{j_1}  & \bar{\Psi}_{j_2}  & \dots & \bar{\Psi}_{j_k}\\
    \bar{\partial}\bar{\Psi}_{j_1} & 
    \bar{\partial}\bar{\Psi}_{j_2} & \dots & \bar{\partial}\bar{\Psi}_{j_k}\\
    \dots&\hdotsfor{2}&\dots\\
    \dots&\hdotsfor{2}&\dots\\
    \bar{\partial}^{k-1}\bar{\Psi}_{j_1} & 
    \bar{\partial}^{k-1}\bar{\Psi}_{j_2} & \dots & \bar{\partial}^{k-1}\bar{\Psi}_{j_k} 
   \end{vmatrix}.
\end{equation}
The consistency condition is that the product of Wronskians of holomorphic and antiholomorphic solutions equals to unity
\begin{equation}\label{Big_Wron}
    \mathbb{W}[\Psi_1,\dots,\Psi_n]\mathbb{W}[\bar{\Psi}_1,\dots,\bar{\Psi}_n]=1.
\end{equation}
In the case of light exponentials, all currents in  Eq \eqref{sl(n)holomorphic} are equal to zero. In this case functions $\Psi_k$ will be the polynomial of degree $n-1$
\begin{equation}
    \Psi_k=a_k^{(1)}+a_k^{(2)}z+\frac{a_k^{(3)}z^2}{2}+\dots+\frac{a_k^{(n)}z^{n-1}}{n-1}.
\end{equation}
The condition \eqref{Big_Wron} transforms to $SL(n,C)$ constraint for the matrix $a_k^{(j)}$
\begin{equation}
    \det
    \begin{pmatrix}
      a_1^{(1)}  & a_1^{(2)}  & \dots & a_1^{(n)}\\
      a_2^{(1)} & a_2^{(2)} & \dots & a_2^{(n)} \\
      \dots&\hdotsfor{2}&\dots\\
      \dots&\hdotsfor{2}&\dots\\
      a_n^{(1)} & a_n^{(2)} & \dots & a_n^{(n)}
    \end{pmatrix}=1
\end{equation}
Semiclassical limit of the correlation function of the light exponentials described by the integral
\begin{equation}
      \frac{1}{Z_0}\langle V_{b\eta_1}(z_1,\bar{z}_1)
      \dots V_{b\eta_N}(z_N,\bar{z}_N)\rangle\rightarrow
      \int\prod_{k=1}^N
      e^{(\eta_k,\phi(z_k,\bar{z}_k))}
      d\Omega(a_k^{(j)}),
\end{equation}
here $d\Omega(a_k^{(j)})$ is $SL(n,C)$ invariant measure.
We suppose to consider semiclassical calculations, which were done here, for $n>3$ in other publication. 
\section{Minisuperspace limit}\label{MSSL}
\begin{figure}
  \begin{center}
   \epsfig{figure=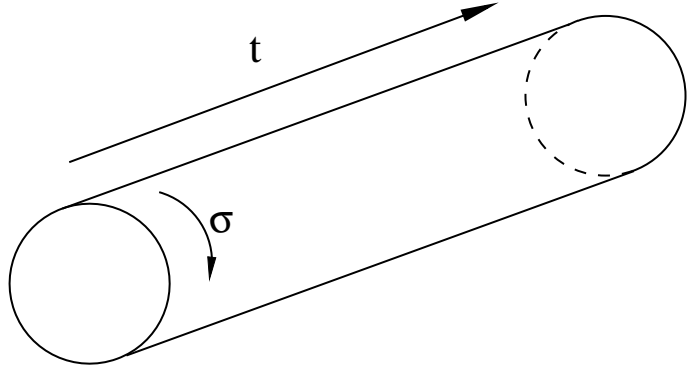,width=.6\textwidth}
   \caption{Cylinder used for minisuperspace approximation.}
   \label{fig:msp}
  \end{center}
\end{figure}

It is interesting to consider another limit of $\mathfrak{sl}(n)$ TFT at $b\rightarrow 0$. In this limit in the Hamiltonian picture, associated with radial quantization\footnote{One should take the geometry of semi-infinite cylinder of circumference $\sigma\in[0,2\pi]$ (fig. \ref{fig:msp})  and consider the states on the circle.}, we take into account only the zero mode dynamics (minisuperspace approach). 
In this approximation the state created by the operator $V_{Q+iP_j}$ corresponds to the wave function
\begin{equation*}
V_{Q+iP_j}\rightarrow \Psi^{(n)}_{P_j}(x),
\end{equation*} 
where $x$ is a zero mode of the field $\varphi$. The function $\Psi^{(n)}_{P}(x)$ ($\mathfrak{sl}(n)$ Whittaker function) satisfies Scr\"odinger equation
\begin{equation}\label{mspShrodinger}
\left(-\nabla_{x}^{2}+2\pi\mu\sum_{i=1}^{n-1} e^{b(e_{i}x)}\right)
\Psi^{(n)}_{P}(x)=P^{2}\Psi^{(n)}_{P}(x),
\end{equation}
and in the region $(e_i,x)<0$ (Weyl chamber) possesses the asymptotic
\begin{equation}\label{mspAss}
   \Psi^{(n)}_{P}(x)\sim\exp
   (i(P,x))+\sum_{\hat{s}\in\mathcal{W}}S_{\hat{s}}(P)\exp (i(\hat{s}(P),x)),
\end{equation}
where the sum runs over all elements of the Weyl group $\mathcal{W}$ besides identical. The coefficients $S_{\hat{s}}(P)$ are known  exactly \cite{Olshanetsky:1981dk} and can be obtained from the reflection amplitude \eqref{ReflAmp} in semiclassical limit $b\rightarrow 0$
\begin{equation*}
     S_{\hat{s}}(P)=\prod_{e>0}\left(\frac{\pi\mu}{b^2}
     \right)^{\frac{i}{2b}(\hat{s}(P)-P,e)}
     \frac{\Gamma\left(-\frac{i(\hat{s}(P),e)}{b}\right)}
     {\Gamma\left(-\frac{i(P,e)}{b}\right)}\;. 
\end{equation*}
One can show, that conditions \eqref{mspShrodinger} and \eqref{mspAss} determine the Whittaker function unambiguously. The minisuperspace approximation is valid if $P_j/b$ are fixed at the limit $b\rightarrow0$. If we take $\alpha_3=ibq$ and $P_k=bp_k$ for $k=1,2$,  then the minisuperspace limit of the three-point correlation function  \eqref{Ctotal} can be represented by the integral
\begin{equation}\label{mspLimit}
   C(Q+ibp_1,Q+ibp_2,ibq)\longrightarrow
   \int d\vec{x}\,\Psi^{(n)}_{bp_{1}}(x)\Psi^{(n)}_{bp_{2}}(x)e^
   {ib(q,x)}\;. 
\end{equation}

The theory of the $\mathfrak{sl}(n)$ Whittaker functions has some long history \cite{Kostant}. In particular, different explicit integral representations for these functions exist \cite{Kharchev:2000ug,Stade:1990} (we will use here representation, which can be extracted from the paper \cite{Stade:1990}). All these functions are build from the Macdonald function $K_{\nu}(y)$ (which can be given by the integral \eqref{Kint}) by the recursive integral representation. To make sense the future statements we  define $\Psi^{(0)}_{P}(x)=\Psi^{(1)}_{P}(x)=1$. It is useful also to introduce the variables
\begin{equation}
    y_k=b^{-1}\sqrt{\pi\mu}e^{b(e_k,x)/2},
\end{equation}
and new function $\widetilde{\Psi}_P^{(n)}(y_1,\dots,y_{n-1})$ through the relation
\begin{equation}
    \Psi^{(n)}_{P}(x)=\frac{2^{n(n-1)/2}}
    {\prod\limits_{e>0}\Gamma(-ib^{-1}(P,e))}
    \left(\frac{\pi\mu}{b^2}\right)^{-i\frac{(P,\rho)}{b}}\;
    \prod_{k=1}^{n-1}\left(\frac{y_k}{y_{n-k}}\right)^
    {\frac{i(P,\omega_k-\omega_{n-k})}{2b}}\;
    \widetilde{\Psi}_P^{(n)}(y_1,\dots,y_{n-1}).
\end{equation}
The recursive relation  connects function $\widetilde{\Psi}^{(n)}_{P}$ with function $\widetilde{\Psi}^{(n-2)}_{P'}$
\begin{multline}\label{Int_Rec_Relat}
    \widetilde{\Psi}^{(n)}_{P}(y_1,\dots,y_{n-1})=\\=
    \int_0^{\infty}\dots\int_0^{\infty}\;
    \widetilde{\Psi}^{(n-2)}_{P'}\left(
    y_2\frac{t_1}{t_2},y_3\frac{t_2}{t_3},\dots,
    y_{n-2}\frac{t_{n-3}}{t_{n-2}}\right)
    K_{\frac{i(P,e_0)}{b}}\left(2y_{n-1}
    \sqrt{(1+t_{n-2}^2)}\right)
    \times\\\times
    \prod_{k=1}^{n-2}\left[\;
    t_k^{ib^{-1}\sum_{j=1}^k(P,e_{n-j}-e_j)}
    K_{\frac{i(P,e_0)}{b}}\left(2y_k
    \sqrt{(1+t_{k-1}^2)(1+t_k^{-2})}\right)\;\right]
    \frac{dt_1}{t_1}\dots\frac{dt_{n-2}}{t_{n-2}}\;,
\end{multline}
where by definition $t_0=0$. Vector $P'$ in Eq \eqref{Int_Rec_Relat} defined in a following way. If vector $P$ has a components $P_1,P_2,\dots,P_{n-1}$ in the basis of fundamental weights of the Lie algebra  $\mathfrak{sl}(n)$ (i.~e. $(P,e_k)=P_k$, where $e_k$ are the simple roots of $\mathfrak{sl}(n)$), then vector $P'$ has components $P_2,P_3,\dots,P_{n-2}$ in the basis of fundamental weights of the Lie algebra $\mathfrak{sl}(n-2)$ (i.~e. $(P',e_k)=P_{k+1}$, where $e_k$ are the simple roots of $\mathfrak{sl}(n-2)$). To clarify Eq \eqref{Int_Rec_Relat} and definition of the vector $P'$ we give an expression for the  function $\widetilde{\Psi}^{(4)}_{P}(y_1,y_2,y_3)$ in the appendix \ref{Ex-Rec-Relat} (Eq \eqref{Psi4}).

It is useful to introduce also Whittaker function in the momentum representation
\begin{equation}
    \hat{\Psi}_P^{(n)}(q)=\int\Psi_P^{(n)}(x)
    e^{i(q,x)}\,d\vec{x}.
\end{equation}
The integral \eqref{mspLimit} describing the asymptotic of three-point correlation function  transforms to
\begin{equation}
    C(Q+ibp_1,Q+ibp_2,ibq)\rightarrow
    \frac{1}{(2\pi)^{n-1}}\int\,d\vec{q'}\,
    \hat{\Psi}^{(n)}_{bp_{1}}(q')\hat{\Psi}^{(n)}_{bp_2}(q-q').
\end{equation}

Let us consider the simplest examples of $\mathfrak{sl}(2)$ and $\mathfrak{sl}(3)$ TFT  here. For the $\mathfrak{sl}(2)$  case (Liouville theory) we derive from Eq \eqref{Int_Rec_Relat}
\begin{subequations}
  \begin{align}
     &\Psi^{(2)}_{bp}(x)=\frac{2}{\Gamma(-ip\sqrt{2})}
     \left(\frac{\pi\mu}{b^2}\right)^{-\frac{ip}{\sqrt{2}}}
     K_{ip\sqrt{2}}\left(
     2b^{-1}\sqrt{\pi\mu}e^{\frac{bx}{\sqrt{2}}}\right),\\
  \intertext{
  Fourier-transformed Whittaker function can be easily find using Eq \eqref{KInt}}
     &\hat{\Psi}^{(2)}_{bp}(bq)=\frac{1}{b\sqrt{2}}
     \left(\frac{\pi\mu}{b^2}\right)^{-\frac{i(p+q)}{\sqrt{2}}}
     \frac{\Gamma\left(\frac{i}{\sqrt{2}}(q+p)\right)
     \Gamma\left(\frac{i}{\sqrt{2}}(q-p)\right)}
     {\Gamma\left(-ip\sqrt{2}\right)}
  \end{align}
\end{subequations}
The integral \eqref{mspLimit} in this case can be evaluated using the formula \eqref{KKInt}
\begin{multline}\label{sl2xsl2}
   \int dx\;\Psi^{(2)}_{bp_{1}}(x)\Psi^{(2)}_{bp_{2}}(x)
   e^{ibs/\sqrt{2}x}=\\=
   \frac{4\sqrt{2}b^{-1}}
   {\Gamma(-ip_1\sqrt{2})\Gamma(-ip_1\sqrt{2})}
   \left(\frac{\pi\mu}{b^2}\right)^
   {-i\left(\frac{p_1+p_2}{\sqrt{2}}\right)-is}
   \int_0^{\infty}y^{is}K_{ip_1\sqrt{2}}(2y)K_{ip_2\sqrt{2}}(2y)
   \;\frac{dy}{y}=\\=\frac{1}{b}
   \left(\frac{\pi\mu}{b^2}\right)^
   {-i\left(\frac{p_1+p_2}{\sqrt{2}}\right)-is}
   \frac{\prod_{\varepsilon_1=\pm}\prod_{\varepsilon_2=\pm}
   \Gamma\left(\frac{is}{2}+\varepsilon_1\frac{ip_1}{\sqrt{2}}+\varepsilon_2
   \frac{ip_2}{\sqrt{2}}\right)}
   {\Gamma(is)\Gamma(-ip_1\sqrt{2})\Gamma(-ip_2\sqrt{2})}\;.
\end{multline}
This function coincides with minisuperspace limit of the three-point correlation function for the $\mathfrak{sl}(2)$ TFT.

In the $\mathfrak{sl}(3)$ case we obtain from Eq \eqref{Int_Rec_Relat} the expression for Whittaker function\footnote{This function was firstly derived in Ref \cite{Takh-Vin}.}
\begin{subequations}
\begin{multline}
      \Psi_{bp}^{(3)}(x)=\frac{8\left(
      \frac{\pi\mu}{b^2}\right)^{-i(p,\rho)}}
      {\prod_{e>0}\Gamma(-i(p,e))}\left(\frac{y_1}{y_2}\right)^
      {i(p,\omega_1-\omega_2)}\times\\\times
      \int_0^{\infty}\frac{dt}{t}\;t^{i(p,e_2-e_1)}
      K_{i(p,e_0)}\left(2y_1\sqrt{1+1/t^2}\right) 
      K_{i(p,e_0)}\left(2y_2\sqrt{1+t^2}\right)
\end{multline}
Fourier-transformed Whittaker function can be easily found using Eqs \eqref{KInt} and \eqref{Int(t)}. The result is expressed again in terms of gamma-functions \cite{Olshanetsky:1981dk}
\begin{equation}
     \hat{\Psi}^{(3)}_{bp}(bq)=\frac{1}{b^2\sqrt{3}}
     \frac{\left(\frac{\pi\mu}{b^2}\right)^{-i(p+q,\rho)}}
     {\prod_{e>0}\Gamma\bigl(-i(p,e)\bigr)}\;
     \frac{\prod_{k=1}^3\Gamma\bigl(i(q,\omega_1)+i(p,h_k)\bigr)
     \Gamma\bigl(i(q,\omega_2)-i(p,h_k)\bigr)}
     {\Gamma\bigl(i(q,\rho)\bigr)}
\end{equation}
\end{subequations}
The integral \eqref{mspLimit} is much more complicated in this case. It is better to write it in the momentum representation. As a result, for the asymptotic of the three-point correlation function in $\mathfrak{sl}(3)$ TFT we obtain Barnes-like integral
\begin{multline}
    C(Q+ibp_1,Q+ibp_2,ibq)\longrightarrow
    \frac{1}{6\pi^2b^2}\frac{\left(\frac{\pi\mu}{b^2}\right)^
    {-i(p_1+p_2+q,\rho)}}
    {\prod_{e>0}\Gamma\bigl(-i(p_1,e)\bigr)\Gamma\bigl(-i(p_2,e)\bigr)}\times\\\times
    \int\frac{d^2q'}
    {\Gamma\bigl(i(q',\rho)\bigr)\Gamma\bigl(i(q-q',\rho)\bigr)}
    \prod_{k=1}^3\Bigl[\Gamma\bigl(i(q',\omega_1)+i(p_1,h_k)\bigr)
     \Gamma\bigl(i(q',\omega_2)-i(p_1,h_k)\bigr)
     \times\\\times
     \Gamma\bigl(i(q-q',\omega_1)+i(p_2,h_k)\bigr)
     \Gamma\bigl(i(q-q',\omega_2)-i(p_2,h_k)\bigr)\Bigr].
\end{multline}
This integral can be calculated exactly in terms of gamma functions if $q=s\omega_1$ or $q=s\omega_2$, as was first noticed  in \cite{Stade:1993}.

In the case of higher $n$, Whittaker function is more involved object. The problem of finding the Fourier transform of the product of two Whittaker functions was considered in \cite{Stade:2002}. The most general  situation, when the answer can be expressed in terms of Gamma functions, is  $q=s\omega_1$ or $q=s\omega_{n-1}$.
The generalization of the explicit $\mathfrak{sl}(2)$ formula \eqref{sl2xsl2} for the case of general $n$
has a form
\begin{multline}\label{slnxsln}
   \int d\vec{x}\;\Psi^{(n)}_{bp_{1}}(x)\Psi^{(n)}_{bp_{2}}(x)
   e^{ibs(\omega_{n-1},x)}=\\=
   \frac{1}{b^{n-1}}
   \left(\frac{\pi\mu}{b^2}\right)^
   {-i\left(s\frac{n-1}{2}+(p_1+p_2,\rho)\right)}
   \frac{\prod\limits_{ij}
   \Gamma\left(\frac{is}{n}+i(p_1,h_i)+i(p_2,h_j)\right)}
   {\Gamma(is)\prod\limits_{e>0}\Gamma\bigl(-i(p_1,e)\bigr)
   \Gamma\bigl(-i(p_2,e)\bigr)}\;.
\end{multline}
We note that expression \eqref{slnxsln} coincides exactly with the minisuperspace limit of the three-point function \eqref{C}.
\section*{Conclusion}\addcontentsline{toc}{section}{Conclusion}
In this paper we have considered in details particular examples of three-point correlation functions $\langle V_{\alpha_1}(z_1,\bar{z}_1)V_{\alpha_2}(z_2,\bar{z}_2)V_{\alpha_3}(z_3,\bar{z}_3)\rangle$ in $\mathfrak{sl}(n)$ conformal Toda field theory, which can be expressed in terms of known in mathematics special functions. If any vector parameter $\alpha_1$, $\alpha_2$ or $\alpha_3$ is proportional to the first or to the last fundamental weights ($\omega_1$ or $\omega_{n-1}$) of the Lie algebra $\mathfrak{sl}(n)$, for example $\alpha_3=\varkappa\omega_{n-1}$, then three-point correlation function can be expressed in terms product of so called $\Upsilon$-functions (see Eq \eqref{C}). Unfortunately, general situation is much more complicated. For example, if one shifts slightly parameter $\varkappa\omega_{n-1}\rightarrow\varkappa\omega_{n-1}-b\omega_1$ then corresponding three-point correlation function can be expressed only in terms of Coulomb integral (see Eq \eqref{C-shifted}) or equivalently in terms of higher hypergeometric functions \eqref{C-shifted-1}. It is difficult to expect that general three-point correlation function can be expressed in terms of known functions.

As we see from the results of sections \ref{CL-heavy}, \ref{CL-light} and \ref{MSSL}, where the semiclassical and minisuperspace approaches to TFT were considered, three-point correlation function is already nontrivial in these cases. For example, in "heavy" semiclassical limit, developed in section \ref{CL-heavy}, a problem of finding it is rather difficult due to the presence of accessory parameters. These accessory parameters disappear  only for the case of the Lie algebra $\mathfrak{sl}(2)$, which corresponds to the Liouville field theory. This is one of the reasons why the three-point correlation function can be found in quantum LFT exactly. For $\mathfrak{sl}(n)$ TFT with $n>2$ there is no (to our knowledge) simple  regular procedure to obtain accessory parameters. In the "light" semiclassical limit (section \ref{CL-light}) and in the minisuperspace limit (section \ref{MSSL}) it is possible to derive the expression for the three-point correlation function in terms of finite dimensional integrals. Generally speaking it is not evident, that in quantum case it is also true. The only thing, which can be done, is to find all cases when the quantum three-point correlation function can be expressed in terms of finite dimensional integrals. We will study this problem in the second part of this paper \cite{Part-Deux}.
\acknowledgments
This work was supported, in part, by Russian Foundation for Basic Research under the grant RBRF 07-02-00799-a, by Russian Ministry of Science and Technology under the Scientific Schools grant 2044.2003.2 and by RAS program "Elementary particles and the fundamental nuclear physics". Work of V.~F. was supported also by the European Committee under contract EUCLID HRPN-CT-2002-00325. Work of A.~L. was supported  by DOE grant DE-FG02-96ER40949. 
An important part of this paper has been made during the visits of A.~L. at the Laboratoire de Physique Th\'eorique et Astroparticules Universit\'e Montpellier~II within ENS-LANDAU program. A.~L. acknowledges also the Abdus Salam International Centre for Theoretical Physics and specially professor S.~Randjbar-Daemi for the hospitality during the visit the center in August 2007, where this paper was meanwhile finished.
\appendix
\section{The Coulomb integrals}\label{AppIntegrals}
Here we will discuss the problem of calculation of the $\mathfrak{sl}(n)$ Coulomb integrals. They appear in the theory of massless $n-1$ component scalar field $\varphi$, as expressions for the correlation functions of the exponential fields $V_{\alpha}=e^{(\alpha,\varphi)}$. We will concentrate ourself on three-point correlation functions
\begin{equation}\label{Coulomb1}
  I_{s_1\dots s_{n-1}}(\alpha_1,\alpha_2,\alpha_3)=
  \Bigl\langle  V_{\alpha_1}(\infty)V_{\alpha_2}(1)V_{\alpha_3}(0)\prod_{k=1}^{n-1}\mathcal{Q}_k^{s_k}
  \Bigr\rangle,
\end{equation}
here $\mathcal{Q}_k$ is a screening field $\mathcal{Q}_k=\int e^{b(e_k,\varphi)}d^2z$ and $s_k$ are some non-negative integers. Correlation function \eqref{Coulomb1} is non-zero only if the screening condition 
\begin{equation*}
   \alpha_1+\alpha_2+\alpha_3+b\sum_{k=1}^{n-1}s_ke_k=2Q
\end{equation*}
is satisfied. In this case correlation function \eqref{Coulomb1} can be rewritten using the Wick rules
\begin{multline}\label{Coulomb-I}
    I_{s_1\dots s_{n-1}}(\alpha_1,\alpha_2,\alpha_3)=\\=
    \int
    \prod_{k=1}^{n-1}d\mu_{s_k}(t_k)\vert\vec{t}_k\vert^{-2b(\alpha_1,e_k)}
    \vert\vec{t}_k-1\vert^{-2b(\alpha_2,e_k)}\mathcal{D}_{s_k}^{-2b^2}(t_k)
    \prod_{l=1}^{n-2}\vert\vec{t}_l-\vec{t}_{l+1}\vert^{2b^2},
\end{multline}
where $\mathcal{D}_{s_k}(t_k)$ is defined by 
\begin{equation}\label{D-A-2}
  \mathcal{D}_{s_k}(t_k)=\prod_{i<j}^{s_k}|t_k^{(i)}-t_k^{(j)}|^2.
\end{equation}
In Eq \eqref{Coulomb-I} we have used the notations $\vec{t}_k=(t_k^{(1)},\dots,t_k^{(s_k)})$ with $t_k^{(j)}$ being the coordinate of the $j$-th screening $e^{b(e_k,\varphi)}$ and we denote
\begin{equation}\label{measure}
  \begin{gathered}
   |\vec{t}_k|=\prod_{j=1}^{s_k}|t_k^{(j)}|;\;\;\;
   |\vec{t}_k-1|=\prod_{j=1}^{s_k}|t_k^{(j)}-1|;\;\;\;
    d\mu_{s_k}(t_k)=\frac{1}{\pi^{s_k}s_k!}\prod_{j=1}^{s_k}d^2t_k^{(j)};\\
   |\vec{t}_k-\vec{t}_l|=
    \prod_{i=1}^{s_k}\prod_{j=1}^{s_l}|t_k^{(i)}-t_l^{(j)}|\;\;\text{if}\;k\neq l
  \end{gathered}
\end{equation}
We will study particular case of the integral \eqref{Coulomb-I}, which corresponds to the value of parameter $\alpha_3=\varkappa\omega_{n-1}$
\begin{multline}\label{Japp}
    J_{s_1\dots s_{n-1}}(a_1\dots a_{n-1}\vert c)=\\=
    \int d\mu_{s_1}(t_1)\dots d\mu_{s_{n-1}}(t_{n-1})\;
    \vert\vec{t}_{n-1}\vert^{2c}  
    \prod_{k=1}^{n-1}\vert\vec{t}_k-1\vert^{2a_k}\mathcal{D}_{s_k}^{-2b^2}(t_k)
    \prod_{l=1}^{n-2}\vert\vec{t}_l-\vec{t}_{l+1}\vert^{2b^2}
\end{multline}
with
\begin{equation}\label{c-a_k->alpha}
  c=-b\varkappa,\quad a_k=-b(\alpha_2,e_k).
\end{equation}
This integral can be calculated using the following identity between integrals of the dimension $2m$ and $2n$ \cite{Baseilhac:1998eq,Fateev:2006:JETP,Fateev:2007:TMPH}
\begin{multline}\label{Fateev_Identity}
\int d\mu_n(w)\mathcal{D}_n(w)\prod_{i=1}^n\prod_{j=1}^{n+m+1}
\vert w_i-x_j\vert^{2p_j}=\\=
\frac{\prod\limits_{i=1}^{n+m+1}\gamma(1+p_i)}
{\gamma(1+n+\sum\limits_{i}p_i)}
\prod_{i<j}\vert x_i-x_j\vert^{2+2p_i+2p_j}
\int d\mu_m(u)\mathcal{D}_m(u)\prod_{i=1}^m\prod_{j=1}^{n+m+1}
\vert u_i-x_j\vert^{-2p_j-2},
\end{multline}
where $\mathcal{D}_n(t)$ is defined similar to Eq \eqref{D-A-2} and equals 
\begin{equation}
  \mathcal{D}_n(t)=\prod_{i<j}|t_i-t_j|^2.
\end{equation}
Measure of integration is defined similar to Eq \eqref{measure} and equal $d\mu_n(w)=\frac{1}{\pi^nn!}\prod_{j=1}^nd^2w_j$. 

Below we list the main steps of calculation. Using integral relation \eqref{Fateev_Identity} function \eqref{Japp} can be calculated as follows
\begin{itemize}
    \item First, it is convenient to represent the factor $\mathcal{D}_{s_1}^{-2b^2}(t_1)$ 
    in Eq \eqref{Japp} as 
   $$\mathcal{D}_{s_1}^{-2b^2}(t_1)=\mathcal{D}_{s_1}(t_1)\mathcal{D}_{s_1}^{-1-2b^2}(t_1)$$ 
   and substitute the factor $\mathcal{D}_{s_1}^{-1-2b^2}(t_1)$ using Eq \eqref{Fateev_Identity} 
   with $n=s_1-1$ and $m=0$
     \begin{equation}\label{Bazeyaka-1}
       \mathcal{D}_{s_1}^{-1-2b^2}(t_1)=
       \frac{\gamma(-s_1b^2)}{\gamma^{s_1}(-b^2)}
       \int\;\mathcal{D}_{s_1-1}(y_1)\;
       \vert\vec{y}_1-\vec{t}_1\vert^{-2b^2-2}d\mu_{s_1-1}(y_1).
     \end{equation}
     Note that the number of variables $\vec{y}_1$ is equal to $s_1-1$.
\end{itemize}
\begin{itemize}    
    \item Second, the integral over variables $\vec{t}_1$ should be converted 
     using Eq \eqref{Fateev_Identity} to the form
     \begin{multline}\label{Third-step}
       \int\mathcal{D}_{s_1}(t_1)\;
       \vert\vec{t}_1-1\vert^{2a_1}\;
       \vert\vec{t}_1-\vec{y}_1\vert^{-2b^2-2}\;
       \vert\vec{t}_1-\vec{t}_2\vert^{2b^2}d\mu_{s_1}(t_1)=\\=
       \frac{\gamma^{s_1-s_2-1}(-b^2)\gamma(1+a_1)}
       {\gamma(2+(s_2-s_1+1)b^2+a_1)}\;
       \vert\vec{y}_1-1\vert^{-2b^2+2a_1}
       \vert\vec{t}_2-1\vert^{2+2b^2+2a_1}
       \;\mathcal{D}_{s_2}^{1+2b^2}(t_2)\times\\\times
       \mathcal{D}_{s_1-1}^{-1-2b^2}(y_1)
       \int\mathcal{D}_{s_2-1}(y_2)\;
       \vert\vec{y}_2-1\vert^{-2a_1-2}\;
       \vert\vec{y}_1-\vec{y}_2\vert^{2b^2}\;
       \vert\vec{y}_2-\vec{t}_2\vert^{-2b^2-2}d\mu_{s_2-1}(y_2).
     \end{multline}
     The number of integrations over variables $\vec{y}_2$ is equal to $s_2-1$. 
      We note that factor $\mathcal{D}_{s_1-1}^{-1-2b^2}(y_1)$ in the r.~h.~s. of Eq \eqref{Third-step} 
      combines with a factor $\mathcal{D}_{s_1-1}(y_1)$ in Eq \eqref{Bazeyaka-1} to the standart  
      combination  $\mathcal{D}_{s_1-1}^{-2b^2}(y_1)$ and interaction between $s_1-1$ points $\vec{y}_1$ 
      and $s_2-1$ points $\vec{y}_2$ has a standart form $|\vec{y}_1-\vec{y}_2|^{2b^2}$.
\end{itemize}
\begin{itemize}
    \item Third, we note that the factor $\mathcal{D}_{s_2}^{1+2b^2}(t_2)$, appearing after the 
     second step, combines with the factor $\mathcal{D}_{s_2}^{-2b^2}(t_2)$ in Eq \eqref{Japp}.  
     Hence we can take the integral over variables $\vec{t}_2$ in a way 
     similar to the second step
     \begin{multline}
       \int\mathcal{D}_{s_2}(t_2)\;
       \vert\vec{t}_2-1\vert^{2+2b^2+2a_1+2a_2}\;
       \vert\vec{t}_2-\vec{y}_2\vert^{-2b^2-2}\;
       \vert\vec{t}_2-\vec{t}_3\vert^{2b^2}d\mu_{s_2}(t_2)=\\=
       \frac{\gamma^{s_2-s_3-1}(-b^2)\gamma(2+b^2+a_1+a_2)}
       {\gamma(3+(s_3-s_2+2)b^2+a_1+a_2)}\;
       \vert\vec{y}_2-1\vert^{2+2a_1+2a_2}
       \vert\vec{t}_3-1\vert^{4+4b^2+2a_1+2a_2}
       \;\mathcal{D}_{s_3}^{1+2b^2}(t_3)\\
       \mathcal{D}_{s_2-1}^{-1-2b^2}(y_2)
       \int\mathcal{D}_{s_3-1}(y_3)\;
       \vert\vec{y}_3-1\vert^{-2a_1-2a_2-2b^2-4}\;
       \vert\vec{y}_2-\vec{y}_3\vert^{2b^2}\;
       \vert\vec{y}_3-\vec{t}_3\vert^{-2b^2-2}d\mu_{s_3-1}(y_3).
     \end{multline}
\end{itemize}
\begin{itemize}
  \item Repeating this procedure we will lower the number 
  of integrations at every step. The last integral over variables $\vec{t}_{n-1}$ will 
  be different from the integrals appearing at the previous steps. Namely,
  \begin{multline}
       \int\mathcal{D}_{s_{n-1}}(t_{n-1})\;
       \vert \vec{t}_{n-1}\vert^{2c}\;
       \vert\vec{t}_{n-1}-1\vert^{2(n-2)(1+b^2)+2\sum a_k}\;
       \vert\vec{t}_{n-1}-\vec{y}_{n-1}\vert^{-2b^2-2}d\mu_{s_{n-1}}(t_{n-1})\\=
       \gamma^{s_{n-1}-1}(-b^2)\;
       \frac{\gamma(1+c)\gamma(n-1+(n-2)b^2+a_1+\dots+a_{n-1})}
       {\gamma(n+c+a_1+\dots+a_{n-1}+(n-1-s_{n-1})b^2)}\times\\\times
       \mathcal{D}_{s_{n-1}}^{-1-2b^2}(y_{n-1})
       |\vec{y}_{n-1}|^{2c-2b^2}|\vec{y}_{n-1}-1|^{2+2(n-3)(1+b^2)+2(a_1+\dots+a_{n-1})}
   \end{multline}
\end{itemize}
\begin{itemize}
  \item As result, we obtain the recurrent relation
    \begin{multline}\label{FateevIdentity2}
      J_{s_1,\dots, s_{n-1}}(a_1,a_2,\dots, a_{n-1}\vert c)=\\=
      K(a_1,a_2,\dots, a_{n-1}\vert c)\;
      J_{s_1-1,\dots, s_{n-1}-1}
      (a_1-b^2,a_2\dots a_{n-1}\vert c-b^2)
    \end{multline}
   with
    \begin{multline*}
      K(a_1,a_2,\dots, a_{n-1}\vert c)=\frac{\gamma(-s_1b^2)}{\gamma^{n-1}(-b^2)}
      \frac{\gamma(1+c)\gamma(n-1+a_1+\dots+a_{n-1}+(n-2)b^2)}
      {\gamma(n+c+a_1+\dots+a_{n-1}+(n-1-s_{n-1})b^2)}
      \\\times\prod_{j=1}^{n-2}
      \frac{\gamma(j+a_1+\dots+a_{j}+(j-1)b^2)}
      {\gamma(1+j+a_1+\dots+a_j+(s_{j+1}-s_{j}+j)b^2)}
    \end{multline*}
\end{itemize}
We note, that if we substitute parameters $a_k$ and $c$ from Eq \eqref{c-a_k->alpha}, then the solution to the recurrent relation \eqref{FateevIdentity2} gives the expression for the integral \eqref{I-equals}.
We note also, that recurrent relation \eqref{FateevIdentity2} can be used to continue integral 
$J_{s_1,\dots, s_{n-1}}(a_1,a_2,\dots, a_{n-1}\vert c)$ to the non-integer values $s_k$ (it gives the expression for the three-point correlation \eqref{C}). 
\section{Simplification of the integral \protect\eqref{Iwasawa1}}\label{Eval_Iwasawa}
We start with the integral \eqref{Iwasawa1}
\begin{equation}\label{Semiclass_Int}
  \mathfrak{J}(\lambda_1,\lambda_2;\kappa_1,\kappa_2;\sigma_1,\sigma_2)=4^{\sigma_1+\sigma_2}
  \int\frac{\varrho^{2\delta}\nu^{2\Delta}}
  {Z_1^{\lambda_1}Z_2^{\lambda_2}Z_3^{\kappa_1}Z_4^{\kappa_2}}\;
  \frac{d\varrho}{\varrho}\frac{d\nu}{\nu}\,d^2a\,d^2b\,d^2c,
\end{equation}
where $Z_k$ is given by Eq \eqref{Zk}. First, we use Feinman representation
\begin{equation}
  \frac{1}{Z_1^{\lambda_1}Z_2^{\lambda_2}Z_3^{\kappa_1}Z_4^{\kappa_2}}=
  \frac{\Gamma(\upsilon)}
  {\Gamma(\lambda_1)\Gamma(\lambda_2)\Gamma(\kappa_1)\Gamma(\kappa_2)}\int
  \frac{\tau^{\lambda_1}\xi^{\lambda_2}t^{\kappa_1}}
  {(Z_4+Z_3t+Z_2\xi+Z_1\tau)^{\upsilon}}
  \frac{d\tau}{\tau}\frac{d\xi}{\xi}\frac{dt}{t}
\end{equation}
with
\begin{equation*}
  \upsilon=\lambda_1+\lambda_2+\kappa_1+\kappa_2.
\end{equation*}
After that, we can calculate integral over $d^2a$, $d^2b$ and over $d\varrho$ with the result
\begin{equation}\label{Semiclass-1}
 \pi^24^{\sigma_1+\sigma_2}\frac{\Gamma(\delta)\Gamma(\upsilon-\delta-2)}
 {\Gamma(\lambda_1)\Gamma(\lambda_2)\Gamma(\kappa_1)\Gamma(\kappa_2)}\int
 \frac{r^{2+\delta-\upsilon}\bar{r}^{\delta-1}}{P^{\delta}}\,\,
 \nu^{2\Delta}\tau^{\lambda_1}\xi^{\lambda_2}t^{\kappa_1}\,\,
 d^2c\,\frac{d\nu}{\nu}\,
 \frac{d\tau}{\tau}\,\frac{d\xi}{\xi}\,\frac{dt}{t}
\end{equation}
with
\begin{equation*}
   \begin{gathered}
     r=1+t+\tau+\xi+\tau|c|^2+t|c+\nu|^2,\;\;\;\bar{r}=(1+\xi)r+t\tau\nu^2;\\
     P=\frac{\tau}{4}\Bigl(1+t\bigl(1+|c+\nu|^2\bigr)\Bigr)+
     \frac{t\xi}{4}\Bigl(\xi+\tau\bigl(1+|c|^2\bigr)\Bigr)+
     \frac{r\xi}{\nu^2}\Bigl(1+|c+\frac{\nu}{2}|^2\Bigr).
   \end{gathered}
\end{equation*}
The problem is that the quantity $P$ is not quadratic in the variable $c$. In order to proceed simplification, we use the following trick. We can multiply our expression
\eqref{Semiclass-1} by
\begin{equation}\label{Identity}
   1=\frac{1}{2\pi}\int_{-\infty}^{\infty}\frac{ds}{s}\,\int_{-\infty}^{\infty}dp\,s^{ip}=
   \int_{-\infty}^{\infty} ds\,\delta(s-1)
\end{equation}
and insert $s$ somewhere into \eqref{Semiclass-1}. More exactly, we will need to do that four times. We just show the places of insertion of different $s_k$
\begin{equation*}
   \begin{gathered}
     r\rightarrow s_4\Bigl(\xi+\tau(1+|c|^2)\Bigr)+\Bigl(1+t(1+|c+\nu|^2)\Bigr),
     \;\;\;\bar{r}\rightarrow(1+\xi)r+s_2t\tau\nu^2;\\
     P\rightarrow s_1\biggl(\frac{\tau}{4}\Bigl(1+t\bigl(1+|c+\nu|^2\bigr)\Bigr)+
     s_3\frac{t\xi}{4}\Bigl(\xi+\tau\bigl(1+|c|^2\bigr)\Bigr)\biggr)+
     \frac{r\xi}{\nu^2}\Bigl(1+|c+\frac{\nu}{2}|^2\Bigr).
   \end{gathered}
\end{equation*}
The integrals over $s_k$ can be calculated  exactly (first over $s_1$, second over $s_2$ etc). After that, the integrals over $t$, $\tau$ and $\xi$ will be of the type \eqref{Int(t)} and  also can be calculated. The remaining integral over $d\nu$ and $d^2c$ will be
\begin{equation}\label{dc_dnu}
  \int \nu^{2(\delta+\Delta-s_1-s_2)}\bigl(1+|c|^2\bigr)^{s_1+s_2-s_3-\lambda_1}
  \bigl(1+|c+\nu|^2\bigr)^{s_2+s_3-\kappa_1}
  \Bigl(1+\bigl|c+\frac{\nu}{2}\bigr|^2\Bigr)^{s_1-\delta}\,\,
  \frac{d\nu}{\nu}\,d^2c.
\end{equation}
Using technique, described above, one can reduce integral \eqref{dc_dnu} to one dimensional integral. As result, the integral \eqref{Semiclass_Int} can be reduced to five dimensional Barnes-like integral. By using the first and the second Barnes lemmas \eqref{Barnes1} and \eqref{Barnes2}, one can reduce it to three dimensional integral \eqref{3_Fold_Int}. 

The integral \eqref{3_Fold_Int} can be also rewritten in a different form in terms of Tricomi functions, which are defined by the integral representation
\begin{equation}
   \Psi(a,c|x)=\frac{1}{\Gamma(a)}\int_0^{\infty}\,dt\,e^{-xt}t^{a-1}(1+t)^{c-a-1}.
\end{equation}
This function can be expressed through the confluent hypergeometric function 
\begin{equation}
  \Phi(a,c|x)=1+\frac{a}{c}x+\frac{1}{2!}\frac{a(a+1)}{c(c+1)}x^2+\dots
\end{equation}
as
\begin{equation}
  \Psi(a,c|x)=\frac{\Gamma(1-c)}{\Gamma(a-c+1)}\,\Phi(a,c|x)+
   \frac{\Gamma(c-1)}{\Gamma(a)}x^{1-c}\,\Phi(a-c+1,2-c|x)
\end{equation}
with $\Psi(a,c;0)=\frac{\Gamma(1-c)}{\Gamma(a-c+1)}$. Tricomi function satisfies the following relation
\begin{equation}
   \Psi(a,c|x)=x^{1-c}\,\Psi(a-c+1,2-c|x).
\end{equation}
The integral \eqref{3_Fold_Int} can be rewritten as
\begin{multline}
  \mathfrak{J}(\lambda_1,\lambda_2;\kappa_1,\kappa_2;\sigma_1,\sigma_2)=
   4^{\lambda_1+\kappa_1+\sigma_1-\Delta}
  \times\\\times
  \frac{\Gamma(\lambda_1+\kappa_1+\sigma_1-\Delta-2)
   \Gamma(\lambda_2+\kappa_2+\sigma_2+\Delta-2)}
  {\Gamma(\lambda_1)\Gamma(\lambda_2)\Gamma(\lambda_1+\lambda_2-1)
  \Gamma(\kappa_1)\Gamma(\kappa_2)\Gamma(\kappa_1+\kappa_2-1)
  \Gamma(\sigma_1)\Gamma(\sigma_2)\Gamma(\sigma_1+\sigma_2-1)}
  \times\\\times\int\,du\,ds\,dy\,\,t^{\lambda_1-1}(1-t)^{\kappa_1-1}\,
  u^{\lambda_1+\kappa_1-\Delta-2}(1-u)^{\lambda_2+\kappa_2+\Delta-2}\,
  e^{-s}s^{\lambda_1+\kappa_1-\sigma_2-\Delta-1}\times\\\times
  F_1(4t(1-t)s)\,F_2(sut)\,F_3(su(1-t)),
\end{multline}
where
\begin{equation}
  \begin{aligned}
    &F_1(x)=\Gamma(\sigma_2)\Gamma(\sigma_2+\Delta)\,\Psi(\sigma_2+\Delta,1+\Delta|x),\\
    &F_2(x)=\Gamma(\kappa_1+\kappa_2-1)\Gamma(\sigma_1+\kappa_1-\Delta-1)\,
    \Psi(\kappa_1+\kappa_2-1,1-\sigma_1+\kappa_2+\Delta|x),\\
    &F_3(x)=\Gamma(\lambda_1+\lambda_2-1)\Gamma(\sigma_1+\lambda_1-\Delta-1)\,
    \Psi(\lambda_1+\lambda_2-1,1-\sigma_1+\lambda_2+\Delta|x).
  \end{aligned}
\end{equation}
This form of the integral \eqref{Iwasawa1} is very convenient to obtain its limit at $\sigma_1\rightarrow-m$ and $\sigma_2\rightarrow-n$ considered in section \ref{CL-light}.
\section{Properties of the $\mathfrak{sl}(3)$ Coulomb integral}\label{SL(3)-Int-def}
In this appendix we study the properties of the $\mathfrak{sl}(3)$ integral
\begin{multline}\label{SL(3)-Int}
  \mathcal{I}_{k,l}(\alpha_1,\alpha_2,\beta_1,\beta_2)=\\=
  \int\prod_{i=1}^k\prod_{j=1}^l|t_i-s_j|^{2b^2}
  \mathcal{D}^{-2b^2}_k(t)\mathcal{D}^{-2b^2}_l(s)
  \prod_{i=1}^k |t_i|^{2\alpha_1}
  |t_i-1|^{2\beta_1}d^2t_i\prod_{j=1}^l|s_j|^{2\alpha_2}|s_j-1|^{2\beta_2}d^2s_j,
\end{multline}
where $\mathcal{D}_k(t)$ is defined by Eq \eqref{D-A-2}.
Using the integral identity \eqref{Fateev_Identity} one can show, that function $\mathcal{I}_{k,l}(\alpha_1,\alpha_2,\beta_1,\beta_2)$ satisfies the set of functional relations, which are generated by two basic relations (we suppose, that $l\geq k$):
\begin{equation}\label{Int-Relat1}
  \mathcal{I}_{k,l}(\alpha_1,\alpha_2,\beta_1,\beta_2)=\Xi_{k,l}^{(1)}(\alpha_1,\alpha_2,\beta_1,\beta_2)
  \mathcal{I}_{k,l}(\alpha_1,\tilde{\beta}_1,\tilde{\alpha}_2,\beta_2),
\end{equation}
where $\tilde{\beta}_1=\beta_1+(l-k)b^2$, $\tilde{\alpha}_2=\alpha_2-(l-k)b^2$ and
\begin{multline*}
  \Xi^{(1)}_{k,l}(\alpha_1,\alpha_2,\beta_1,\beta_2)=
  \prod_{j=0}^{l-k-1}\frac{\gamma(1+\alpha_2-jb^2)}
  {\gamma(1+\tilde{\beta}_1-jb^2)}\times\\\times
  \prod_{j=0}^{k-1}\frac{\gamma(2+\alpha_1+\alpha_2-(j-1)b^2)}
  {\gamma(2+\alpha_1+\tilde{\beta}_1-(j-1)b^2)}
  \prod_{j=0}^{l-1}\frac{\gamma(2+\beta_1+\beta_2-(j-1)b^2)}
  {\gamma(2+\tilde{\alpha}_2+\beta_2-(j-1)b^2)}
\end{multline*}
and by the relation
\begin{multline}\label{Int-Relat2}
  \mathcal{I}_{k,l}(\alpha_1,\alpha_2,\beta_1,\beta_2)=
  \Xi^{(2)}_{k,l}(\alpha_1,\alpha_2,\beta_1,\beta_2)\times\\\times
  \mathcal{I}_{k,l}(\alpha_1,-2-\alpha_1-\alpha_2+(l-2)b^2,\beta_1,-2-\beta_1-\beta_2+(l-2)b^2)
\end{multline}
with 
\begin{multline*}
  \Xi^{(2)}_{k,l}(\alpha_1,\alpha_2,\beta_1,\beta_2)=\\
  \prod_{j=0}^{l-1}
  \frac{\gamma(1+\alpha_1-jb^2)\gamma(1+\beta_1-jb^2)\gamma(2+\alpha_1+\alpha_2-(j-1)b^2)
  \gamma(2+\beta_1+\beta_2-(j-1)b^2)}
  {\gamma(2+\alpha_1+\beta_1-(l-k-1+j)b^2)\gamma(3+\alpha_1+\alpha_2+\beta_1+\beta_2-(k-2+j)b^2)}.
\end{multline*}
Relations \eqref{Int-Relat1} and \eqref{Int-Relat2} can be used for the analytical continuation and sometimes for the simplification of the integral \eqref{SL(3)-Int}. 

Integral \eqref{SL(3)-Int} can be calculated exactly if $k=0$ or $l=0$ and also if one of the parameters $\alpha_k$ or $\beta_k$ equals to zero (see Appendix \ref{AppIntegrals}). In the case $k=1$ (or $l=1$) it also can be reduced to known functions. To show it we apply integral relation \cite{Fateev:2006:JETP,Fateev:2007:TMPH}
\begin{multline}\label{n->1}
  \frac{1}{\pi^ll!}
    \int\prod_{j=1}^l |s_j|^{2\alpha_2}|s_j-1|^{2\beta_2}|s_j-t|^{2b^2}\mathcal{D}_l^{-2b^2}(s)d^2s_1\dots 
    d^2s_l=\\=
   \prod_{j=0}^{l-2}\frac{\gamma(-(j+2)b^2)}{\gamma(-b^2)}
   \frac{\gamma(1+\alpha_2-jb^2)\gamma(1+\beta_2-jb^2)}{\gamma(2+\alpha_2+\beta_2-(l-1+j)b^2)}
   \times\\\times\frac{1}{\pi}
   \int|u|^{2\alpha_2-2(l-1)b^2}|u-1|^{2\beta_2-2(l-1)b^2}|u-t|^{2lb^2}d^2u.  
\end{multline}
Relation \eqref{n->1} allows to reduce integral \eqref{SL(3)-Int} to the four-dimensional integral
\begin{equation*}
   \int|t|^{2\alpha_1}|t-1|^{2\beta_1}
   |u|^{2\alpha_2-2(l-1)b^2}|u-1|^{2\beta_2-2(l-1)b^2}|u-s|^{2lb^2}
   \,d^2u\,d^2s,
\end{equation*}
which can be expressed in terms hypergeometric function of the type $(3,2)$ using Eq \eqref{C-shifted-1}.
For $k>1$ integral \eqref{SL(3)-Int} can be reduced to $4k$-dimensional Coulomb integral. We will give the explicit expression for this integral in Ref \cite{Part-Deux}. Here we give two different asymptotics at $b\rightarrow0$ of the meromorphic function defined by the integral \eqref{SL(3)-Int}.
First asymptotic is (we assume that $l\geq k$)
\begin{multline}\label{SL(3)-Int-ass1}
   \mathcal{I}_{k,l}(-\lambda_1b^2,-\lambda_2b^2,-\kappa_1b^2,-\kappa_2b^2)
   \underset{b\rightarrow0}{\longrightarrow}
   (-\pi b^2)^{k+l}\times\\\times
   \frac{(-1)^k(\lambda_2)_{l-k}(\kappa_2)_{l-k}}
   {(\lambda_1+\lambda_2+\kappa_1+\kappa_2+l-2)_{k}(\lambda_1+\kappa_1+k-l-1)_{k}
   (\lambda_2+\kappa_2+l-k-1)_{l}}\times\\\times
   \sum_{s_1,s_2,s_3\geq 0}^k4^{-s_1-s_2-s_3}
   \frac{(-k)_{s_1+s_2+s_3}(-l)_{s_1+s_2+s_3}(1-2k-\lambda_1-\kappa_1+s_1+s_2+s_3)_{s_1}}
   {s_1!s_2!s_3!}\times\\\times
   (\lambda_1)_{k-s_1-s_3}(\kappa_1)_{k-s_1-s_2}(\lambda_1+\kappa_1+k-l-1)_{s_2+s_3}
   (3-\lambda_1-\lambda_2-\kappa_1-\kappa_2-k)_{k-s_2-s_3}\times\\\times
   (\kappa_1+\kappa_2-1)_{s_2}(\lambda_1+\lambda_2-1)_{s_3}(l-k+\lambda_2)_{k-s_2}(l-k+\kappa_2)_{k-s_3}.
\end{multline}
Second asymptotic is
\begin{multline}
   \mathcal{I}_{k,l}(-1-\lambda_1b^2,-1-\lambda_2b^2,-1-\kappa_1b^2,-1-\kappa_2b^2)
   \underset{b\rightarrow0}{\longrightarrow}
   \left(-\frac{\pi}{b^2}\right)^{k+l}\times\\\times
   \sum_{s_1=0}^k\sum_{s_2=0}^l C_{k}^{s_1}C_{l}^{s_2}\;
   \frac{(-1+\lambda_1+\lambda_2+l-s_2)_{k-s_1}}
   {(\lambda_1)_{k-s_1}(\lambda_2)_{l-s_2}(\lambda_1+\lambda_2-1)_{k-s_1}}\;
   \frac{(-1+\kappa_1+\kappa_2+s_1)_{s_2}}
   {(\kappa_1)_{s_1}(\kappa_2)_{s_2}(\kappa_1+\kappa_2-1)_{s_2}},
\end{multline}
where $C_k^j$ are the binomial coefficients.
\section{Useful formulae}\label{KBessel}
Here we collect some basic facts concerning Macdonald function $K_{\nu}(y)$
\begin{itemize}
\item Integral representation
  \begin{equation}\label{Kint}
     K_{\nu}(y)=\frac{1}{2}\int_{0}^{\infty}\frac{dt}{t}\;
     t^{\nu}\exp(-y(t+1/t)/2).
  \end{equation}
\item Asymptotic formula
 \begin{equation}
    K_{\nu}(2y)\rightarrow\frac{1}{2}
    \left(\Gamma(-\nu)y^{\nu}+\Gamma(\nu)y^{-\nu}\right)\quad
    \text{at}\quad y\rightarrow 0 
 \end{equation}
\item Mellin transformation of single Macdonald function
 \begin{equation}\label{KInt}
   \int_0^{\infty}y^{\mu}K_{\nu}(2ay)\;\frac{dy}{y}=\frac{1}{4a^{\mu}}\;
   \Gamma\left(\frac{\mu+\nu}{2}\right)
   \Gamma\left(\frac{\mu-\nu}{2}\right)
 \end{equation}
\item Mellin transformation of the product of two Macdonald functions
 \begin{multline}\label{KKInt}
  \int_0^{\infty}y^{\lambda}K_{\mu}(2ay)K_{\nu}(2cy)\;\frac{dy}{y}=
  \frac{c^{\nu}}{8a^{\nu+\lambda}\Gamma(\lambda)}\;
  \Gamma\left(\frac{\lambda+\mu+\nu}{2}\right)
  \Gamma\left(\frac{\lambda+\mu-\nu}{2}\right)\times\\\times
  \Gamma\left(\frac{\lambda-\mu+\nu}{2}\right)
  \Gamma\left(\frac{\lambda-\mu-\nu}{2}\right)
  F\left(\genfrac{}{}{0pt}{1}{\frac{\lambda+\mu+\nu}{2}
  \:\frac{\lambda-\mu+\nu}{2}}
    {\lambda}\biggl|1-\frac{c^2}{a^2}\right).
\end{multline}
here $F$ denotes the hypergeometric function of the  type $(2,1)$.
\end{itemize}
Beta-like integral
\begin{equation}\label{Int(t)}
   \int_0^{\infty}\;\frac{dt}{t}\;t^{A}\;(1+t^2)^B=
   \frac{1}{2}\;\frac{\Gamma\left(\frac{A}{2}\right)
   \Gamma\left(-B-\frac{A}{2}\right)}{\Gamma(-B)}\;.
\end{equation}
Barnes first lemma
\begin{equation}\label{Barnes1}
    \frac{1}{2\pi i}
    \int_{-i\infty}^{i\infty}\Gamma(\alpha+s)\Gamma(\beta+s)
    \Gamma(\gamma-s)\Gamma(\delta-s)=
    \frac{\Gamma(\alpha+\gamma)\Gamma(\alpha+\delta)
    \Gamma(\beta+\gamma)\Gamma(\beta+\delta)}
    {\Gamma(\alpha+\beta+\gamma+\delta)}.
\end{equation}
Barnes second lemma states that
\begin{multline}\label{Barnes2}
   \frac{1}{2\pi i}\int\frac{\Gamma(\alpha_1+s)
   \Gamma(\alpha_2+s)\Gamma(\alpha_3+s)\Gamma(1-\beta_1-s)
   \Gamma(-s)ds}{\Gamma(\beta_2+s)}=\\=
   \frac{\Gamma(\alpha_1)\Gamma(\alpha_2)\Gamma(\alpha_3)
   \Gamma(1-\beta_1+\alpha_1)\Gamma(1-\beta_1+\alpha_2)
   \Gamma(1-\beta_1+\alpha_3)}{\Gamma(\beta_2-\alpha_1)
   \Gamma(\beta_2-\alpha_2)\Gamma(\beta_2-\alpha_3)}
\end{multline}
provided that $\beta_1+\beta_2=\alpha_1+\alpha_2+\alpha_3+1$.
\section{Example of application of the recursive relation \protect\eqref{Int_Rec_Relat}}\label{Ex-Rec-Relat}
In this appendix we explain how to use recursive relation \eqref{Int_Rec_Relat}. For example, using Eq \eqref{Int_Rec_Relat} one obtains for function $\widetilde{\Psi}^{(4)}_{P}(y_1,y_2,y_3)$ exact expression 
\begin{multline}\label{Psi4}
 \widetilde{\Psi}^{(4)}_{P}(y_1,y_2,y_3)=\int_{0}^{\infty}\int_{0}^{\infty}
 t_1^{ib^{-1}(P,e_3-e_1)}t_2^{ib^{-1}(P,e_3-e_1)}\times\\\times
 K_{\frac{i(P,e_2)}{b}}\left(2y_{2}\frac{t_1}{t_2}\right)
 K_{\frac{i(P,e_0)}{b}}\left(2y_{1}\sqrt{(1+t_{1}^{-2})}\right)\times\\\times
 K_{\frac{i(P,e_0)}{b}}\left(2y_{2}\sqrt{(1+t_{1}^2)(1+t_{2}^{-2})}\right)
 K_{\frac{i(P,e_0)}{b}}\left(2y_{3}\sqrt{(1+t_{2}^{2})}\right)\;
 \frac{dt_1}{t_1}\,\frac{dt_2}{t_2}.
\end{multline}
In Eq \eqref{Psi4} we substitute
\begin{equation}\label{Psi4-Explain}
   \widetilde{\Psi}^{(2)}_{P'}\left(y_2\frac{t_1}{t_2}\right)=
   K_{\frac{i(P,e_2)}{b}}\left(2y_{2}\frac{t_1}{t_2}\right).
\end{equation}
As we see from Eq \eqref{Psi4-Explain}, it is convenient to think that $P'=P$, but vector $P'$ lives on a lattice with cutted-off ends. Symbolically it can be pictured as  
\begin{equation*}
\setlength{\unitlength}{1.18pt}
  \begin{picture}(362,22)(-15,-1)
    \thicklines
    \matrixput(0,0)(20,0){3}(0,0){2}{\circle*{3}}
    \matrixput(0,0)(20,0){2}(0,0){1}{\line(1,0){30}}
    \matrixput(80,0)(20,0){3}(0,0){2}{\circle*{3}}
    \matrixput(70,0)(20,0){2}(0,0){1}{\line(1,0){30}}
    \matrixput(55,0)(5,0){3}(0,0){2}{\circle*{1}}
    \put(-2,-7){\makebox{\small{$e_1$}}}
    \put(18,-7){\makebox{\small{$e_2$}}}
    \put(38,-7){\makebox{\small{$e_3$}}}
    \put(78,-7){\makebox{\small{$e_{n-3}$}}}
    \put(98,-7){\makebox{\small{$e_{n-2}$}}}
    \put(118,-7){\makebox{\small{$e_{n-1}$}}}
    \put(150,0){\vector(1,0){60}}
    \matrixput(240,0)(20,0){2}(0,0){2}{\circle*{3}}
    \matrixput(240,0)(20,0){1}(0,0){1}{\line(1,0){30}}
    \matrixput(300,0)(20,0){2}(0,0){2}{\circle*{3}}
    \matrixput(290,0)(20,0){1}(0,0){1}{\line(1,0){30}}
    \matrixput(275,0)(5,0){3}(0,0){2}{\circle*{1}}
    \put(238,-7){\makebox{\small{$e_2$}}}
    \put(258,-7){\makebox{\small{$e_3$}}}
    \put(298,-7){\makebox{\small{$e_{n-3}$}}}
    \put(318,-7){\makebox{\small{$e_{n-2}$}}}
  \end{picture}
\vspace*{12pt}
\end{equation*}
Using function \eqref{Psi4} we can reconstract function $\widetilde{\Psi}^{(6)}_{P}(y_1,y_2,y_3,y_4,y_5)$ and so on.


\begin{thebibliography}{99}
\bibitem{Polyakov:1981rd}
  A.~M.~Polyakov,
  Quantum geometry of bosonic strings,
  Phys.\ Lett.\ B {\bf 103}, 207 (1981).
\bibitem{Belavin:1984vu}
  A.~A.~Belavin, A.~M.~Polyakov and A.~B.~Zamolodchikov,
  Infinite conformal symmetry in two-dimensional quantum field theory,
  Nucl.\ Phys.\ B {\bf 241} (1984) 333.
\bibitem{Zamolodchikov:1985wn}
  A.~B.~Zamolodchikov,
  Infinite additional symmetries in two-dimensional conformal quantum  
  field theory,
  Theor.\ Math.\ Phys.\  {\bf 65} (1985) 1205
  [Teor.\ Mat.\ Fiz.\  {\bf 65} (1985) 347].
\bibitem{Gervais:1993yh}
  J.~L.~Gervais,
  Introduction to differential W geometry, Published in Strings 1993:0397-415,
  arXiv:hep-th/9310116.
\bibitem{Razumov:1993pv}
  A.~V.~Razumov and M.~V.~Saveliev,
  Differential geometry of Toda systems,
  Commun.\ Anal.\ Geom.\  {\bf 2} (1994) 461
  [arXiv:hep-th/9311167].
\bibitem{Pope:1992mi}
  C.~N.~Pope,
  A Review of W strings, Contribution to Proc. of Int. Symp. on Black Holes, Worm Holes, 
  Membranes and Superstrings, Woodlands, TX, Jan 16-18, 1992,
  arXiv:hep-th/9204093.
\bibitem{West:1993np}
  P.~C.~West,
  A Review of W strings, Published in Salamfest 1993:0451-477,
  arXiv:hep-th/9309095.
\bibitem{Fateev:1985mm}
  V.~A.~Fateev and A.~B.~Zamolodchikov,
  Parafermionic currents in the two-dimensional conformal quantum field
  theory and selfdual critical points in Z(N) invariant 
  statistical   systems,
  Sov.\ Phys.\ JETP {\bf 62}, 215 (1985)
  [Zh.\ Eksp.\ Teor.\ Fiz.\  {\bf 89}, 380 (1985)].
\bibitem{Date:1987vf}
  E.~Date, M.~Jimbo, T.~Miwa and M.~Okado,
  Solvable Lattice Models,
  Invited lectures delivered at the AMS Summer Institute on Theta Functions, July 1987.
\bibitem{Jimbo:1987ra}
  M.~Jimbo, T.~Miwa and M.~Okado,
  Solvable Lattice Models Related To The Vector Representation Of Classical
  Simple Lie Algebras,
  Commun.\ Math.\ Phys.\  {\bf 116}, 507 (1988).
\bibitem{Ahn:1999dz}
  C.~r.~Ahn, V.~A.~Fateev, C.~j.~Kim, C.~Rim and B.~Yang,
  Reflection amplitudes of ADE Toda theories and thermodynamic Bethe
  ansatz,
  Nucl.\ Phys.\  B {\bf 565}, 611 (2000)
  [arXiv:hep-th/9907072].
\bibitem{Ahn:2000ki}
  C.~r.~Ahn, P.~Baseilhac, V.~A.~Fateev, C.~j.~Kim and C.~Rim,
  Reflection amplitudes in non-simply laced Toda theories and  thermodynamic
  Bethe ansatz,
  Phys.\ Lett.\  B {\bf 481}, 114 (2000)
  [arXiv:hep-th/0002213].
\bibitem{Fateev:2000pi}
  V.~A.~Fateev,
  Normalization factors in conformal field theory and their applications,
  Mod.\ Phys.\ Lett.\  A {\bf 15}, 259 (2000).
\bibitem{Feher:1992yx}
  L.~Feher, L.~O'Raifeartaigh, P.~Ruelle, I.~Tsutsui and A.~Wipf,
  On Hamiltonian reductions of the Wess-Zumino-Novikov-Witten theories,
  Phys.\ Rept.\  {\bf 222} (1992) 1.
\bibitem{Dorn:1992at}
  H.~Dorn and H.~J.~Otto,
  On correlation functions for noncritical strings with $c\leq 1$  but $d\geq1$,
  Phys.\ Lett.\ B {\bf 291} (1992) 39
  [arXiv:hep-th/9206053].
\bibitem{Dorn:1994xn}
  H.~Dorn and H.~J.~Otto,
  Two and three point functions in Liouville theory, 
  Nucl.\ Phys.\ B {\bf 429} (1994) 375
  [arXiv:hep-th/9403141].
\bibitem{Zamolodchikov:1995aa}
  A.~B.~Zamolodchikov and Al.~B.~Zamolodchikov,
  Structure constants and conformal bootstrap in Liouville 
  field theory,
  Nucl.\ Phys.\ B {\bf 477} (1996) 577
  [arXiv:hep-th/9506136].
\bibitem{Teschner:2003en}
  J.~Teschner,
  A lecture on the Liouville vertex operators,
  Int.\ J.\ Mod.\ Phys.\  A {\bf 19S2} (2004) 436
  [arXiv:hep-th/0303150].
\bibitem{Bowcock:1993wq}
  P.~Bowcock and G.~M.~T.~Watts,
  Null vectors, three point and four point functions in conformal field
  theory,
  Theor.\ Math.\ Phys.\  {\bf 98} (1994) 350
  [Teor.\ Mat.\ Fiz.\  {\bf 98} (1994) 500]
  [arXiv:hep-th/9309146].
\bibitem{Fateev:2005gs}
  V.~A.~Fateev and A.~V.~Litvinov,
  On differential equation on four-point correlation function in the
  conformal Toda field theory,
  JETP Lett.\  {\bf 81} (2005) 594
  [Pisma Zh.\ Eksp.\ Teor.\ Fiz.\  {\bf 81} (2005) 728]
  [arXiv:hep-th/0505120].
\bibitem{Teschner:1995yf}
  J.~Teschner,
  On the Liouville three point function,
  Phys.\ Lett.\ B {\bf 363} (1995) 65
  [arXiv:hep-th/9507109].
\bibitem{Part-Deux}
V.~A.~Fateev and A.~V.~Litvinov,
Correlation functions in conformal Toda field theory II, in preparation.
\bibitem{Fateev:1987zh}
  V.~A.~Fateev and S.~L.~Lukyanov,
  The models of two-dimensional conformal quantum field theory with Z(N)
  symmetry,
  Int.\ J.\ Mod.\ Phys.\ A {\bf 3}, 507 (1988).
\bibitem{Fateev:2001mj}
V.~A.~Fateev,
Normalization factors, reflection amplitudes and integrable systems, 
arXiv:hep-th/0103014.
\bibitem{Goulian:1990qr}
  M.~Goulian and M.~Li,
  Correlation functions in Liouville theory,
  Phys.\ Rev.\ Lett.\  {\bf 66} (1991) 2051.
\bibitem{Selberg}
  A.~Selberg,
  Bemerkninger om et multiplet integral, 
  Norsk Mat.\ Tidsskr.\ \textbf{26} (1944) 71-78.
\bibitem{Dotsenko:1984nm}
  V.~S.~Dotsenko and V.~A.~Fateev,
  Conformal algebra and multipoint correlation functions in  2D statistical
  models,
  Nucl.\ Phys.\ B {\bf 240} (1984) 312.
\bibitem{Dotsenko:1984ad}
  V.~S.~Dotsenko and V.~A.~Fateev,
  Four point correlation functions and the operator algebra in the
  two-dimensional conformal invariant theories with the central 
  charge $c<1$,
  Nucl.\ Phys.\ B {\bf 251} (1985) 691.
\bibitem{Fateev:2006:JETP}
    V.~A.~Fateev and A.~V.~Litvinov,
    Coulomb integrals in Liouville theory and Liouville gravity,
    JETP Lett.\  {\bf 84}, 531 (2007)
    [Pisma Zh.\ Eksp.\ Teor.\ Fiz.\  {\bf 84} (2006) 625].
\bibitem{Fateev:2007:TMPH}
    V.~A.~Fateev and A.~V.~Litvinov,
    Multipoint correlation functions in Liouville field theory and minimal Liouville gravity,
    Contribution to the proceedings of the International Workshop on Classical and Quantum Integrable
    Systems, Dubna, Russia, January 22--25, 2007, arXiv:0707.1664 [hep-th].
\bibitem{Barnes}
E.~W.~Barnes, The genesis of the double gamma function, Proc.~London.~Math.~Soc., \textbf{31} 358 (1899),
The theory of the double gamma function, Phil.~Trans.~Roy.~Soc., \textbf{A196} 265 (1901).
\bibitem{Fateev:1987vh}
  V.~A.~Fateev and A.~B.~Zamolodchikov,
  Conformal quantum field theory models in two-dimensions having Z(3)
  symmetry,
  Nucl.\ Phys.\ B {\bf 280} (1987) 644.
\bibitem{Bajnok:1992nj}
  Z.~Bajnok, L.~Palla and G.~Takacs,
  A(2) Toda theory in reduced WZNW framework and the representations of the W
  algebra,
  Nucl.\ Phys.\  B {\bf 385}, 329 (1992)
  [arXiv:hep-th/9206075].
\bibitem{Bowcock:1992gt}
  P.~Bowcock and G.~M.~T.~Watts,
  Null vectors of the W(3) algebra,
  Phys.\ Lett.\  B {\bf 297}, 282 (1992)
  [arXiv:hep-th/9209105].
\bibitem{Fateev:2000ik}
  V.~Fateev, A.~B.~Zamolodchikov and A.~B.~Zamolodchikov,
  Boundary Liouville field theory. I: Boundary state and boundary  
  two-point function,
  arXiv:hep-th/0001012.
\bibitem{Seiberg:1990eb}
  N.~Seiberg,
  Notes on quantum Liouville theory and quantum gravity,
  Prog.\ Theor.\ Phys.\ Suppl.\  {\bf 102} (1990) 319.
\bibitem{Zograf}
P.~Zograf and L.~Takhtajan, 
Action of the Liouville equation is a generating function for the accessory parameters and the potential of the Weil-Petersson metric on the Teichm\"uller space,
Funct.\ Anal.\ Appl. \textbf{19} (1986) 219.
\bibitem{Olshanetsky:1981dk}
  M.~A.~Olshanetsky and A.~M.~Perelomov,
  Classical integrable finite dimensional systems related to 
  Lie algebras,
  Phys.\ Rept.\  {\bf 71}, 313 (1981).
\bibitem{Kostant}
B.~Kostant,
The Solution to a Generalized Toda Lattice and Representation Theory, Adv.\ in Math.\ {\bf34}, 195 (1979).
\bibitem{Kharchev:2000ug}
  S.~Kharchev and D.~Lebedev,
  Eigenfunctions of $GL(N,R)$ Toda chain: The Mellin-Barnes
  representation,
  Pisma Zh.\ Eksp.\ Teor.\ Fiz.\  {\bf 71}, 338 (2000)
  [JETP Lett.\  {\bf 71}, 235 (2000)]
  [arXiv:hep-th/0004065].
\bibitem{Stade:1990}
  E.~Stade, On explicit integral formulas for $\mathfrak{gl}(n)$ 
  Whittaker functions,
  Duke.\ Math.\ J. {\bf 60}, 313 (1990). 
\bibitem{Takh-Vin}
L.~Takhtajan and A.~Vinogradov, Theory of the Eisenstein series for the group $SL(3,R)$ and its application to the binary problem I, Notes of the LOMI seminars, \textbf{76} 5 (1978).
\bibitem{Stade:1993}
  E.~Stade, Hypergeometric series and Euler  factor at infinity for 
  $L$-functions on $\mathfrak{gl}(3)\times
  \mathfrak{gl}(3)$,
  Am.\ J.\ Math {\bf 115}, 371 (1993).
\bibitem{Stade:2002}
  E.~Stade, Archimedean $L$-factor on 
  $\mathfrak{gl}(n)\times\mathfrak{gl}(n)$ and generalized Barnes 
  integrals,
  Isr.\ J.\ Math {\bf 127}, 201 (2002).
\bibitem{Baseilhac:1998eq}
  P.~Baseilhac and V.~A.~Fateev,
  Expectation values of local fields for a two-parameter family of
  integrable models and related perturbed conformal field theories,
  Nucl.\ Phys.\ B {\bf 532} (1998) 567
  [arXiv:hep-th/9906010].
\end{thebibliography}
\end{document}